\documentclass[aps,pra,twocolumn,superscriptaddress,longbibliography]{revtex4-1}%
\usepackage{graphicx}
\usepackage{float}
\usepackage{dcolumn}
\usepackage{bm,color}
\usepackage{amsmath,amssymb,dsfont,amstext,amsfonts}
\usepackage{extarrows}
\usepackage[colorlinks=true,linkcolor=blue,urlcolor=blue,citecolor=blue]{hyperref}
\usepackage{xcolor}
\usepackage{wasysym}
\usepackage{mathtools}
\usepackage{bbold}
\usepackage{tikz}
\usepackage[normalem]{ulem} 
\usepackage{color}

\def\mod{\:\mathrm{mod}\:}

\usepackage{pifont}% http://ctan.org/pkg/pifont

\makeatletter
\newcommand{\subalign}[1]{%
  \vcenter{%
    \Let@ \restore@math@cr \default@tag
    \baselineskip\fontdimen10 \scriptfont\tw@
    \advance\baselineskip\fontdimen12 \scriptfont\tw@
    \lineskip\thr@@\fontdimen8 \scriptfont\thr@@
    \lineskiplimit\lineskip
    \ialign{\hfil$\m@th\scriptstyle##$&$\m@th\scriptstyle{}##$\hfil\crcr
      #1\crcr
    }%
  }%
}
\makeatother

\usepackage{soul}

\begin{document}
\title{Topological correspondence between magnetic space group representations and subdimensions}
\author{Adrien Bouhon}
\email{adrien.bouhon@gmail.com} 
\affiliation{Nordic Institute for Theoretical Physics (NORDITA), Stockholm, Sweden}
\affiliation{Department of Physics and Astronomy, Uppsala University, Box 516, SE-751 21 Uppsala, Sweden}
\author{Gunnar F. Lange}
\affiliation{TCM Group, Cavendish Laboratory, University of Cambridge, J.J.~Thomson Avenue, Cambridge CB3 0HE, United Kingdom}
\author{Robert-Jan Slager}
\email{rjs269@cam.ac.uk} 
\affiliation{TCM Group, Cavendish Laboratory, University of Cambridge, J.J.~Thomson Avenue, Cambridge CB3 0HE, United Kingdom}
%\affiliation{Department of Physics, Harvard University, Cambridge MA 02138, USA}

%%TC:ignore
\date{\today}
\begin{abstract}
The past years have seen rapid progress in the classification of topological materials. These diagnostical methods are increasingly getting explored in the pertinent context of magnetic structures. We report on a general class of electronic configurations within a set of anti-ferromagnetic-compatible space groups that are necessarily topological. Interestingly, we find a systematic correspondence between these anti-ferromagnetic phases to necessarily nontrivial topological ferro/ferrimagnetic counterparts that are readily obtained through physically motivated perturbations. Addressing the exhaustive list of magnetic space groups in which this mechanism occurs, we also verify its presence on planes in 3D systems that were deemed trivial in existing classification schemes. This leads to the formulation of the concept of {\it subdimensional topologies}, featuring non-triviality within part of the system that coexists with stable Weyl points away from these planes, thereby uncovering novel topological materials in the full 3D sense that have readily observable features in their bulk and surface spectrum.
\end{abstract}
\maketitle
%%TC:endignore 

%Interestingly, these anti-ferromagnetic cases relate to ferro/ferrimagnetic counterparts, the correspondence of which manifests itself by guaranteeing topological non-triviality in the latter and can be tuned by physically relevant perturbations. 

\section{Introduction}
With the advent of topological insulators (TIs) -- gapped quantum matter having a topological entity by virtue of symmetry -- the past years have seen a reinvigorated interest in band theory. Time reversal symmetry (TRS) has played a major role in these developments, standing at the basis of the developments of the first models of the general notion of symmetry protected states~\cite{Rmp1,Rmp2}. More recently, the interplay with crystalline symmetries has provided a plethora of topological characterizations~\cite{Clas2,Clas1, codefects2,Wi2, Nodal_chains, HolAlex_Bloch_Oscillations, BJYNogaoseDSM,Wannierfloquet2020, Unal19_PRR,ShiozakiSatoGomiK,  Chenprb2012,  Borgnia, Vanderbilt_WCC, vdbilt1, probes_2D,  scheurer, vdbilt2, Codefects1,regdef, UnifiedBBc}. In particular, it was found that a substantial fraction of topological materials can be diagnosed by refined symmetry eigenvalue methods. Heuristically this pertains to considering combinatorial constraints between high symmetry momenta in the Brillouin zone (BZ), which  can be shown to reveal classes of band structures that actually match the full machinery of K-theory analysis in certain cases~\cite{Clas3}, and then comparing them to real space atomic limits in order to define non-triviality with respect to this reference~\cite{Clas4, Clas5}.

Despite the crucial role of TRS, arguably the most paradigmatic TI model actually involves the formulation of TRS-breaking Chern bands~\cite{Haldane1988}, manifesting the original inspiration of these pursuits by Quantum Hall effects.  Hence, it is of natural interest to consider the role of magnetism in combination with the above recent developments. While the interplay of topology and magnetism entails a vast and established literature, ranging from spin liquids to axion insulators \cite{Axion1,Axion2,vanderbilt_axion,Axion3,Axion4,Axion5,Axion6}, there have been rather fruitful results on both essential symmetry eigenvalues indicated schemes \cite{xiaoxing}, that is symmetry indicators \cite{mSI} and, very recently, topological quantum chemistry \cite{mtqc}. 

%\textcolor{black}{by expanding the Hilbert space to include (appropriately chosen) trivial bands}
Already in the non-magnetic case the refined evaluations resulted in new insights. In particular, the discrepancies between different approaches culminated in the formulation of the concept of fragile topology~\cite{Ft1}. These are topological invariants that, unlike stable counterparts, characterize \textcolor{black}{band-subspaces separated by energy gaps from the other bands that can be trivialized upon the closing of the gaps \cite{bouhon2019wilson,Peri797,song2019fragile,BYJ_fragile}. Of particular interest are systems with $PT$ or $C_2T$ symmetry that were early characterized through a stable $\mathbf{Z}_2$ invariant \cite{Fu_3D_C2T,Fu_C2T_nodal,BJY_2D_weyls,Zhao_PT}, and more recently through a fragile $\mathbf{Z}$ invariant \cite{bouhon2019wilson,Tomas_tables_2017} given by the Euler class \cite{BJY_linking,BJY_nielsen,bouhonGeometric2020} for which new physical effects have been predicted \cite{Eulerdrive}. In fact, taking into account multiple gap conditions~\cite{bouhonGeometric2020}, these} phases go beyond any symmetry eigenvalue indicated notion and relate to the momentum space braid trajectories of non-Abelian frame-charge characterized spectral nodes~\cite{Wu1273,BJY_nielsen,bouhon2019nonabelian}. \textcolor{black}{The role of $C_{2}T$ symmetry has also been pointed out in the nontrivial topology of the low energy bands of twisted bilayer graphene \cite{Song_TBLG,Potwisted}.}

Here we revert to the question what physical implications the extension to magnetic space groups (MSG) symmetries can bring within the above context. 
To this end we start by a case study in space group family (SG) 75 and find that within a magnetic background some Wyckhoff positions necessarily imply non-triviality. Turning to anti-ferromagnetic case we, for the first time,  find a model exhibiting fragile and Euler class topology in a MSG and identify the protecting symmetries as well as defining quantities. The magnetic symmetry defining anti-ferromagnetic (AFM) order is however broken upon adding a generic Zeeman term, giving rise to a ferro/ferrimagnetic-compatible (FM) phase within the same family. The correspondence subsequently manifests itself  by conveying that the fragile topological nature has to translate into bands of finite Chern number in the FM counterpart. Going beyond fragile topology, we find that the other remaining possibility of this configuration entails a symmetry protected Weyl semi-metallic phase phase, characterized by a quantized $\pi$-Berry phase. The stable nodal phase corresponds to finite $\mathbf{Z}_2$ symmetry indicator \cite{mSI} that is protected by the combination of the $C_4$ and $C_2T$ symmetries. We moreover show that this phase \textcolor{black}{possesses a systematic} correspondence to \textcolor{black}{a nontrivial Chern insulating} FM counterpart \textcolor{black}{at half-filling, characterized by an even Chern number $\mathcal{C} = 2\mod 4$}. We then generalize our findings by formulating an exhaustive list of tetragonal MSGs featuring this necessarily present topological configurations and their systematic correspondences relating AFM and FM counterparts. \textcolor{black}{We moreover address the effect of adding and removing unitary symmetries leading to the identification of magnetic Dirac points \cite{Wieder_magnetic_Dirac}, and the generalization of the $C_2T$ protected Weyl semi-metallic phases to numerous MSGs}. Most importantly, we find that this mechanism can occur on planes in 3D systems that were previously diagnosed as trivial. At the crux of the argument lies that the in-plane topology must \textcolor{black}{coexist with symmetry indicated} nodes away from the subdimensional regions, such that the 3D conditions appear trivial. Nonetheless, these {\it subdimensionally enriched topological nodal topologies} exhibit robust topological features, such as \textcolor{black}{corner modes plus Fermi arcs in the subdimensional gapped fragile AFM case \cite{lange2021subdimensional}, or Fermi arcs plus Fermi arcs in the subdimensional Weyl nodal AFM case}, and thus pinpoint to a new class of \textcolor{black}{gapped-nodal} topological materials to explore.

\begin{figure}
    \centering
    \includegraphics[width=\linewidth]{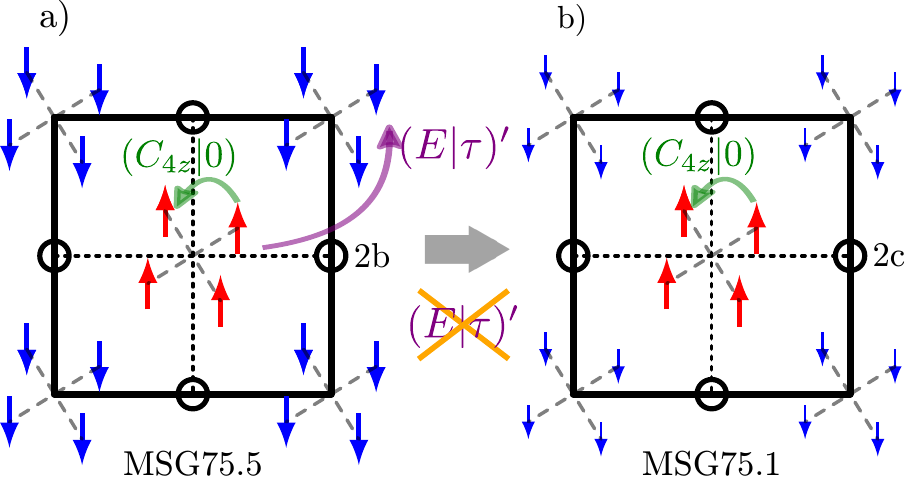}
    \caption{Background magnetic structures for MSG75.5 and 75.1 with Wyckoff positions of interest indicated in standard notation. The latter MSG is effectively realized by making the oppositely-oriented moments (red/blue spins \textcolor{black}{aligned with the $\hat{z}$-axis perpendicular to the plane}) in the former of unequal magnitude. This thus relates an antiferromagnetic-compatible configuration of the SG75-family with a ferro/ferrimagnetic one.}
    \label{fig:mag_PT}
\end{figure}

\section{Magnetic space group and magnetic structure: a first case study in SG P4 (No. 75)}
To concretize matters we depart from a simple model for the tetragonal magnetic space group \textcolor{black}{$P_C4$ (MSG No. 75.5 using the BNS convention)} \cite{BradCrack,MSGtables}. The MSG can be decomposed into left cosets as $\mathcal{G}_{75.5}/\mathcal{G}_{75.1} = (E\vert 0) \mathcal{G}_{75.1} + (E\vert \tau)' \mathcal{G}_{75.1}$, where the space group $\mathcal{G}_{75.1} = C_4 \times \mathbf{T}$ (\textcolor{black}{$P4$, No. 75.1 in the BNS convention}) has no anti-unitary symmetry \cite{BradCrack} ($E$ is the identity, the prime $(\cdot)'$ stands for time reversal, and $\tau=\boldsymbol{a}_1/2+\boldsymbol{a}_2/2$). MSG75.1 \textcolor{black}{(P4)} has point group $C_4$ with the normal subgroup of translations $\mathbf{T}$ corresponding to the primitive tetragonal Bravais lattice $ \{ n_1 \boldsymbol{a}_1 + n_2 \boldsymbol{a}_2+n_3 \boldsymbol{a}_3\}_{n_1,n_2,n_3\in \mathbb{Z}}$ where the primitive vectors are $\boldsymbol{a}_1 = a (1,0,0)$, $\boldsymbol{a}_2 = a (0,1,0)$, and $\boldsymbol{a}_3 = c (0,0,1)$. In the following we first focus on the two-dimensional projection $z=0$, namely we study the corresponding magnetic layer group \textcolor{black}{$p_c4$} (denoted MLG49.4.357 in \cite{MSGtables}). In addition, we use the fact that MSG75.5 \textcolor{black}{($P_C4$)} is generated by $(C_{4z}\vert 0) \mathbf{T}$ ($C_{4z}$ is the rotation by $\pi/2$ around the $z$-axis in the positive trigonometric orientation) and $(E\vert \tau)' \mathbf{T}$ (we will call $(E\vert \tau)'$ a \textit{non-symmorphic}  time reversal symmetry (TRS) as it contains a fractional translation). 

Generally, magnetic space groups with {\it non-symmorphic TRS}, called Shubnikov space groups of type IV \cite{BradCrack}, correspond to AFM structure. Writing $(\boldsymbol{r},\boldsymbol{m})$ for a magnetic moment $\boldsymbol{m}$ located at $\boldsymbol{r}$, the action of $(E\vert \tau)'$ gives $^{(E\vert \tau)'}(\boldsymbol{r},\boldsymbol{m}) = (\boldsymbol{r}+\tau,-\boldsymbol{m})$, and the square $^{[(E\vert \tau)']^2}(\boldsymbol{r},\boldsymbol{m}) = (\boldsymbol{r}+\boldsymbol{a}_1+\boldsymbol{a}_2,\boldsymbol{m})$, i.e. the moment is conserved under translation by a Bravais lattice vector while it is flipped under a fractional translation. Hence MSG75.5 \textcolor{black}{($P_C4$)} is compatible with the AFM structure drawn over one unit cell in Fig.~\ref{fig:mag_PT}a), where all the moments of equal sign \textcolor{black}{(pointing in the direction of the vertical $\hat{z}$-axis)} are obtained under the action of elements generated by $(C_{4z}\vert 0)\mathbf{T}$. In the following, we assume the existence of a magnetic background and describe its effect on the band structure's topology of itinerant electrons.  We note that such magnetism can be obtained directly as localized atomic magnetism within density functional theory frameworks \cite{Kubler_magnetism}, or as the solution of an effective spin Hamiltonian mapped from the Green's functions of interacting electrons \cite{Eriksson_spin_dynamics,ASD}. \textcolor{black}{Alternatively, effective electronic tight-binding Hamiltonians were derived from the double exchange model, i.e.~non self-interacting electrons coupled through Hund's coupling to local classical magnetic moments that interact via super-exchange coupling \cite{ALONSO2001587}, where the electron spins anti-align with the local moments, and for which line-nodal semimetallic phases were recently discussed \cite{Geilhufe_MNL}.}

\subsection{Necessary crystalline fragile antiferromagnetic topology}
%In the following we consider Wyckoff position $2b$ of MSG75.5 spanned by two sub-lattice sites ($A$ and $B$) \cite{MSGtables}, where each hosts one $s$-electronic orbital and two spin-$z$-$1/2$ components 
Adopting \textcolor{black}{maximal} Wyckoff position $2b$ \cite{MSGtables}, spanned by the sub-lattice sites $\boldsymbol{r}_A = \boldsymbol{a}_1/2$ and $\boldsymbol{r}_B = \tau-\boldsymbol{r}_A = \boldsymbol{a}_2/2$, and setting one $s$-electronic orbital and both spin-$z$-$1/2$ components per site, we define the corresponding Bloch orbital basis functions 
\begin{equation}
    \vert \varphi_{\alpha,\sigma} , \boldsymbol{k}\rangle = \sum_{\boldsymbol{R}\in \mathbf{T}} \mathrm{e}^{\mathrm{i} \boldsymbol{k} \cdot (\boldsymbol{R}+\boldsymbol{r}_{\alpha})} \vert w_{\alpha,\sigma} , \boldsymbol{R}+\boldsymbol{r}_{\alpha} \rangle,
\end{equation}
with $\alpha = A,B$ and $\sigma=\uparrow, \downarrow$ (taking $\hat{z}$ as the quantization axis of the spins). Ordering the degrees of freedom as $\boldsymbol{\varphi} = (\varphi_{A,\uparrow},\varphi_{A,\downarrow},\varphi_{B,\uparrow},\varphi_{B,\downarrow})$, the generators of MSG75.5 \textcolor{black}{($P_C4$)}, i.e.~$C_{4z}$ rotation and non-symmorphic time reversal, are then represented through 
\begin{equation}
\begin{aligned}
    \langle \boldsymbol{\varphi} , D_{\pi/2}\boldsymbol{k} \vert ^{(C_{4z}\vert 0 )}\vert \boldsymbol{\varphi} , \boldsymbol{k} \rangle &=  (\sigma_x \otimes M_4), \\
    \langle \boldsymbol{\varphi} , -\boldsymbol{k} \vert ^{(E\vert \tau )'}\vert \boldsymbol{\varphi} , \boldsymbol{k} \rangle &= \mathrm{e}^{\mathrm{i} \boldsymbol{k}\cdot \tau}  (\sigma_x \otimes -\mathrm{i}\sigma_y ) \mathcal{K},
\end{aligned}
\end{equation}
where $M_4 = \mathrm{diag}[\mathrm{e}^{-\mathrm{i} \pi/4},\mathrm{e}^{\mathrm{i} \pi/4}]$, and $D_{\pi/2}$ is the 3D rotation matrix by an angle $\pi/2$ around the $k_z$-axis, $\{\sigma_i\}_{i=x,y,z}$ are the Pauli matrices, and $\mathcal{K}$ is complex conjugation. Combining the two generators, we also obtain 
\begin{equation}\langle \boldsymbol{\varphi} , -D_{\pi/2}\boldsymbol{k} \vert ^{(C_{4z}\vert \tau )'}\vert \boldsymbol{\varphi} , \boldsymbol{k} \rangle = \mathrm{e}^{\mathrm{i} D_{\pi/2}\boldsymbol{k}\cdot \tau}  (\mathbb{1} \otimes -\mathrm{i}\sigma_y M_4^* ) \mathcal{K}.
\end{equation}
It follows that the orbit of the action of the symmetries $\{(E\vert 0), (C_{4z}\vert 0 ), (E\vert \tau )', (C_{4z}\vert \tau )'\}$ on $\varphi_{A,\uparrow}$ is $\{\varphi_{A,\uparrow},\varphi_{A,\downarrow},\varphi_{B,\uparrow},\varphi_{B,\downarrow}\}$, i.e.~all four degrees of the freedom are intertwined by the symmetries of MSG75.5 \textcolor{black}{($P_C4$)}. It can be easily checked that this remains true for any change of basis (i.e.~under any general rotation among the sub-lattice and spinor components).  \textcolor{black}{Furthermore, it can be verified that the atomic orbitals cannot be moved to any other Wyckoff position without breaking the symmetries.} \textcolor{black}{We therefore conclude that} $\{\vert \boldsymbol{\varphi},\boldsymbol{k} \rangle\}_{\boldsymbol{k}\in\mathrm{BZ}}$ defines a four dimensional \textit{elementary band representation} (EBR) \cite{Zak_EBR1,Zak_EBR2,Zak_EBR3, Clas5} as it is formed by the \textit{minimal} set of localized (atomic like) orbitals at the sites $2b$ that is compatible with the magnetic space group symmetries, and we denote it EBR$^{2b}_{75.5}$. \textcolor{black}{This agrees with Ref.~\cite{mtqc} which lists $2b$ as a maximal Wyckoff position and excludes this EBR from the exceptional composite EBRs.}

A minimal tight-binding model for EBR$^{2b}_{75.5}$ is given by
\begin{equation}
\label{model_75_5}
\begin{aligned}
    &H(\boldsymbol{k}) = t_1 f_1(\boldsymbol{k}) \sigma_z\otimes\sigma_z \\
    &+ t_2 f_2(\boldsymbol{k}) \sigma_y\otimes\mathbb{1}+ t_3 f_3(\boldsymbol{k})   \sigma_x\otimes\mathbb{1} \\
    &+ \lambda_1 g_1(\boldsymbol{k}) \mathbb{1}\otimes\sigma_+ + \lambda_1^*   g_1^*(\boldsymbol{k}) \mathbb{1}\otimes\sigma_- \\
    &+
    \lambda_2 g_2(\boldsymbol{k}) \sigma_x\otimes\sigma_+ +
    \lambda_2^*  g_2^*(\boldsymbol{k}) \sigma_x\otimes\sigma_- ,
\end{aligned}
\end{equation}
with $\sigma_{\pm} = (\sigma_x \pm \mathrm{i} \sigma_y)/2$ and the lattice form factors
\begin{equation}
    \begin{array}{ll}
        f_1  = \cos \boldsymbol{a}_1\boldsymbol{k} -\cos \boldsymbol{a}_2\boldsymbol{k} ,&
        g_1  = \sin \boldsymbol{a}_1 \boldsymbol{k} - \mathrm{i}\sin \boldsymbol{a}_2 \boldsymbol{k},\\
        f_2  = \cos \boldsymbol{\delta}_1 \boldsymbol{k}-
        \cos \boldsymbol{\delta}_2 \boldsymbol{k} ,
        & g_2  = \sin \boldsymbol{\delta}_1 \boldsymbol{k} - \mathrm{i}\sin \boldsymbol{\delta}_2 \boldsymbol{k} ,\\
         f_3  = \cos \boldsymbol{\delta}_1 \boldsymbol{k} + \cos \boldsymbol{\delta}_2 \boldsymbol{k}, &
    \end{array}
\end{equation}
defined in terms of the bond vectors $\boldsymbol{\delta}_{\left(\substack{1\\2}\right)} = (\boldsymbol{a}_1 \left(\substack{-\\+}\right) \boldsymbol{a}_2)/2$. It is assumed that $\{t_1,t_2,t_3\}$ are real, while $\{\lambda_1,\lambda_2\}$ can be complex. In the following, we first set $t_1, t_2, t_3 = 1$, and $\lambda_1,\lambda_2 = (1/2) \mathrm{e}^{\mathrm{i} \pi/5} $.

Of importance for the analysis of the band structure and its topology are the squares of the twofold anti-unitary symmetries \cite{Shiozaki14,Shiozaki15}. The non-symmorphic time reversal squares as
\begin{equation}
    \langle \boldsymbol{\varphi} , \boldsymbol{k} \vert ^{[(E\vert \tau)']^2} \vert \boldsymbol{\varphi} , \boldsymbol{k} \rangle = - \mathrm{e}^{-\mathrm{i} \boldsymbol{k}\cdot 2\tau} \mathbb{1}_{4\times 4}, 
\end{equation}
from which we get 
\begin{equation}
        \langle \boldsymbol{\varphi} , \Gamma \vert ^{[(E\vert \tau)']^2} \vert \boldsymbol{\varphi} , \Gamma \rangle = \langle \boldsymbol{\varphi} , \text{M} \vert ^{[(E\vert \tau)']^2} \vert \boldsymbol{\varphi} , \text{M} \rangle = -\mathbb{1}_{4\times4} .
\end{equation}
We thus conclude that there must be a twofold Kramers degeneracy at $\Gamma$ and M, and we call them TRIM (time reversal invariant momentum) in the following. Combining the non-symmorphic time reversal with $C_{2z}$, we get $(C_{2z}\vert \tau)'$ that is represented through
\begin{equation}
    \langle \boldsymbol{\varphi},-C_{2z}\boldsymbol{k}\vert^{(C_{2z}\vert \tau)'}\vert \boldsymbol{\varphi},\boldsymbol{k}\rangle =   \mathrm{e}^{\mathrm{i} D_{\pi} \boldsymbol{k} \cdot \tau} \left(\sigma_x \otimes \mathrm{i} \sigma_x\right) \mathcal{K},
\end{equation}
and squares as
\begin{equation}
    \langle \boldsymbol{\varphi},\boldsymbol{k}\vert^{[(C_{2z}\vert \tau)']^2}\vert \boldsymbol{\varphi},\boldsymbol{k}\rangle = \mathbb{1}_{4\times 4}.
\end{equation}
The existence of such an antiunatry symmetry that leaves the momentum invariant and squares to $+1$ implies that there exists a change of orbital basis in which the Hamiltonian is real symmetric \cite{bouhon2019nonabelian}. This is here achieved through $\widetilde{H}(\boldsymbol{k}) = V\cdot H(\boldsymbol{k}) \cdot V^{\dagger} $ where $V = \sqrt{\sigma_x \otimes \mathrm{i} \sigma_x}$. We symbolically refer to this symmetry as the ``$C_2T$'' symmetry.

{\def\arraystretch{1.3}  
 \begin{table}[t]
 \caption{\label{coIRREPs} Character table for the  magnetic space group IRREPs of MSG75.1 \textcolor{black}{($P4$)}, and coIRREPs of the unitary symmetries of MSG75.5 \textcolor{black}{($P_C4$)}, at $\Gamma$, M, and X, with $\omega = \text{e}^{\text{i}\pi/4}$. \textcolor{black}{The coIRREPs of MSG75.5 \textcolor{black}{($P_C4$)} are given by the pairing of the two IRREPs of MSG75.1 \textcolor{black}{($P4$)} within the same column (e.g. $\overline{\Gamma}_5\overline{\Gamma}_7= \overline{\Gamma}_5\oplus\overline{\Gamma}_7$).} Retrieved from the Bilbao Crystallographic Server \cite{Bilbao}. \textcolor{black}{The second column gives the spin components located at the Wyckoff position $2c$ of MSG75.1 ($P4$) from which the IRREPs are induced.}\footnote{We note that for any maximal Wyckoff position (WP) of MSG75.1 ($P4$) $\{1a,1b,2c\}$ \cite{Bilbao,mtqc} the spin-$z$ components are good quantum numbers at $\Gamma$ and M, which originates from the fact that the vertical $C_{4z}$-axes (for WPs $1a$ and $1b$) and $C_{2z}$-axes (for WPs $2c$) give natural quantization axes for the spins. At the Wyckoff position $2c$ (compatible with the Wyckoff position $2b$ of MSG75.5), the spin-$z$ $+1/2$ ($+3/2$) induces $\{\overline{\Gamma}_5,\overline{\Gamma}_6,\overline{M}_7,\overline{M}_8 \}$, and the spin-$z$ $-1/2$ ($-3/2$) induces $\{\overline{\Gamma}_7,\overline{\Gamma}_8,\overline{M}_5,\overline{M}_6 \}$.}}%At the Wyckoff position $1a$, the spin-$z$ $+1/2$, $+3/2$, $-1/2$, and $-3/2$ induce, respectively, $\{\overline{\Gamma}_8,\overline{\Gamma}_8\}$, $\{\overline{\Gamma}_7,\overline{\Gamma}_7\}$, $\{\overline{\Gamma}_6,\overline{\Gamma}_6\}$, and $\{\overline{\Gamma}_5,\overline{\Gamma}_5\}$. 
\begin{tabular}{ c |c | c | c|c|c|c|c }
 \hline 
 \hline 
 	  & WP & $\overline{\Gamma}_5$  & $\overline{\Gamma}_6$ &
 	  $\overline{\text{M}}_8$ &
 	  $\overline{\text{M}}_7$ &
 	  $\overline{\text{X}}_3$ &
 	   $\overline{\text{X}}_4$ \\
 	   & $2c$ & $\overline{\Gamma}_7$  & $\overline{\Gamma}_8$ &
 	  $\overline{\text{M}}_5$ &
 	  $\overline{\text{M}}_6$ &
 	   &
 	   \\
 	\hline 
 	 $C_{4z}$ & $\begin{array}{l} \uparrow_z \\ \downarrow_z \end{array}$ & $\begin{array}{l}-\omega^* \\ -\omega \end{array}$ & 
 	 $\begin{array}{l}\omega^* \\ \omega \end{array}$ &
 	 $\begin{array}{r}\omega~\;\\ -\omega^*  \end{array}$ &
 	 $\begin{array}{r}-\omega~\; \\ \omega^*  \end{array}$ &
 	 &
 	 \\
 	\hline
 	$C_{2z}$ & $\begin{array}{l} \uparrow_z \\ \downarrow_z \end{array}$ & $\begin{array}{r}-\text{i} \\ \text{i} \end{array}$ & 
	$\begin{array}{r}-\text{i} \\ \text{i} \end{array}$ &
 	$\begin{array}{r}\text{i} \\ -\text{i} \end{array}$ &
	$\begin{array}{r}\text{i} \\ -\text{i} \end{array}$ &
	-\text{i}&\text{i}
	\\
	\hline
	\hline
 \end{tabular}
\end{table} 
}

The bands are then effectively analyzed using the (co-)irreducible representations at the $\Gamma$, $M$ and $X$ points. These are summarized in Table \ref{coIRREPs} (and discussed further in Appendix \ref{ap:SI_analysis_75.5}).

\begin{figure}
    \centering  
\includegraphics[width=\linewidth]{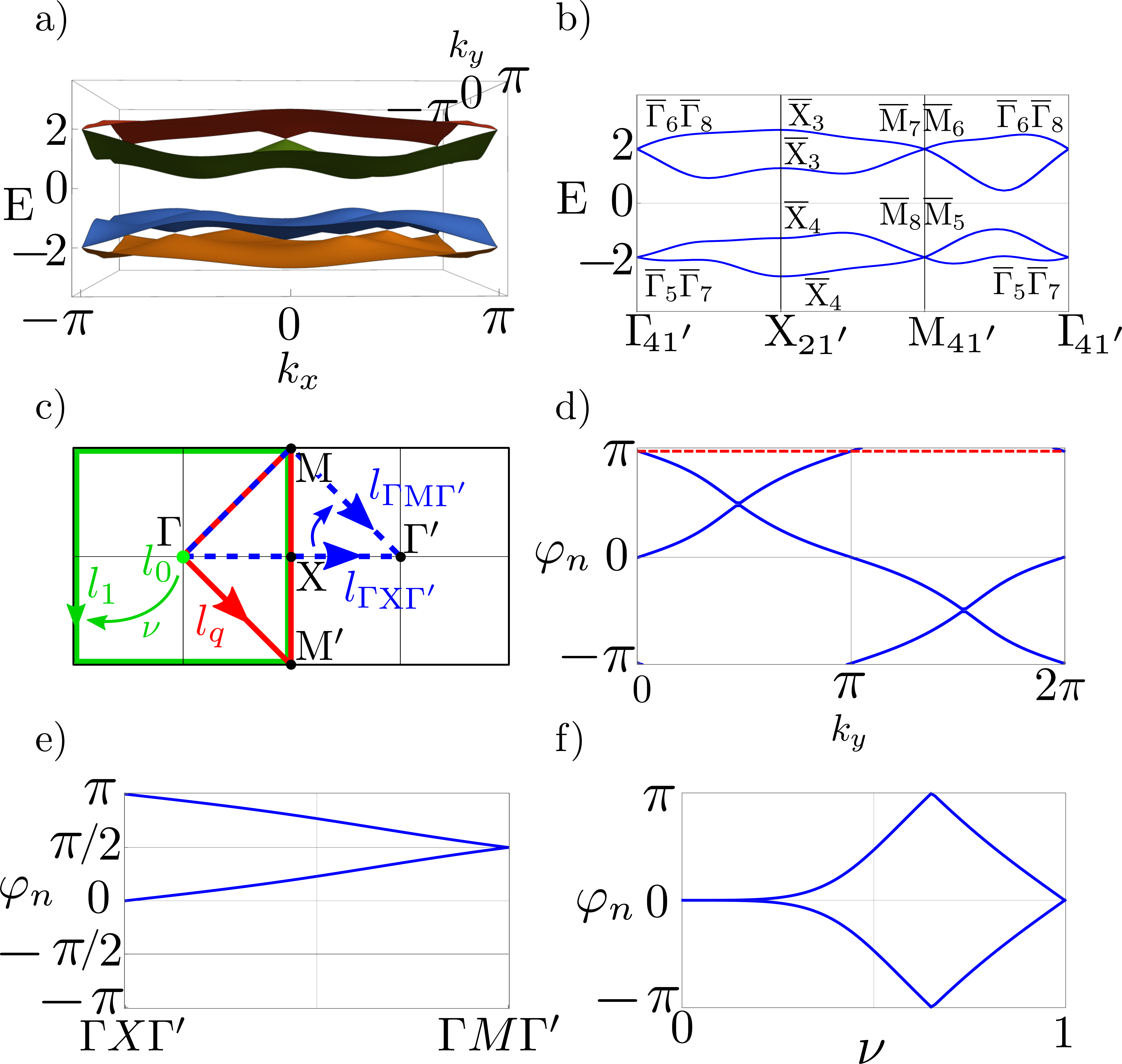}\\
\caption{Non-triviality in MSG75.5.  a) Full band structure of model defined in Eq. \eqref{model_75_5} and b) along the high symmetry directions with coIRREPs indicated. c) Symmetry-based paths within the Brillouin zone used as base loops for the patch Wilson loop (dashed blue), the $C_4$-symmetric Wilson loop (green), and the symmetry indicated Berry phase (red). The small arrows show the direction of flow (deformation of base loops). d) Two-band Wilson loop (blue lines), integrated along $k_x\in[0,2\pi]$, and total Berry phase (red dashed line) for the valence (equivalently, conduction) bands of EBR$^{2b}_{75.5}$. 
Integrating along $k_y\in[0,2\pi]$ gives equivalent results.
e)  Wilson loop flow over a patch from the base loop $\Gamma\text{X}\Gamma'$ to $\Gamma\text{M}\Gamma'$. f) $C_4$-symmetric Wilson loop flow from the point $l_0=\Gamma$ to the boundary of the Brillouin zone, $l_1=\partial \mathrm{BZ}$. Notation follows conventions of the main text.}
\label{fig:wilson75.5}
\end{figure}

Whenever a band structure of an EBR may be split by an energy gap, at least one band subspace must be topological\textcolor{black}{, namely either both band subspaces are stable or fragile topological, or one is trivial and the other must be fragile topological} \cite{Clas5,Ft1,bouhon2019wilson}. Heuristically this is the case because there must be an obstruction forbidding the mapping of  Bloch eigenstates of  EBR subspaces to localized Wannier functions (i.e.~atomic limits) as a result of the space group symmetries, since the dimensionality of any band subspace's Wannier basis (here two) is necessarily smaller than the dimensionality of the by definition minimal EBR (here four). As a consequence, the Wannier functions representing an EBR's subspace are either delocalized if we impose all symmetry constraints, or are incompatible with the space group symmetries. From the induced irreducible co-representations (coIRREPs) and the compatibility relations among these \cite{Clas3,EvarestovSmirnov,Bilbao}, we conclude that EBR$^{2b}_{75.5}$ can be split over all high symmetry regions of the Brillouin zone. We actually obtain a gapped band structure over the whole Brillouin zone in our minimal model, see Fig.~\ref{fig:wilson75.5}a) and b) that gives the ordering in energy of the induced coIRREPs\footnote{The coIRREPs of MSG75.5 are obtained from the IRREPs of MSG75.1 with the degeneracies imposed by the extra anti-unitary symmetries (this is practically determined by the Herring rule \cite{BradCrack}).} (defined in Table \ref{coIRREPs}). The rational behind the splitting of the EBR will be explained when we address the symmetry indicator, which turns out to be trivial for the IRREPs ordering of Fig.~\ref{fig:wilson75.5}b). In the following we refer to the lower (higher) two-band subspace as the \textit{valence} (\textit{conduction}) subspace \textcolor{black}{at half-filling}. The question is then to determine the topology of each gapped subspace as \textcolor{black}{in that case} there is no \textcolor{black}{stable symmetry indicated} topology.

%Importantly, the system has the antiunitary symmetry $(C_{2z}\vert \tau)'$ that leaves the momentum invariant and squares to $+1$, which implies that the Hamiltonian can be made real symmetric through the appropriate change of basis (see Methods) \cite{bouhon2019nonabelian}. We symbolically refer to this symmetry as the ``$C_2T$'' symmetry. 

As pointed out above, the $C_2T$ symmetry implies that the Bloch eigenvectors can be made real through the appropriate change of basis. It follows that (oriented) two-band subspaces of such Hamiltonians are topologically classified by Euler class $\chi \in \mathbb{Z}$ that can be computed as a winding of Wilson loop \cite{BJY_linking,bouhon2019wilson,BJY_nielsen, bouhon2019nonabelian,bouhonGeometric2020}. 
We find that the Wilson loop ($\mathcal{W}[l]$, see Appendix \ref{m:patch_WL}) of each two-band subspace winds along both directions (i.e.~integrating along the base path $l_{k_y} = \{(k_x,k_y)\vert k_x\in[0,2\pi]\}$ and scanning through $k_y\in [0,2\pi]$, and similarly if we exchange $k_x \leftrightarrow k_y$), see Fig.~\ref{fig:wilson75.5}d). We moreover find that the Berry phase ($\mathrm{e}^{\mathrm{i}\gamma_B[l]} = \mathrm{Det}\mathcal{W}[l]$) of both two-band subspaces is $\pi$ along both directions, see Fig.~\ref{fig:wilson75.5}d) (red dashed line), pointing to the non-orientability of the subspaces' frames of Bloch eigenvectors \cite{bouhonGeometric2020}. While the Euler class is not defined strictly speaking for an \textit{un}orientable band subspace \cite{Hatcher_2}, we still obtain the winding of Wilson loop as an element of $\pi_1(\mathrm{O}(2)) = \mathbb{Z}$ since $\mathcal{W}[l]\in \mathrm{O}(2)$ when computed in the real gauge (i.e.~using the Bloch eigenvectors of the real symmetric form) \footnote{Strictly speaking there is reduction $\mathbb{Z}\rightarrow\mathbb{N}$ of the classification as a consequence of the facts that band subspaces are \textit{orientable} and not \textit{oriented}, and that the homotopy classes of gapped Hamiltonians have no fixed base points \cite{bouhonGeometric2020}.}. We thus conclude that each two-band subspace has a non-orientable nontrivial fragile topology \cite{bouhonGeometric2020}. Moreover, we also point out that the non-trivial $\pi$ Berry-phases are actually appealing from a bulk-boundary perspective \cite{Zakphase}. Indeed, they culminate in-gap edge states, reflecting a physical signature\textcolor{black}{, see \cite{lange2021subdimensional} for a detailed analysis}.

We further derive in Appendix \ref{m:patch_WL} the necessary non-triviality of the split EBR$^{2b}_{75.5}$. Following Ref.~\cite{bouhon2019wilson} we show that the crystal symmetries impose a finite fractional winding of Wilson loop over one quarter of the Brillouin zone, i.e.~the patch bounded by the paths $\Gamma\text{X}\Gamma'$ and $\Gamma\text{M}\Gamma'$ (blue dashed lines) in Fig.~\ref{fig:wilson75.5}c). This results from the difference in the symmetry protected quantizations of the Wilson loops over the two base paths, i.e.~$\mathrm{Arg}\left[\mathrm{eig}\{\mathcal{W}[l_{\Gamma\text{X}\Gamma'}]\}\right] = [0,\pi]$ and $\mathrm{Arg}\left[\mathrm{eig}\{\mathcal{W}[l_{\Gamma\text{M}\Gamma'}]\}\right] = [\pi/2,\pi/2]$, which depends on both the IRREPs and the spinor structure of the bands (i.e.~spin-parallel vs. spin-flip parallel transports, see Appendix \ref{m:patch_WL}), as is also verified through direct numerical evaluation of the Wilson loop over the patch in Fig.~\ref{fig:wilson75.5}e). Then, by $C_{4}$ symmetry, the Wilson loop must have a finite integer winding over the whole Brillouin zone, as confirmed by Fig.~\ref{fig:wilson75.5}d). We later refer to it as the \textit{crystalline Euler fragile topology} (written CEF in Table \ref{list}) when we address the generalization to other MSGs. 

We furthermore compute the $C_4$-symmetric Wilson loop flow \cite{bouhon2019wilson,Barry_fragile, ortix} from the point $l_0=\Gamma$ to the contour of the Brillouin zone $l_1 = \partial \text{BS}$, shown in green in Fig.~\ref{fig:wilson75.5}c), and between which we extrapolate by taking the scaled contour $\nu \partial \text{BZ}$ for $\nu\in [0,1]$. This also exhibits a full winding shown in Fig.~\ref{fig:wilson75.5}f). The $C_4$-symmetric Wilson loop winding alludes to the persistence of nontrivial fragile topology after breaking $(C_{2z}\vert 0)'$ symmetry, i.e.~without Euler class. We refer to it as the \textit{crystalline fragile topology} (written CF in Table \ref{list}). %We note that CF topology does not host quantized Wilson loops over the paths $\Gamma\text{X}\Gamma'$ and $\Gamma\text{M}\Gamma'$, contrary to CEF topology. 

%[HERE]We  present additional numerical evaluations of Wilson flows in MSG75.5. In particular, we consider the Wilson loop flow from the base loop $\Gamma\text{X}\Gamma'$ to $\Gamma\text{M}\Gamma'$ and $C_4$-symmetric Wilson loop flow from the point $l_0=\Gamma$ to the contour of the Brillouin zone $l_1 = \partial \text{BS}$, between which we extrapolate by taking the scaled contour $\nu \partial \text{BZ}$ for $\nu\in [0,1]$ \cite{bouhon2019wilson,Barry_fragile, ortix}, see Fig.~\ref{fig:wilson75.5}c). As shown in Fig.~\ref{fig:WL} the former indicates winding as for the patch covering a quarter of the BZ this flow adds to $\pi/2$. Similarly, the $C_4$-symmetric flow shows the anticipated winding indicating non-triviality as described in the main text. 

%\begin{figure}[ht!]
%    \centering
%    \includegraphics[width=\linewidth]{figure_9_combined.pdf}
%    \caption{a)  Wilson loop flow from the base loop $\Gamma\text{X}\Gamma'$ to $\Gamma\text{M}\Gamma'$. b) $C_4$-symmetric Wilson loop flow from the point $l_0=\Gamma$ to the boundary of the Brillouin zone, $l_1=\partial \mathrm{BZ}$. Notation follows conventions of the main text.}
%    \label{fig:WL}
%\end{figure}

These results thus constitute three complementary ways to reveal the {\it necessary non-triviality} of the crystalline fragile topology of the split EBR$^{2b}_{75.5}$.

\subsection{Stable nodal topological phases}

\begin{figure}[t!]
    \centering
    \includegraphics[width=\linewidth]{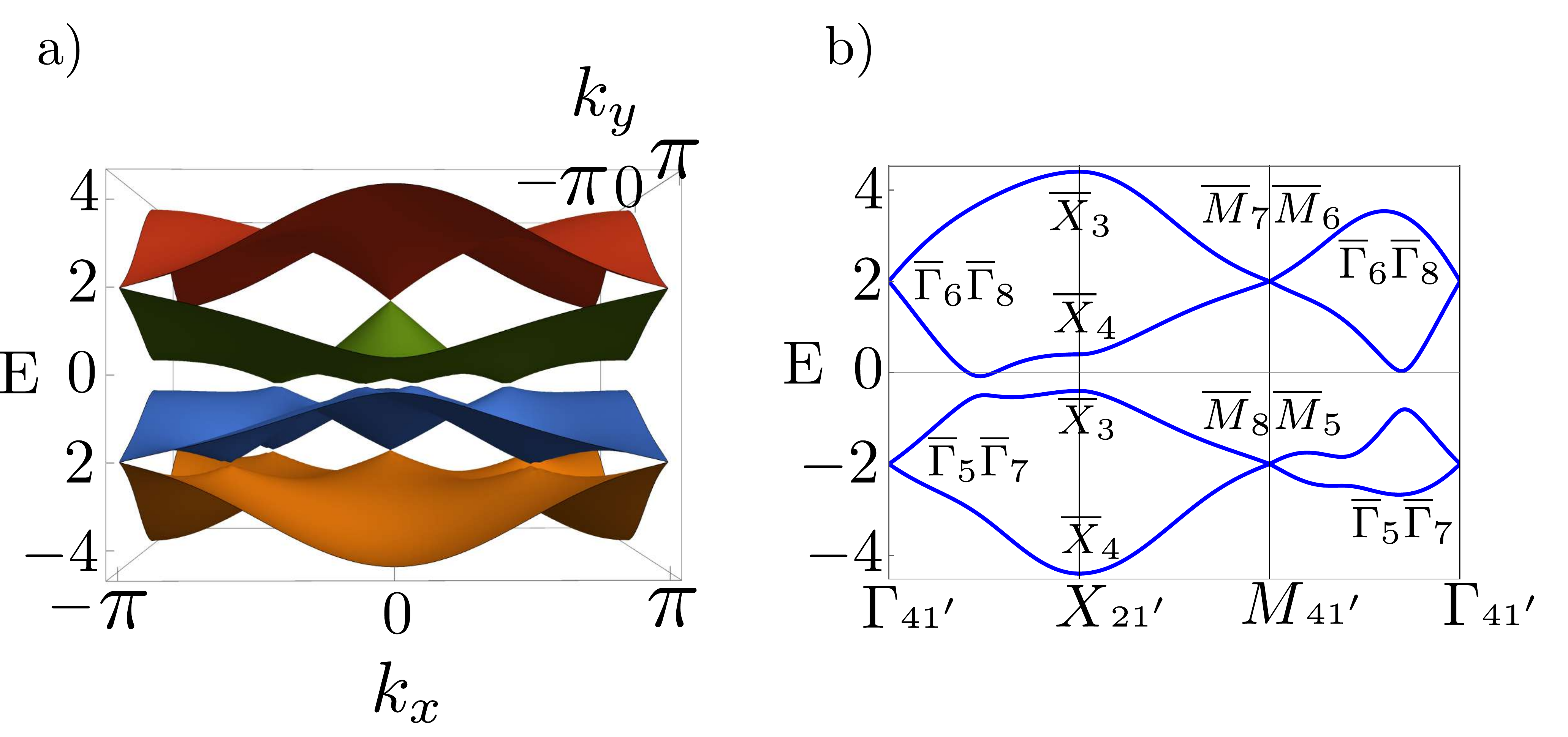}
    \caption{Stable topological semimetallic phase of MSG75.5 \textcolor{black}{($P_C4$)} indicated by $z_2 = 1 \mod 2$, \textcolor{black}{here represented by EBR$^{2b}_{75.5}$ at half-filling}. Full band structure over a) BZ and b) along high symmetry lines where two coIRREPs at X have been inverted when compared to figure \ref{fig:wilson75.5}b).}
    \label{fig:stable_nodal}
\end{figure}

We can characterize the symmetry indicators of a given band structure by using the matrix containing all allowed magnetic EBRs. This results in a $\mathbb{Z}_2$ indicator for MSG75.5 \textcolor{black}{($P_C4$)}. As detailed in Appendix~\ref{ap:SI_analysis}, the explicit expression for this indicator is
\begin{equation}\label{eq:::indicator}
    z_2 = n_{\mathrm{\overline{X}}_3}\ \mathrm{mod\ }2,
\end{equation}
where $n_{\mathrm{\overline{X}_3}}$ is the number of occupied bands at the X-point with the IRREP $\mathrm{\overline{X}_3}$. \textcolor{black}{In agreement with the discussion in the previous section, the indicator is trivial for the fragile phase of EBR$^{2b}_{75.5}$ at half-filling, as can be verified from the coIRREPs of Fig.~\ref{fig:wilson75.5}b). We emphasize that this symmetry indicator readily generalizes for an arbitrary even number of occupied bands, i.e.~at a filling $\nu\in 2\mathbb{Z}+2$}. 

We thus conclude that a stable topological phase can be reached through a band inversion at X. This is achieved for the model Eq.~(\ref{model_75_5}) by taking $ \vert \lambda_2 \vert > \sqrt{2} \vert t_1 \vert$. Setting $\lambda_2 = (6/5)\sqrt{2}$, we obtain the band structure of Fig.~\ref{fig:stable_nodal} that exhibits a semimetallic phase with four nodal points around $\Gamma$ at half-filling. We find that the stable symmetry indicator $z_2$ corresponds to a $\pi$-Berry phase for the valence (conduction) bands along the path $l_q$ [see Fig.~\ref{fig:wilson75.5}(c)], i.e. (see derivation in Appendix \ref{m:derivation_eq4})
\begin{eqnarray}
    \gamma^{(1::2)}_B[l_q] &=&  -\mathrm{i} \log\left[\text{Det}\left(\mathcal{W}^{(1::2)}[l_q]\right)\right] \nonumber\\
    &=&  -\mathrm{i} \log \left[\dfrac{\xi^{\Gamma}_4(1)\xi^{\Gamma}_4(2)   \xi^{\mathrm{M}}_2(1)\xi^{\mathrm{M}}_2(2)}{
    \xi^{\mathrm{M}}_4(1)\xi^{\mathrm{M}}_4(2) \xi^{\mathrm{X}}_2(1)\xi^{\mathrm{X}}_2(2)}\right] \nonumber\\
    &=&  -\mathrm{i} \log \left[\dfrac{(+1)   (+1) }{
    (-1) \xi^{\mathrm{X}}_2(1)\xi^{\mathrm{X}}_2(2)}\right] \nonumber\\
    &=&  -\mathrm{i} \log \left[
    (-1) \xi^{\mathrm{X}}_2(1)\xi^{\mathrm{X}}_2(2)\right] \nonumber\\
    &=& \left\{\begin{array}{ll} 0 \mod 2\pi,& \text{if}~z_2=0,\\
    \pi \mod 2\pi,& \text{if}~z_2=1.
    \end{array}\right.
\label{eq:indicated_berry}
\end{eqnarray}
\textcolor{black}{Let us first note that, similarly to Eq.~(\ref{eq:::indicator}), Eq.~(\ref{eq:indicated_berry}) can also be generalized for an arbitrary even number of occupied bands (i.e.~a filling $\nu\in 2\mathbb{Z}+2$). Importantly, $C_2T$ symmetry (with $[C_2T]^2=+1$) imposes the vanishing of the $U(1)$ Berry curvature over the two-band occupied eigen-subspaces, since within the real basis we have $\mathcal{F} = \mathrm{Pf}[\mathcal{F}] \mathrm{i}\sigma_y$ \cite{bouhon2019nonabelian} and thus $\mathrm{tr}\mathcal{F} \equiv 0$. As a consequence, the Chern number of the gapped AFM phase at half-filling is identically zero. From there results that the nontriviality of the Berry phase indicates a nodal phase (i.e.~it indicates the obstruction to define a smooth projector on the occupied bands over the whole Brillouin zone due to the presence of topologically stable band crossings with the unoccupied bands), i.e.~the necessary existence of an odd number of nodal points inside the domain bounded by $l_q$. Upon the breaking of the non-symmorphic TRS, $C_{2z}T$ is also broken, and the $\pi$ Berry phase indicates a $C_4$-symmetry protected Chern number at half-filling, or more generally at a filling $\nu\in 2\mathbb{Z}+2$,
\begin{equation}
\label{eq_chern_halffilling}
    \mathcal{C} = 2z_2\mod 4 .
\end{equation}
The nontrivial Chern phases are discussed in detail in the next section.}

%We note that Eq.~(\ref{eq:sym_chern}) follows by reducing this expression to a single band-subspace. 

We emphasize that the nodal points at general momenta are  not indicated by the compatibility relations. Indeed, these are stabilized by the $(C_{2z}\vert \tau)'$-symmetry ($C_2T$) for which there is not an eigenvalue structure. Instead, the $C_2T$ symmetry quantizes the Berry phase to the values $\{0,\pi\}$, with $\pi$ indicating an odd number of nodes encircled by $l_q$. Embedded in 3D the nodal points correspond to single Weyl points that are \textit{pinned} on the \textcolor{black}{$C_{2}T$ invariant plane (i.e.~at $k_z=0$ where $- C_{2z} \boldsymbol{k}=IC_{2z} \boldsymbol{k} = m_z\boldsymbol{k} = \boldsymbol{k} $) by virtue of the chirality-preserving property of $C_2T$. Indeed, any Weyl point leaving the $k_z=0$ plane must have a mirror symmetric image with equal chirality by $C_2T$ symmetry. It is therefore forbidden for a single node to leave the plane by conservation of Chern number.} We  refer to these phases in the 3D context as the \textit{crystalline Weyl topology} (written CW in Table \ref{list}).

\section{AFM-FM correspondence}
We now turn to ferro/ferrimagnetic (FM) phases associated with SG75 obtained from the fragile and stable nodal AFM phases discussed above through the breaking of the antiunitary symmetry $(E\vert \tau)'$, thereby effectively realizing MSG75.1 (\textcolor{black}{P4,} Shubnikov type I). This is done in Eq.\eqref{model_75_5} by adding a Zeeman coupling term $\epsilon_{\mathrm{Z}} (\mathbb{1} \otimes \sigma_z)$. We find that the topology of the FM-compatible phases are necessarily nontrivial, exhibiting Chern numbers constrained by crystalline symmetries that intricately relate to the topology of the AFM counterparts.

\textcolor{black}{We note that the correspondence discussed here must be contrasted from the Chern phases obtained under an external magnetic field \cite{Biao_LL} which are in general not symmetry indicated.}

\subsection{General mechanism}
\begin{figure}
\centering
\includegraphics[width=\linewidth]{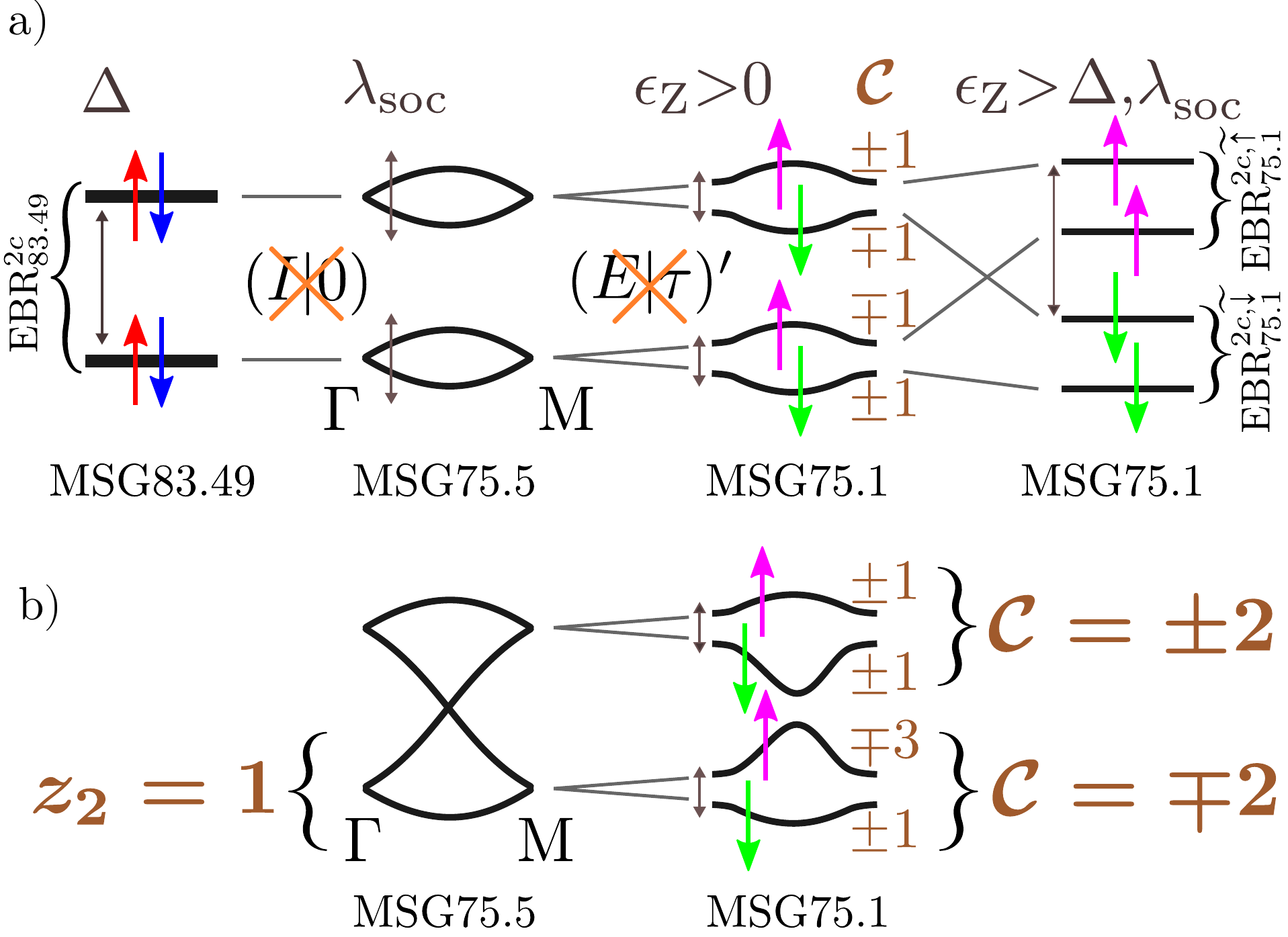}
\caption{a) Splitting and ordering of the EBR's energy levels induced by the successive breaking of inversion $(I\vert 0)$ and non-symmorphic time reversal $(E\vert\tau)'$ symmetries. The split ($t_1,t_2,t_3=\Delta $) four-dimensional EBR of MSG83.49 \textcolor{black}{($P_C4/m$)}, and MSG75.5 \textcolor{black}{($P_C4$)} \textcolor{black}{for $z_2=0$}, separates into \textit{two} split EBRs of MSG75.1 \textcolor{black}{($P4$)} under the combined effect of Dresselhaus-Rashba spin-orbit ($\lambda_1,\lambda_2=\lambda_{\mathrm{soc}} $) and Zeeman ($\epsilon_{\mathrm{Z}}$) couplings, giving rise to energy ordered pseudo-spin-polarized Chern bands ($\mathcal{C} = \mathrm{sign}[t_2]$, brown). \textcolor{black}{Pure spin components are drawn in red and blue (for MSG83.49), pseudo-spin components are drawn in magenta and green for MSG75.1.} Under a strong Zeeman splitting, the valence (conduction) subspaces become fully \textcolor{black}{pseudo-spin-polarized (see text)} while conserving (at fixed $\lambda_{\mathrm{soc}}$) the Chern characters of the bands. \textcolor{black}{b) Zeeman splitting when $z_2=1$ for the EBR of MSG75.5 \textcolor{black}{($P_C4$)} leading to a symmetry indicated nontrivial even Chern insulator ($\mathcal{C}=2\mod4$) at half-filling}.}
\label{fig:mag_energy_levels}
\end{figure}

Let us first generally address the AFM-FM correspondence and its physical mechanisms. For this purpose it is worth starting from MSG83.49 \textcolor{black}{($P_C4/m$)} which is obtained from MSG75.5 by simply adding inversion symmetry, i.e.~SG83 has point group $C_{4h}$. The presence of $(I\vert \tau)'$ symmetry which squares to $-1$ leads to the twofold Kramers degeneracy of the bands over the whole Brillouin zone. The parent EBR, which we write EBR$^{2c}_{83.49}$, also splits \textcolor{black}{with a topology characterized by symmetry indicated mirror Chern numbers \cite{mtqc}}. We readily obtain the corresponding Hamiltonian by taking $\lambda_1,\lambda_2 = 0$ in Eq.~(\ref{model_75_5}). The full splitting of EBR$^{2c}_{83.49}$ requires $\vert t_1 \vert ,\vert t_2\vert, \vert t_3 \vert >0$ (which we symbolize by a single variable $\Delta$ in Fig.~\ref{fig:mag_energy_levels}), where $t_1$ is a spin-$z$-preserving spin-orbit coupling that acts as a delocalized Zeeman coupling on each sub-lattice orbital but changes sign between sub-lattice sites, and $t_2$ and $t_3$ are spin-preserving inter-sub-lattice site couplings. \textcolor{black}{Due to the basal mirror symmetry $\langle\boldsymbol{\varphi},m_z\boldsymbol{k} \vert ^{m_z} \vert \boldsymbol{\varphi},\boldsymbol{k}\rangle=  \mathbb{1}\otimes -\mathrm{i}\sigma_z$, each band-doublet can be separated into the $-\mathrm{i}$ and $\mathrm{i}$ mirror-eigenvalue sectors, matching with the spin-$\uparrow$ and spin-$\downarrow$ components (i.e.~the spin-$z$-components are good quantum numbers over the whole Brillouin zone).} 

The terms in $\{\lambda_1,\lambda_2\}$ in Eq.~\eqref{model_75_5} break inversion symmetry and correspond to combined Dresselhaus and Rashba spin-orbit couplings. The effect of the latter (symbolized by $\lambda_{\mathrm{soc}}$) is to split the Kramers degeneracy away from $\Gamma$ and M, as represented schematically in Fig.~\ref{fig:mag_energy_levels}\textcolor{black}{a) for $z_2=0$ in Eq.~(\ref{eq:::indicator})}. The conservation of Kramers doublets at $\Gamma$ and M is due to the non-symmorphic time reversal which still squares to $-1$ at these points, as derived above. \textcolor{black}{While the bands now have pseudo-spin components at generic momenta, the pure spin-$\uparrow$ and spin-$\downarrow$ components are still good quantum number at $\Gamma$ and M since the terms in $\lambda_{1,2}$ vanish there.} 

The AFM-FM transition can then be modeled through a Zeeman term ($\epsilon_{\mathrm{Z}}$) that breaks the non-symmorphic TRS leading to the splitting of the ($\Gamma$ and M) Kramers doublets. \textcolor{black}{This leads to the pure spin polarization of the bands at $\Gamma$ and M since, for any maximal Wyckoff position of MSSG75.1 ($P$4) \cite{Bilbao,mtqc}, spin-$\uparrow$ and spin-$\downarrow$ induce distinct sets of IRREPs at $\Gamma$ and at M (see the footnote of Table \ref{coIRREPs}). In the following we thus refer to the pseudo-spin-polarizations $\widetilde{\uparrow}$ and $\widetilde{\downarrow}$ of the bands in the sense that the band $\widetilde{\uparrow}$ ($\widetilde{\downarrow}$) at $\boldsymbol{k}$ has the pure spin component $\uparrow$ ($\downarrow$) at $\Gamma$ and M. This does not exclude the case, for dominant spin-flip terms as compared to Zeeman splitting, of a band subspace with an opposite pure spin configuration at $\Gamma$ (say spin-$\uparrow$) and M (spin-$\downarrow$), see the discussion around Eq.~(A6) in Appendix A. Such a configuration typically requires long-range spin-flip terms [36].} This results in energy ordered \textcolor{black}{pseudo-spin-polarized} Chern bands (column $\epsilon_{\mathrm{Z}}>0$ in Fig.~\ref{fig:mag_energy_levels}, Fig.~\ref{fig:MSG75.1}, and Fig.~\ref{fig:berry} in Appendix~\ref{ap:numerical_appendix_75.1}) with the relative chirality of the minimal model set by sign[$t_2$]. Further increasing Zeeman coupling (while keeping $\lambda_{\mathrm{soc}}$ fixed, see below) leads to fully \textcolor{black}{pseudo-spin-polarized} valence and conduction subspaces illustrated in the right column in Fig.~\ref{fig:mag_energy_levels}a).

\textcolor{black}{As a next step, by switching off the spin-flip $\lambda_{1,2}$-terms, while maintaining a dominant Zeeman splitting, we adiabatically map the fully pseudo-spin-polarized bands into pure spin-polarized split EBRs of MSG75.1 \textcolor{black}{($P4$)}.} Indeed, the sub-lattice sites $A$ and $B$ still span a single \textcolor{black}{maximal} Wyckoff position, now labeled $2c$ for MSG75.1 \textcolor{black}{($P4$)} \cite{ITCA,Bilbao}, and from the absence of spin-mixing symmetry we may form spin polarized EBR$^{2c,\uparrow}_{75.1}$, and EBR$^{2c,\downarrow}_{75.1}$, from the orbital basis $(\varphi_{A,\uparrow},\varphi_{B,\uparrow})$, and $(\varphi_{A,\downarrow},\varphi_{B,\downarrow})$, respectively. \textcolor{black}{We write the adiabatic mapping of the fully pseudo-spin-polarized valence and conduction bands into spin-polarized split EBRs as $\mathrm{EBR}^{2c,\widetilde{\uparrow}}_{75.1} \sim \mathrm{EBR}^{2c,\uparrow}_{75.1}$ and $\mathrm{EBR}^{2c,\widetilde{\downarrow}}_{75.1}\sim \mathrm{EBR}^{2c,\downarrow}_{75.1}$.} The symmetry breaking term (Zeeman) thus induces the following phase transition from \textit{one} four-dimensional split EBR (of MSG75.5) to \textit{two} two-dimensional split EBRs (of MSG75.1), 
\begin{equation}
    \mathrm{EBR}^{2b}_{75.5} \longrightarrow \mathrm{EBR}^{2c,\widetilde{\uparrow}}_{75.1} + \mathrm{EBR}^{2c,\widetilde{\downarrow}}_{75.1}
    \sim\mathrm{EBR}^{2c,\uparrow}_{75.1} + \mathrm{EBR}^{2c,\downarrow}_{75.1}.
\end{equation}  
\textcolor{black}{We emphasize that this mapping is model independent, in the sense that it continues to exist when we add any extra term in Eq.~(\ref{model_75_5}) that satisfies the symmetries of MSG75.5 ($P_C 4$) and MSG75.1 ($P4$).} The symmetry broken band structure is now split into four separated bands and the question is to characterize the topology of each single band. 

The symmetry indicated Berry phase for Band $n$ (using the bottom-up labeling of the energy eigenvalues, i.e.~$E_{n}\leq E_{n+1}$) along the path $l_q=\Gamma$M$'$XM$\Gamma$ (M$'$=M-$\boldsymbol{b}_2$), see (in red)  Fig.~\ref{fig:wilson75.5}a), is \cite{Chenprb2012,Wi2,hourglass} (one-band reduction of Eq.~(\ref{eq:indicated_berry}), see derivation in Appendix \ref{m:derivation_eq4}) 
\begin{equation}
\label{eq:sym_chern}
    \gamma^{(n)}_B[l_q] = -\mathrm{i} \log [\xi^{\Gamma}_4(n)
    \xi^{\text{M}}_4(n)^{-1}  
    \xi^{\mathrm{M}}_2(n) \xi^{\mathrm{X}}_2(n)^{-1}],
\end{equation}
where $\xi^{\bar{\boldsymbol{k}}}_4(n)$ and $\xi^{\bar{\boldsymbol{k}}}_2(n)$ are the $C_4$- and $C_2$-eigenvalues, respectively, at the high-symmetry point $\bar{\boldsymbol{k}}$ listed in Table \ref{coIRREPs}. The Chern number of Band $n$, given through $\mathrm{e}^{-\mathrm{i} 2\pi \mathcal{C}(n)} = (\mathrm{e}^{\mathrm{i} \gamma^{(n)}_B[l_q]})^4$, is thus 
\begin{equation}
\label{eq:sym_chernb}
    \mathcal{C}(n) = -(2/\pi) \gamma^{(n)}_B[l_q] \mod 4,
\end{equation}
\textcolor{black}{see also \cite{mtqc}}. We show below that whenever the FM phase is obtained from one of the (necessarily) nontrivial AFM phases of EBR$^{2b}_{75.5}$, it must be made of nontrivial Chern bands. \textcolor{black}{Remarkably, the Zeeman splitting of the stable nodal phase of MSG75.5 ($P_C4$), i.e.~with $z_2=1$ in Eq.~(\ref{eq:::indicator}), necessarily generates a nontrivial Chern FM phases at half-filling, with $\mathcal{C} = 2 \mod 4$ according to Eq.~(\ref{eq_chern_halffilling}), see Fig.~\ref{fig:mag_energy_levels}b). This is discussed in detail below where we show that Eq.~(\ref{eq_chern_halffilling}) matches with Eq.~(\ref{eq:sym_chern},\ref{eq:sym_chernb}) for Band 1 and 2. Below we also use the pseudo-spin polarization to predict single bands of higher Chern number (i.e.~$\mathcal{C} = \pm3\mod4$) in some regime.} In the following we refer to these symmetry indicated Chern phases as the \textit{crystalline Chern topology} (we call it CC topology in the following).

\textcolor{black}{Before we study the AFM to FM phases correspondence for the model Eq.~(\ref{model_75_5}) in more detail, we importantly note that the same nontrivial AFM phases, as well as the AFM to FM correspondence, can be obtained from the following EBRs} \textcolor{black}{(for MSG75.5 ($P_C$4) $\rightarrow$ MSG75.1 ($P$4)):} \textcolor{black}{
\begin{equation}
\begin{aligned}
    \mathrm{EBR}_{2a}^{\uparrow}\oplus \mathrm{EBR}_{2a}^{\downarrow} &\rightarrow \mathrm{EBR}_{1a}^{\uparrow}\oplus \mathrm{EBR}_{1b}^{\downarrow} \oplus \mathrm{EBR}_{1a}^{\downarrow}\oplus \mathrm{EBR}_{1b}^{\uparrow},\\
    \mathrm{EBR}_{4c}^{\rightarrow} &\rightarrow \mathrm{EBR}_{4d}^{\rightarrow}  ,
\end{aligned}
\end{equation}
where $\mathrm{EBR}^{\rightarrow}$ is an EBR formed with non-colinear in-plane spinors $\{({\rightarrow}), C_{4z}({\rightarrow)}, C_{2z}({\rightarrow)}, C^{-1}_{4z}({\rightarrow)} \}$, i.e.~with a quantization axis that is perpendicular to the vertical $C_4$-axis \footnote{In that case however, there is no spin polarization at $\Gamma$ and M, which we relate to the fact that $4d$ is not a maximal Wyckoff position for MSG75.1, i.e.~the non-colinear in-plane spins at $4d$ can be can superposed on top of each other by moving them to another Wyckoff position with higher symmetry.} (we chose $\hat{y}$ in Table \ref{list} with $\rightarrow_y$).}

\subsection{Small Zeeman splitting}
Here we derive the topology of the FM phases obtained from each of the nontrivial AFM phases of EBR$^{2b}_{75.5}$ when the Zeeman splitting is small compared to the other energy scales (i.e.~$\epsilon_{\mathrm{Z}}<\Delta,\lambda_{\mathrm{SOC}}$ in Fig.~\ref{fig:mag_energy_levels}), underpinning the general mechanism outlined previously. 

\subsubsection{Crystalline Chern ferro/ferrimagnetic topology from fragile AFM phase}

\begin{figure}
    \centering  
\includegraphics[width=\linewidth]{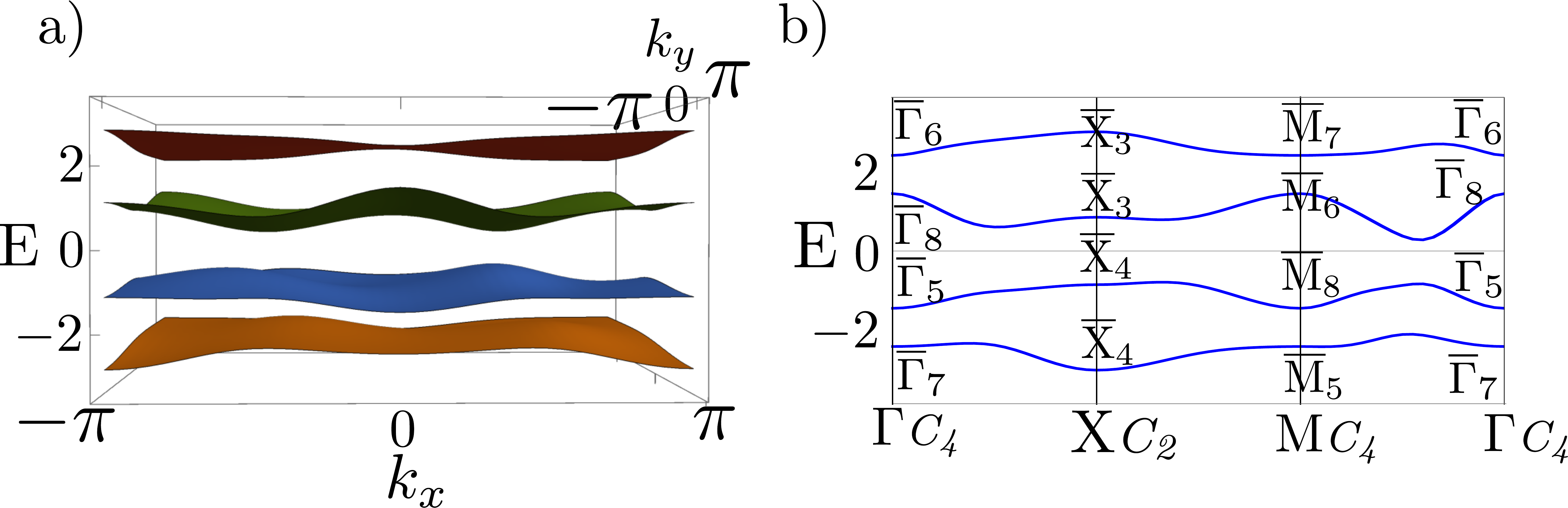}
\caption{Non-triviality in MSG75.1. a) Band structure for MSG75.1 obtained from the model in Eq. \eqref{model_75_5} together with Zeeman coupling, and b) along high-symmetry lines with the IRREPs indicated. We have taken $\epsilon_{\mathrm{Z}} = 1/2$. \textcolor{black}{Applying Eq.~(\ref{eq:sym_chern}) we find that each band hosts a nonzero Chern number}.}
\label{fig:MSG75.1}
\end{figure}

Starting from the gapped fragile phase of EBR$^{2b}_{75.5}$ \textcolor{black}{($z_2=0$)} and given the sign of the Zeeman coupling ($\epsilon_{\mathrm{Z}}>0$, i.e.~$E_{\uparrow_{z}}> E_{\downarrow_{z}} $) we predict the ordering in energy of the IRREPs of each split Kramers degeneracy to be $E(\overline{\Gamma}_7) < E(\overline{\Gamma}_5) < E(\overline{\Gamma}_8) < E(\overline{\Gamma}_6)$, and $E(\overline{\text{M}}_5) < E(\overline{\text{M}}_8) < E(\overline{\text{M}}_6) < E(\overline{\text{M}}_7)$. We show the band structure for MSG75.1 in Fig.~\ref{fig:MSG75.1}a) and b) together with the IRREPs along high-symmetry lines thus confirming the IRREPs ordering.

Substituting the symmetry eigenvalues in Eq.~(\ref{eq:sym_chern}) we then readily find $\mathcal{C}(1) = \mathcal{C}(4) = +1\mod 4$, and $ \mathcal{C}(2) = \mathcal{C}(3) = -1\mod 4$. We conclude that each split EBR$^{2c}_{75.1}$ has a stable Chern class topology. This is confirmed by direct evaluation of the flow of Berry phase for each band, see Fig.~\ref{fig:berry} in Appendix~\ref{ap:numerical_appendix_75.1}.

\subsubsection{Crystalline Chern FM from stable nodal AFM phase}

We now start from the stable nodal phase of EBR$^{2b}_{75.5}$ ($z_2 = 1$). The breaking of non-symmorphic TRS unlocks the nodal (Weyl) points which then become free to leave the basal momentum plane (when embedded in 3D). As for the fragile topological phase this results in a fully gapped band structure at $k_z=0$ where each band acquires a symmetry indicated Chern number given by Eq.~(\ref{eq:sym_chern}). Given the band inversion at X between Band 2 and 3 required in the fragile to stable topological transition (see the IRREPs ordering in Fig.~\ref{fig:stable_nodal}b)), we now find band Chern numbers 
\begin{equation}
\label{eq_high_chern}
    \mathcal{C}(1)= \mathcal{C}(2)=\mathcal{C}(3)=\mathcal{C}(4)=+1\mod 4.
\end{equation}
Then, together with the cancellation sum rule $\sum_{i=1}^{4} \mathcal{C}(i) = 0$, we predict $\mathcal{C}(2) = -3$ and $\mathcal{C}(3) = +1$ (or, equivalently, $\mathcal{C}(2) = +1$ and $\mathcal{C}(3) = -3$). This is confirmed numerically in Fig.~\ref{fig:berry_stable} of Appendix~\ref{ap:numerical_appendix_75.1}. We thus reach the conclusion that for small Zeeman coupling the bands in the vicinity of the half-filling energy must exhibit a higher Chern number. Also, contrary to the gapped FM phase obtained from the fragile topological phase where the valence (conduction) subspace has a trivial summed topology (i.e.~$\mathcal{C}(1)+\mathcal{C}(2)=0$), we here necessarily obtain a nontrivial Chern phase at half-filling with $\mathcal{C}(1)+\mathcal{C}(2)=\pm2$\textcolor{black}{, thus recovering the general prediction of Eq.~(\ref{eq_chern_halffilling})}. %{\color{orange}We thus conclude with the robust physical signature of Fermi arcs being converted into chiral edge banches .}

\subsection{Fully \textcolor{black}{pseudo-spin-polarized} FM phases}
%We have so far characterized the gapped FM phases obtained from the fragile-gapped and stable-nodal AFM phases when the Zeeman splitting is small (i.e.~symbolically $\epsilon_{\mathrm{Z}} < \Delta,\lambda_{\mathrm{soc}}$).
Given the spin-$z$ components associated with the induced IRREPs of the EBR \textcolor{black}{at $\Gamma$ and M} (Table \ref{coIRREPs}), we anticipate that by increasing $\epsilon_{\mathrm{Z}}$ there must be a second transition into a phase with fully \textcolor{black}{pseudo-spin-polarized} valence (conduction) bands (right column of Fig.~\ref{fig:mag_energy_levels}), \textcolor{black}{i.e. the $\widetilde{\uparrow}$-($\widetilde{\downarrow}$-)band has a $\uparrow$-($\downarrow$-)spin component at $\Gamma$ and M}. This phase transition must happen through two band inversions, i.e.~at $\Gamma$ and at M. Assuming $\epsilon_Z>0$, we infer that beyond the transition the IRREPs ordering at $\Gamma$ and M are $E(\overline{\Gamma}_{7})<E(\overline{\Gamma}_{8})<E(\overline{\Gamma}_{5})<E(\overline{\Gamma}_{6})$, and $E(\overline{\text{M}}_{5})<E(\overline{\text{M}}_{6})<E(\overline{\text{M}}_{8})<E(\overline{\text{M}}_{7})$, respectively (\textcolor{black}{importantly}, note the difference with \textcolor{black}{the ordering of the previous section} and Fig.~\ref{fig:MSG75.1}b)). The question of the Chern numbers, as determined by Eq.~(\ref{eq:sym_chern}), is then reduced to the IRREPs ordering at X.

First, without loss of generality we can assume that the lowest (highest) energy level has IRREP $\overline{\text{X}}_4$ ($\overline{\text{X}}_3$), as in Fig.~\ref{fig:wilson75.5}b) and Fig.~\ref{fig:stable_nodal}b), from which we get $\mathcal{C}(1) = \mathcal{C}(4) = +1 \mod 4$. Then, for dominant values of $\epsilon_{Z}$ and $\lambda_{\mathrm{SOC}}$ , we get $\mathcal{C}(2) = \mathcal{C}(3) = -1\mod 4 $. If we assume intermediary values of $\epsilon_{Z}$ and $\lambda_{\mathrm{SOC}}$ (see below), we instead obtain $\mathcal{C}(2) = \mathcal{C}(3) = +1\mod 4 $. In the later case, (similarly to the discussion below Eq.~(\ref{eq_high_chern})) one of the two middle bands must exhibit a high Chern number of $-3$, giving a total Chern number of $\pm2$ at half-filling for the valence/conduction space.  

%The IRREP ordering at $\Gamma$ and M beyond these critical points are thus $E(\overline{\Gamma}_{7})<E(\overline{\Gamma}_{8})<E(\overline{\Gamma}_{5})<E(\overline{\Gamma}_{6})$, and $E(\overline{\text{M}}_{5})<E(\overline{\text{M}}_{6})<E(\overline{\text{M}}_{8})<E(\overline{\text{M}}_{7})$. The Chern numbers, as determined by Eq.~(\ref{eq:sym_chern}), are then determined by the IRREP ordering at X. 
We now detail the phase transition to the fully polarized phase in the context of the model Eq.~(\ref{model_75_5}) to underpin the general scheme outlined above. The fully \textcolor{black}{pseudo-spin-polarized} phase must happen through two band inversions, at $\Gamma$ and M, which are analytically defined for Eq.~(\ref{model_75_5}) by the conditions $\epsilon_{\mathrm{Z}} > 2  t_3 $ and $\epsilon_{\mathrm{Z}} > 2  t_2 $, respectively (assuming $t_{2,3}>0$), and with the IRREPs ordering at $\Gamma$ and M given above for $\epsilon_{\mathrm{Z}}>0$. The general form of the energy eigenvalues at X for Eq.~(\ref{model_75_5}) including the Zeeman term is $\epsilon_{s_1,s_2} = s_1 2t_1 + s_2 \sqrt{2\vert \lambda_2\vert^2 + \epsilon_{Z}^2} $ with $s_{1,2}=\pm1$, and we find $E_{\pm}(\overline{\text{X}}_4) = \epsilon_{-,\pm}$ and $E_{\pm}(\overline{\text{X}}_3) = \epsilon_{+,\pm}$. Fixing $t_1>0$, we note that $E_{-}(\overline{\text{X}}_4) \leq E_{+}(\overline{\text{X}}_4)\leq E_{+}(\overline{\text{X}}_3)$ and $E_{-}(\overline{\text{X}}_4) \leq E_{-}(\overline{\text{X}}_3) \leq E_{+}(\overline{\text{X}}_3)$. Hence, the lowest and highest energy levels are $E_1 = E_{-}(\overline{\text{X}}_4)$ and $E_4 = E_{+}(\overline{\text{X}}_3)$, respectively, from which we get $\mathcal{C}(1) = \mathcal{C}(4) = 1 \mod 4$. 

The topology of the two remaining bands is then determined by the sign of
\begin{equation}
 E_{-}(\overline{\text{X}}_3) - E_{+}(\overline{\text{X}}_4) = 2t_1 - \sqrt{2\vert \lambda_2\vert^2 + \epsilon_{Z}^2}  .
\end{equation}
Let us first we assume $\vert \lambda_2\vert > \sqrt{2}t_1$, for which we find $E_2 = E_{-}(\overline{\text{X}}_3) < E_{+}(\overline{\text{X}}_4) = E_3 $ for all $\epsilon_{\mathrm{Z}}> 2t_2,2t_3$, and $\mathcal{C}(2) = \mathcal{C}(3) = -1\mod 4 $. \textcolor{black}{This case thus has zero Chern number at half-filling.} 

If we take instead $t_1> \vert \lambda_2\vert/ \sqrt{2} $, then either $\epsilon_{\mathrm{Z}} > \sqrt{4t_1^2 - 2\vert \lambda_2\vert^2}, 2t_2,2t_3$, and we reach the same conclusion as before, or $   \sqrt{4t_1^2 - 2\vert \lambda_2\vert^2} > \epsilon_{\mathrm{Z}} > 2t_2,2t_3$, in which case $E_2 = E_{+}(\overline{\text{X}}_4) < E_{-}(\overline{\text{X}}_3) = E_3 $, and we find $\mathcal{C}(2) = \mathcal{C}(3) = +1\mod 4$ \textcolor{black}{which, we have shown, leads to a higher Chern number for Band 2 or 3, thus leading to a finite Chern number at half-filling ($\mathcal{C} = 2\mod 4$)}. 

We conclude this section by noting that the bands of a single split EBR$^{2c,\uparrow(\downarrow)}_{75.1}$ must always carry non-zero Chern numbers irrespectively of the ordering of IRREPs.

\section{3D topology and general MSG}
Having determined the topology of the 2D projection of MSG75.5 \textcolor{black}{($P_C4$)} (i.e.~for the corresponding magnetic layer group), we now address the 3D topology introducing the third momentum component $k_z$. First of all, we note that each Kramers doublet at $\Gamma$, M, Z, and A, are Weyl points carrying a chirality (Chern number). \textcolor{black}{This results from the chirality of any crystal structure with MSG75.5 ($P_C4$), see Section \ref{chiral_structure}.} This nodal topology will be manifested in terms of Fermi arcs on surface spectra only at quarter-filling\textcolor{black}{, and more generally at a filling $\nu \in 2\mathbb{Z}+1$}. In the following we instead focus on the topology at half-filling\textcolor{black}{, and more generally at a filling $\nu \in 2\mathbb{Z}+2$}.

\begin{figure}[t!]
    \centering
   \includegraphics[width=\linewidth]{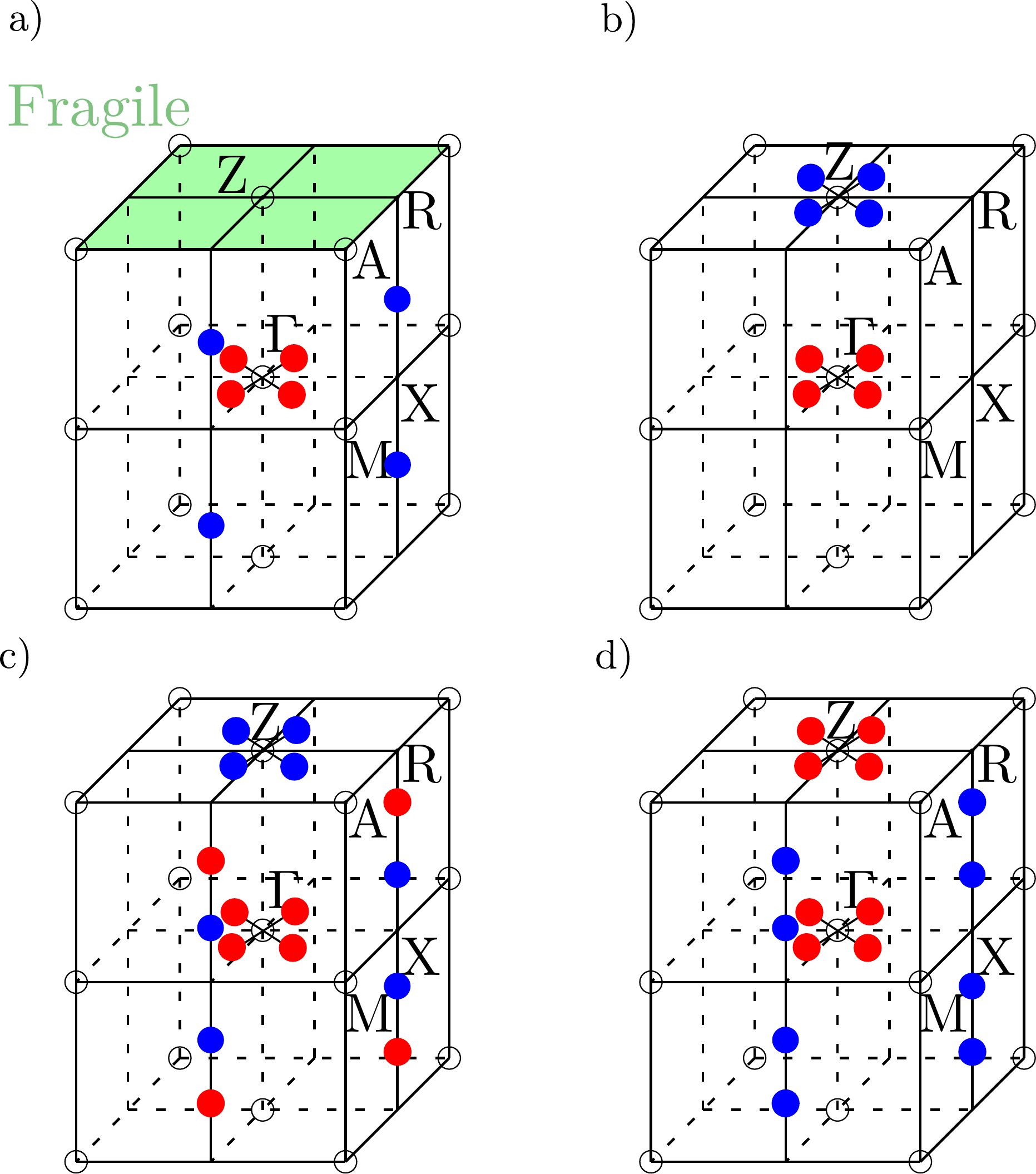}\\
    \caption{Topology at half-filling of the 3D EBR$^{2b}_{75.5}$ for a) $(z_2^0,z_2^\pi)=(1,0)$, and b) $(z_2^0,z_2^\pi)=(1,1)$. %Topology at half-filling of the 3D EBR$^{2a}_{77.18}$ for c) $z_2^0=0$, and d) $z_2^0=1$. 
    The colored dots represent the Weyl points (and their chirality) of the crystalline Weyl (CW) phases. The plane with crystalline Euler fragile (CEF) topology is colored in green. TRIM (time reversal invariant) momenta are indicated as open circles.
    c) Same as b) before the annihilation of the Weyl points on the vertical axes. d) Alternative to b) when the horizontal (vertical) Weyl points all have equal chirality.}
    \label{fig:weyls_755}
\end{figure}

The above results are directly transferable to the $k_z=0$ and $k_z=\pi$ planes of the 3D Brillouin zone, which leads to a $(z_2^{0},z_2^{\pi})\in \mathbf{Z}_2^2$ classification. If the two symmetry indicators are distinct, e.g.~$(z_2^0,z_2^\pi)=(1,0)$, they indicate the presence of $C_{2z}$ protected Weyl points (at half-filling) on the $\overline{\text{X}\text{R}}$ high-symmetry axis, on top of the four Weyl points on the $k_z=0$-plane, while the plane at $k_z=\pi$ has CEF topology. These thus form a $\mathbf{Z}_2$ indicated octuplet of Weyl points (i.e.~the CW topology). It is interesting to note that by $C_4$ symmetry the Weyl points in plane must all have the same chirality, while the Weyl points on the $\overline{\text{X}\text{R}}$ axis must all be of the opposite chirality by $C_2T$ symmetry, which leads to the configuration of Fig.~\ref{fig:weyls_755}a) where the plane with CEF topology is colored in green. \textcolor{black}{If we reverse the indicators, i.e.~$(z_2^0,z_2^{\pi})=(0,1)$, the plane with CEF topology moves to $k_z=0$ and the plane with the Weyl nodes moves to $k_z=\pi$.}

When both symmetry indicators are nonzero (obtained from above through a band inversion at $\text{R}$) both planes are stable nodal and we either obtain two quadruplets of Weyl points of opposite chirality in each horizontal plane as illustrated in Fig.~\ref{fig:weyls_755}b) (after the annihilation on the vertical axes of the nodes with opposite chirality visible in Fig.~\ref{fig:weyls_755}c)), or we have two octuplets of Weyl points with all Weyl points on the horizontal planes with the same chirality and all Weyl points on the vertical axes with the opposite chirality as shown in Fig.~\ref{fig:weyls_755}d).

Only when both symmetry indicators are zero do we retrieve a gapped 3D phase where both planes $k_z=0$ and $k_z=\pi$ are fragile topological. Our classification thus characterizes and refines the earlier prediction of a $\mathbf{Z}_2$ symmetry indicator for the 3D MSG75.5 \textcolor{black}{($P_C4$)} phases \cite{mSI}. As a side remark, we note that planes hosting the four TRIM $\{\Gamma,\text{M},\text{Z},\text{A}\}$ characterized by (non-symmorphic) TRS, i.e. is an anti-unitary symmetry squaring to $-1$ and inverting the momentum, give rise to a non-symmetry indicated $\mathbf{Z}_2$ index of a strong two-dimensional TI \cite{Axion5}. \textcolor{black}{We note that the non-symmorphic TRS squares to $-1$ over the whole plane that contains the four TRIMPs. Therefore, restricting any 3D model on that plane, it can be seen effectively as a 2D system with TRS and the index can be computed in the same way as the Kane-Mele $\mathbf{Z}_2$ invariant \cite{Bernevig_AFM_2013}.}

\textcolor{black}{For completeness, let us also mention the axion insulating phases protected by $C_{2}T$, i.e. the three-dimensional gapped topological phases indicated by the difference in the second Stiefel Whitney class between the two $C_2T$ symmetric planes $k_z=0,\pi$ \cite{Fu_3D_C2T,vanderbilt_axion,Axion3,Ahn2019}. This phase requires that $C_2T$ squares to $+1$ on both planes and, contrary to its parent phases with inversion symmetry \cite{mtqc}, it is not symmetry indicated.}

%\textcolor{orange}{[INTEGRATE HERE]}[As stated in the main text, we here discuss alternative Weyl nodal phases for the 3D embedding of the MSG75.5. In Fig.~\ref{fig:additional75.5}a) we show the configuration of Weyl points for $(z_2^0,z_2^{\pi})=(1,1)$ when the Weyl points of the planes $k_z=0,\pi$ have opposite chiralities and such that the vertical Weyl nodes are not annihilated (this can be compared with Fig.~\ref{fig:weyls_755}a)). In Fig.~\ref{fig:additional75.5}b) we show the configuration of Weyl points, still for $(z_2^0,z_2^{\pi})=(1,1)$, when the Weyl points of the planes $k_z=0,\pi$ have the same chirality.]

The mechanism discussed so far is directly transferable to the other tetragonal AFM candidate MSG81.37 \textcolor{black}{($P_C\bar{4}$)} and its FM counterpart MSG81.33 \textcolor{black}{($P\bar{4}$)}, where the fourfold rotoinversion point symmetry $S_4$ takes the place of $C_4$. The only differences with MSG75.5 \textcolor{black}{($P_C$4)} \textcolor{black}{are the reversal of chirality of the Weyl points under the action of $S_4 = IC_{4z}$ symmetry and}, for the 3D gapped phase, the existence of an additional $z_2'\in \mathbf{Z}_2$ symmetry indicator \cite{mSI} of a strong 3D TI protected by $\mathcal{S}_4$-symmetry and (non-symmorphic) TRS \cite{Khalaf_sum_indicators} (see also Appendix \ref{ap:MSG81.36_sym_ind} where this symmetry indicator is derived for MSG81.36 \textcolor{black}{($P_c\bar{4}$)} for which it is the unique symmetry indicator of the 3D gapped phase, similarly to MSG81.38 \textcolor{black}{($P_I\bar{4}$)}). \textcolor{black}{The nontrivial value of the symmetry indicator $z_2'$ in Eq.~(\ref{eq:SI_axion}),  corresponding to the indicator $z_2$ identified for MSG81.33 ($P\bar{4}$) in Ref.~\cite{mtqc}, indicates a 3D axion topological insulating phase with a non-trivial axion angle $\pi$ \cite{mtqc} and a quantized magnetoelectric response \cite{Axion2,Vanderbuilt_Axion_magneto}.}

\textcolor{black}{We conclude this section by noting the candidate MSSG77.17 ($P_C4_2$) that has a $\mathbf{Z}_2$ symmetry indicator \cite{mSI} which indicates $C_2T$ protected Weyl semi-metallic phases, as in MSG75.5 ($P_C4$), but now with a minimal connectivity of bands of 4, i.e.~the filling must be $\nu\in 4\mathbb{Z}+4$. The 2D gapped phases at $k_z=0,\pi$ are thus either trivial, or host the second Stiefel Whitney topology that is not symmetry indicated, since the non-trivial Euler class topology only exists within two-band subspaces.}

%(see Table \ref{table::complete})

%{\color{orange}ATTENTION mention the Kane-Mele $\mathbf{Z}_2$ from the four TRIMPs, etc. it is not sym indicated.}

%\begin{figure}
%    \centering
%    \includegraphics[width=\linewidth]{Figure5_appendix.pdf}
%    \caption{Additional scenarios pertaining to  Fig.~\ref{fig:weyls_755} for MSG75.5 with $(z_2^0,z_2^{\pi}) = (1,1)$. a) Same as Fig.~\ref{fig:weyls_755}b) before the annihilation of the Weyl points on the vertical axes. b) Alternative to Fig.~\ref{fig:weyls_755}b) when the horizontal (vertical) Weyl points all have equal chirality. }
%    \label{fig:additional75.5}
%\end{figure}

\section{Coexistence of nodal and subdimensional topologies}

In the previous MSG candidates we were guided by the possibility of having a non-trivial symmetry indicator of the 3D gapped phase, signaling the possibility of splitting groups of bands (possibly EBRs) into (fragile) topological bands, see also Appendix~\ref{ap:SI_analysis}. We now address a class of MSGs that host a similar mechanism that nonetheless appear trivial from a standard symmetry indicator or topological quantum chemistry perspective. At the crux of the argument lies the observation that these MSGs host groups of bands (possibly EBRs) that can be split {\it at planes} in the Brillouin zone, hosting the same (stable) topological features, while their {\it total} three-dimensional band structure must be connected. \textcolor{black}{Consequently, within these ``trivial'' groups of bands, i.e.~in the sense that they lack 3D symmetry indicators, the in-plane non-trivial signatures must coexist with symmetry indicated nodal structures located away from the (possibly) gapped planes. We discuss below one example where the connectivity of the three-dimensional EBR by itself indicates the presence of protected Weyl points in the direction perpendicular to the 2D topological planes. It thus has stable 3D signatures, such as Fermi arcs \cite{Thiang_top_semi}, which topological origin is independent of the 2D topologies and their signatures. Since the 3D symmetry indicators are blind to this kind of coexistence, these topological phases can thus only be perceived in this refined context of {\it subdimensional topology}.}

\textcolor{black}{We emphasize that our use of subdimensional topology is distinct from the usual correspondence between the topological charges of a $d$-dimensional node, with codimension $\delta $ within a $d$+$\delta$=$D$-dimensional Brillouin zone, and the $p$-dimensional gapped topologies for $\delta-1 \leq p \leq D-1$ \cite{Tomas_tables_2017,BBS_nodal_lines}. The archetypal example of this usual decent approach is the stability of a Weyl point being captured by the Chern number of a gapped sphere surrounding it \cite{Kiritsis}. We discuss below such as a situation for the case of Weyl nodes protected by the screw axis $4_2$ with a chirality $\chi=\pm2$ captured by gapped Chern planes with $\mathcal{C}=2\mod4$.  Many correspondences of this kind have been formulated recently for new types of crystal-symmetry protected gapped topologies, e.g.~\cite{BBS_nodal_lines,Hugues_quadrupolar_Weyl,Wieder_higher_monopole,Wieder_Dirac_higher}.}

\textcolor{black}{In contrast, the new sub-dimensional topology we are referring to is independent of the charges of the Weyl nodes protected by the screw symmetry, since we show that the Chern number must be zero on the 2D planes that host the sub-dimensional topology. This leads to the prediction of new phases with coexisting topological features, i.e.~the manifestations of the nodal topology in 3D together with the manifestations of the nontrivial sub-dimensional topology.} 

\subsection{Case study of MSG77.18 \textcolor{black}{($P_I4_2$)}}

As an example, we take MSG77.18 \textcolor{black}{($P_I4_2$)} that hosts the mechanism discussed for MSG75.5 \textcolor{black}{($P_C4$)} as a 2D sub-dimensional topology. The coset decomposition of the AFM compatible MSG77.18 \textcolor{black}{($P_I4_2$)} in terms of its FM partner MSG77.13 \textcolor{black}{($P4_2$)} is $\mathcal{G}_{77.18}/\mathcal{G}_{77.13} = (E\vert 0 ) \mathcal{G}_{77.13} + (E\vert \tau_d )' \mathcal{G}_{77.13}$ with $\tau_d = \boldsymbol{a}_1/2+\boldsymbol{a}_2/2+\boldsymbol{a}_3/2$, where $\mathbf{T}$ is the primitive Bravais lattice and $\mathcal{G}_{77.13}$ is generated by $(C_{4z}\vert \tau_3)\mathbf{T}$ with $\tau_3 = \boldsymbol{a}_3/2$. We consider the Wyckoff position (WP) $2a$ \cite{Bilbao} that is spanned by the sub-lattice sites $\boldsymbol{r}_A = \boldsymbol{a}_1/2$ and $\boldsymbol{r}_B = \boldsymbol{a}_2/2+\boldsymbol{a}_3/2$. The same sites correspond to WP $2c$ of MSG77.13. Populating the sites with $s$-electronic orbitals and both spin-$z$-1/2 components we get the Bloch orbital basis $\vert \boldsymbol{\varphi},\boldsymbol{k}\rangle $ with $\boldsymbol{\varphi} =\left( \varphi_{A\uparrow}, \varphi_{A\downarrow},\varphi_{B\uparrow},\varphi_{B\downarrow}\right)$ forming an elementary band representation which we write EBR$^{2a}_{77.18}$. EBR$^{2a}_{77.18}$ resembles EBR$^{2b}_{75.5}$ except that it is \textit{indecomposable} over the 3D Brillouin zone.

%Indeed, MSG77.18 has a $4_2$ screw axis which imposes the exchange of branches between the Kramers doublets at $\Gamma$ and Z, and between M and A, leading to a pair of Weyl nodes at half-filling on both $C_4$-symmetric axes $\overline{\Gamma\text{Z}}$ and $\overline{\text{MA}}$ (see the band structure in Fig.~\ref{fig:MSG77.18_13_BS}a) in Methods and Fig.~\ref{fig:MSG77.18_13}a) and c) in Appendix \ref{ap:numerical_appendix_77.18_77.13}) \cite{global_top_semi}.

An other important difference with MSG75.5 $(P_C4)$ is the algebra of symmetries at $k_z=\pi$. Taking a point on the $k_z=\pi$ plane $\bar{\boldsymbol{k}}=(k_x,k_y,\pi)$, the $C_{2}T$ symmetry in MSG77.18 \textcolor{black}{($P_I4_2$)} is represented for EBR$^{2a}_{77.18}$ by  
\begin{equation}
\begin{aligned}
    %\langle  (C_{2z}\vert \tau_d)' \rangle_{(\boldsymbol{\varphi},\bar{\boldsymbol{k}})} &= 
    \langle \boldsymbol{\varphi},\bar{\boldsymbol{k}}\vert ^{(C_{2z}\vert \tau_d)'}\vert \boldsymbol{\varphi},\bar{\boldsymbol{k}} \rangle 
    &=\mathrm{e}^{\mathrm{i} \boldsymbol{k}C_{2z}\tau_d} \langle \boldsymbol{\varphi},\bar{\boldsymbol{k}}\vert \boldsymbol{\varphi},IC_{2z}\bar{\boldsymbol{k}} \rangle  (\sigma_x \otimes \mathrm{i}\sigma_x) \mathcal{K} \\
    &= \mathrm{e}^{\mathrm{i} \boldsymbol{k}C_{2z}\tau_d} \langle \boldsymbol{\varphi},\bar{\boldsymbol{k}}\vert \boldsymbol{\varphi},\bar{\boldsymbol{k}} -\boldsymbol{b}_3 \rangle  (\sigma_x \otimes \mathrm{i}\sigma_x) \mathcal{K} \\
    &=\mathrm{e}^{\mathrm{i} \boldsymbol{k}C_{2z}\tau_d} \hat{T}(-\boldsymbol{b}_3 ) (\sigma_x \otimes \mathrm{i}\sigma_x) \mathcal{K},
\end{aligned}
\end{equation}
with $\hat{T}(-\boldsymbol{b}_3 ) = \mathrm{diag}(\mathrm{e}^{\mathrm{i} \boldsymbol{r}_A \cdot \boldsymbol{K}},\mathrm{e}^{\mathrm{i} \boldsymbol{r}_A \cdot \boldsymbol{K}},\mathrm{e}^{\mathrm{i} \boldsymbol{r}_B \cdot \boldsymbol{K}},\mathrm{e}^{\mathrm{i} \boldsymbol{r}_B \cdot \boldsymbol{K}})_{\boldsymbol{K}=-\boldsymbol{b}_3}$ $=\mathrm{diag}(1,1,-1,-1)$. We thus find the square to be $\langle  \boldsymbol{\varphi},\bar{\boldsymbol{k}}\vert ^{[(C_{2z}\vert \tau_d)']^2} \vert \boldsymbol{\varphi},\bar{\boldsymbol{k}}\rangle = - \mathbb{1}_{4\times4}$, i.e.~the $C_{2}T$ symmetry squares to $-1$. The bands hence exhibit a twofold Kramers degeneracy over the whole $k_z=\pi$-plane

{\def\arraystretch{1.3}  
 \begin{table}[h]
 \caption{\label{coIRREPs_77} \textcolor{black}{Character table for the magnetic space group IRREPs of MSG77.13 \textcolor{black}{($P4_2$)}, and coIRREPs of the unitary symmetries of MSG77.18 \textcolor{black}{($P_I4_2$)}, at Z, A, and R, with $\omega = \text{e}^{\text{i}\pi/4}$. The (co-)IRREPs at $\Gamma$, M, and X are the same as in Table \ref{coIRREPs}. The coIRREPs of MSG77.18 are given by the pairing of the two IRREPs of MSG77.13 within the same column. Retrieved from the Bilbao Crystallographic Server \cite{Bilbao}.} }
\begin{tabular}{ c |c | c | c|c|c|c|c }
 \hline 
 \hline 
 	  & WP & $\overline{\text{Z}}_5$  & $\overline{\text{Z}}_6$ &
 	  $\overline{\text{A}}_8$ &
 	  $\overline{\text{A}}_7$ &
 	  $\overline{\text{R}}_3$ &
 	   $\overline{\text{R}}_4$ \\
 	   & $2c$ & $\overline{\text{Z}}_8$  & $\overline{\text{Z}}_7$ &
 	  $\overline{\text{A}}_5$ &
 	  $\overline{\text{A}}_6$ &
 	   &
 	   \\
 	\hline 
 	 $(C_{4z}\vert\tau_3)$ & $\begin{array}{l} \uparrow_z \\ \downarrow_z \end{array}$ & $\begin{array}{l}-\omega \\ -\omega^* \end{array}$ & 
 	 $\begin{array}{l}\omega \\ \omega^* \end{array}$ &
 	 $\begin{array}{r}\omega^* \;\\ -\omega ~ \end{array}$ &
 	 $\begin{array}{r}-\omega^*\; \\ \omega~  \end{array}$ &
 	 &
 	 \\
 	\hline
 	$(C_{2z}\vert0)$ & $\begin{array}{l} \uparrow_z \\ \downarrow_z \end{array}$ & $\begin{array}{r}-\text{i} \\ \text{i} \end{array}$ & 
	$\begin{array}{r}-\text{i} \\ \text{i} \end{array}$ &
 	$\begin{array}{r}\text{i} \\ -\text{i} \end{array}$ &
	$\begin{array}{r}\text{i} \\ -\text{i} \end{array}$ &
	-\text{i}&\text{i}
	\\
	\hline
	\hline
 \end{tabular}
\end{table} 
}

\subsubsection{Model and band structure}

We illustrate this with the following minimal 3D extension of Eq.~(\ref{model_75_5}) which we rewrite as $H[f_1,f_2,f_3,g_1,g_2](\boldsymbol{k})$, 
\begin{multline}
\label{eq:model_77}
    H'(\boldsymbol{k}) = H[f_1,f_2',f_3',g_1,g_2'](\boldsymbol{k}) + \\
    \rho_1 h_1(\boldsymbol{k}) \sigma_x \otimes \sigma_z + \rho_2 h_2(\boldsymbol{k}) \sigma_y\otimes \sigma_z ,
\end{multline}
where the new lattice form factors are now extended to 3D momentum space,
\begin{equation}
    \begin{aligned}
        f'_2(\boldsymbol{k}) &= \left(\cos \boldsymbol{\delta}'_1\boldsymbol{k} - 
        \cos \boldsymbol{\delta}'_2\boldsymbol{k} +
        \cos \boldsymbol{\delta}'_3\boldsymbol{k} -
        \cos \boldsymbol{\delta}'_4\boldsymbol{k}
        \right)/2,\\
        f'_3(\boldsymbol{k}) &= \left(\cos \boldsymbol{\delta}'_1\boldsymbol{k} + 
        \cos \boldsymbol{\delta}'_2\boldsymbol{k} +
        \cos \boldsymbol{\delta}'_3\boldsymbol{k} +
        \cos \boldsymbol{\delta}'_4\boldsymbol{k}
        \right)/2,\\
        g'_2(\boldsymbol{k}) &= \left(\sin \boldsymbol{\delta}'_1\boldsymbol{k} - \mathrm{i} 
        \sin \boldsymbol{\delta}'_2\boldsymbol{k} -
        \sin \boldsymbol{\delta}'_3\boldsymbol{k} + \mathrm{i}
        \sin \boldsymbol{\delta}'_4\boldsymbol{k}
        \right)/2,\\
        h_1(\boldsymbol{k}) &= \left(\sin \boldsymbol{\delta}'_1\boldsymbol{k} + 
        \sin \boldsymbol{\delta}'_2\boldsymbol{k} +
        \sin \boldsymbol{\delta}'_3\boldsymbol{k} +
        \sin \boldsymbol{\delta}'_4\boldsymbol{k}
        \right)/2,\\
        h_2(\boldsymbol{k}) &= \left(\sin \boldsymbol{\delta}'_1\boldsymbol{k} - 
        \sin \boldsymbol{\delta}'_2\boldsymbol{k} +
        \sin \boldsymbol{\delta}'_3\boldsymbol{k} -
        \sin \boldsymbol{\delta}'_4\boldsymbol{k}
        \right)/2,
    \end{aligned}
\end{equation}
with $\boldsymbol{\delta}'_{1,2} =   \boldsymbol{\delta}_{1,2}+\boldsymbol{a}_3/2$, and $\boldsymbol{\delta}'_{3,4} = - \boldsymbol{\delta}_{1,2}+\boldsymbol{a}_3/2$, and with the new real parameters $\rho_1,\rho_2 \in \mathbb{R}$ (we take $\rho_1=-1$ and $\rho_2=-2/5$).

\begin{figure}[t!]
    \centering
    \includegraphics[width=0.8\linewidth]{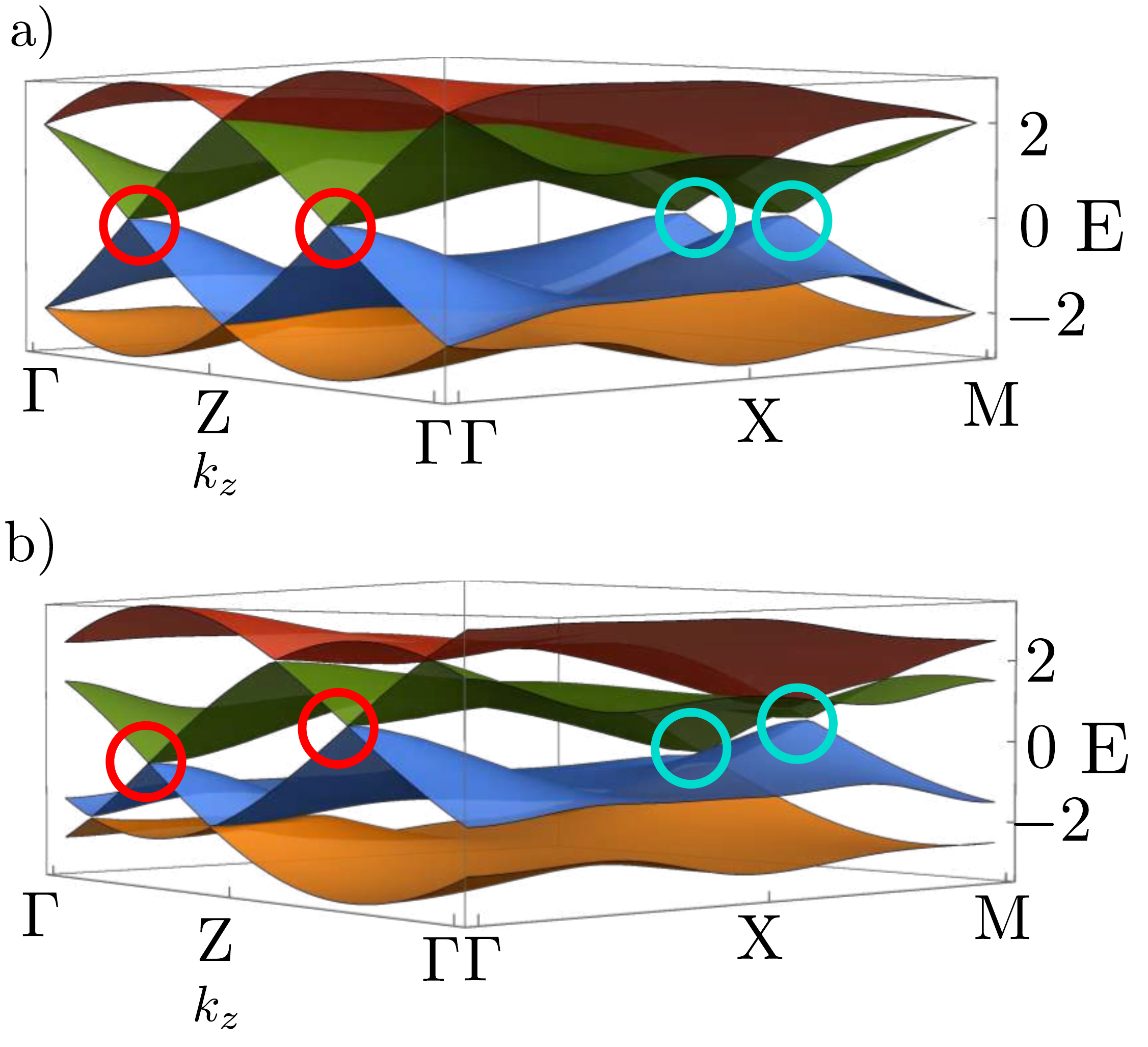}
    \caption{Sub-dimensional gapped phases within a 3D indecomposable EBR, illustrated by a) EBR$^{2a}_{77.18}$, and b) EBR$^{2c}_{77.13}$ obtained by breaking TRS through a Zeeman coupling. The Weyl nodes imposed at half-filling by the screw axis $4_2$ are marked with colored circles indicating the sign of the symmetry-imposed chiralities.}
    \label{fig:MSG77.18_13_BS}
\end{figure}
\begin{figure}[t!]
    \centering
    \includegraphics[width=\linewidth]{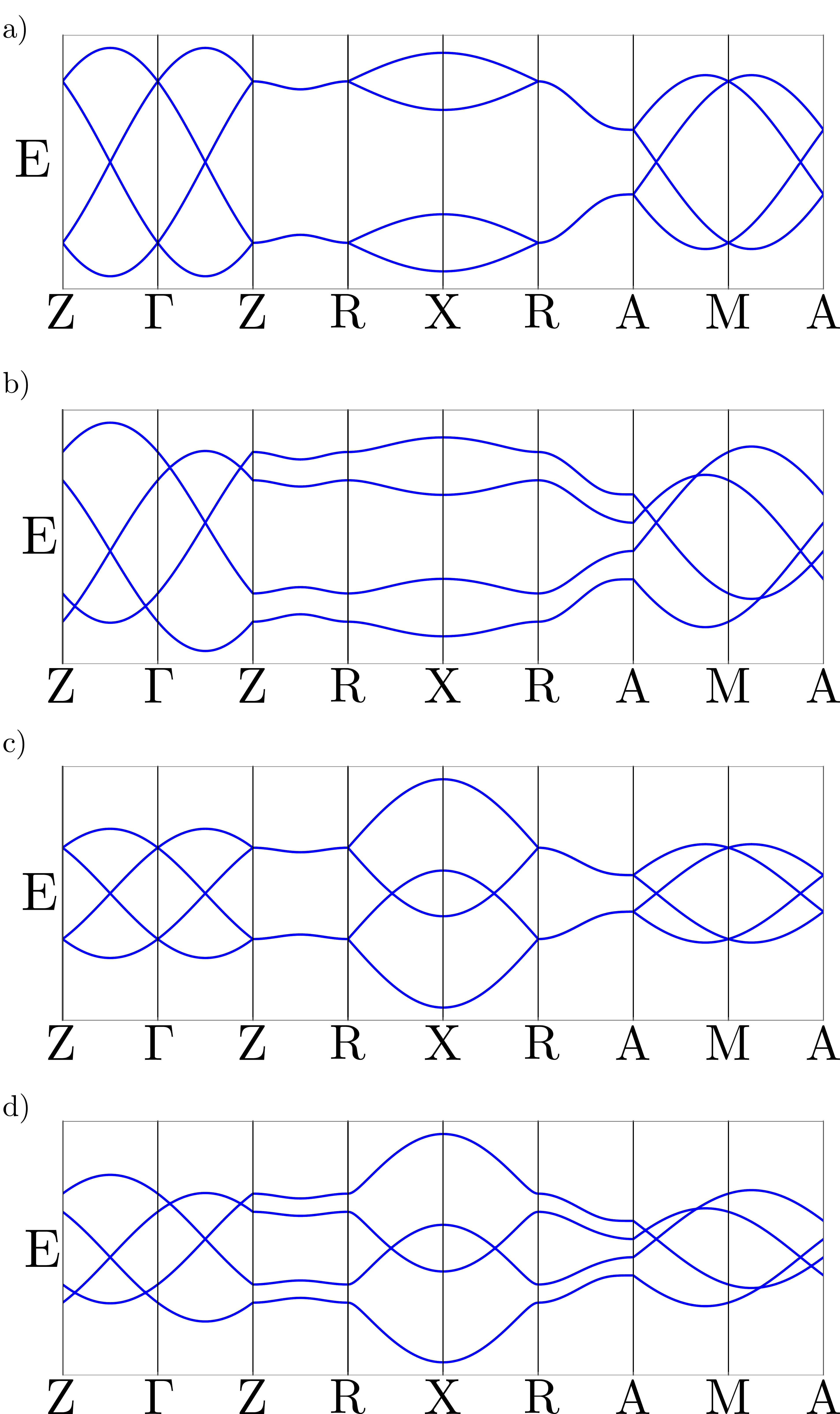}
    \caption{Band structure along high-symmetry lines for EBR$^{2a}_{77.18}$ a) with $z_2^0=0$ and c) with $z_2^{0}=1$, and for EBR$^{2c}_{77.13}$ obtained through Zeeman splitting b) from a), and d) from c). }
    \label{fig:MSG77.18_13}
\end{figure}

We show the band structure of model Eq.~(\ref{eq:model_77}) in Fig.~\ref{fig:MSG77.18_13_BS}a) where the $k_z$-axis covers $[0,2\pi]$, and the other axis corresponds to the successive paths $\Gamma\text{X}$ and $\text{X}\text{M}$ within the plane $k_z=0$. The band structure along the high-symmetry lines is shown in Fig.~\ref{fig:MSG77.18_13}a), and c) after a band inversion at X. We note the twofold degeneracy at $k_z=\pi$ which explains the degeneracies along $\overline{\text{ZR}}$ and $\overline{\text{RA}}$ in Fig.~\ref{fig:MSG77.18_13_BS}a), and in Fig.~\ref{fig:MSG77.18_13}a) and c). 

Importantly, the compatibility relations along the $C_4$-symmetric axes $\overline{\Gamma\text{Z}}$ and $\overline{\text{M}\text{A}}$ imply that the EBR cannot be split \cite{Bilbao}, see Appendix \ref{ap:SI_analysis_77.18}. Indeed, the fourfold screw symmetry $4_2\equiv (C_{4z}\vert \tau_3)$ imposes an exchange of branches between the Kramers doublets of $\Gamma$ and Z (M and A), see the coIRREPs given in Table \ref{coIRREPs_77} retrieved from \cite{Bilbao}. This leads to two $4_2$-protected nodal points on the $\overline{\Gamma \text{Z}}$-line (resp.~the $\overline{\text{MA}}$-line) at half-filling (marked by circles in  Fig.~\ref{fig:MSG77.18_13_BS}a))\textcolor{black}{, and more generally at a filling $\nu \in 4\mathbb{Z}+2$}. \textcolor{black}{We note that this exchange of IRREPs along the $4_2$-axes originates from the monodromy of the irreducible representations of the screw symmetry $4_2$ \cite{MichelZak_monodromy,Wi2,BBS_nodal_lines,Vanderbilt_screw,Bilbao}}. %We hence conclude that each high-symmetry line $\overline{\Gamma\text{Z}}$, and $\overline{\text{MA}}$, host two Weyl nodes at half-filling separated by the horizontal $k_z=0,\pi$-planes. 

\subsubsection{Chirality of Weyl nodes protected by a screw axis $4_2$}\label{weyl_screw}

\begin{figure}[t!]
    \centering
    \includegraphics[width=1.\linewidth]{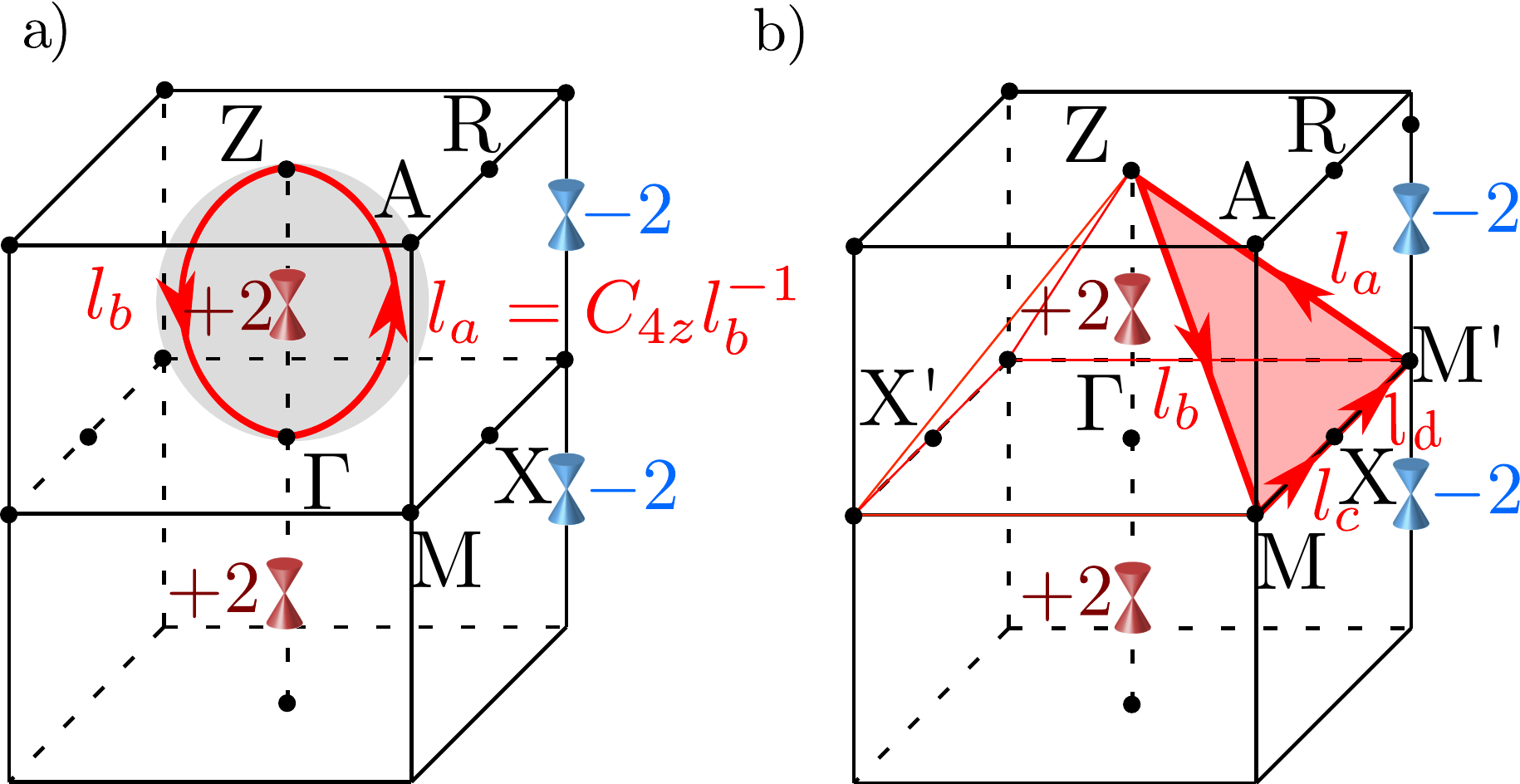}
    \caption{Chirality of $\chi=2\mod4$ for each Weyl point protected by the $4_2$ screw axis on the $\overline{\Gamma\text{Z}}$ and $\overline{\text{MA}}$ lines, at half-filling ($\nu \in 4\mathbb{Z}+2$), derived from the symmetry reduction of the Wilson loop (see text).}
    \label{fig:Wilson_box}
\end{figure}

\textcolor{black}{Following the algebraic argument of Ref.~\cite{Wi2}, see also Ref.~\cite{BBS_nodal_lines}, we now derive the symmetry enforced chirality of $\chi=2\mod 4$ for each Weyl point protected by $4_2$. Let us start with a sphere $\mathbb{S}$ surrounding one of the Weyl points, say the one on the upper half of the $\overline{\Gamma\text{Z}}$ line. We fix the south pole at $\Gamma$ and the north pole at Z, see Fig.~\ref{fig:Wilson_box}a). Then, we divide the sphere in four quarters, one of which, let us call it $\mathcal{S}$, is bounded by an oriented loop $\partial S\equiv l = l_b \circ l_a  $ (which we read as first $l_a$ followed by $l_b$) with $l_a = C_{4z}l^{-1}_b$ (where $l^{-1}_b$ is the reversed oriented path), see Fig.~\ref{fig:Wilson_box}a). Since we can recompose the total sphere through $C_{4z}$ actions, i.e.~$ \mathbb{S} = \mathcal{S} \cup C_{4z} \mathcal{S} \cup C^2_{4z} \mathcal{S} \cup C^3_{4z} \mathcal{S} $, the Chern number of the two occupied bands over the gapped sphere thus reads
\begin{equation}
    \mathrm{e}^{-\mathrm{i}2\pi \mathcal{C}[{S}]} = \left(
        \mathrm{e}^{-\mathrm{i}\gamma_B[l] }
    \right)^4 = \mathrm{e}^{-\mathrm{i}4\gamma_B[l] },    
\end{equation}
by the invariance of the Berry curvature under rotation symmetry \cite{Wi2}, and where $ \gamma_B[l] $ is the Berry phase of the two occupied bands over the loop $l=l_b\circ l_a$, i.e. (see the definition of symmetry transformation of the Wilson loop in Appendix \ref{m:patch_WL})
\begin{equation}
\begin{aligned}
    \mathrm{e}^{-\mathrm{i}\gamma_B[l] } &= \mathrm{det}\mathcal{W}[l] = \mathrm{det} \left( 
        \mathcal{W}[l_b] \cdot \mathcal{W}[l_a]
    \right) \\
    &= \mathrm{det} \left( 
       R^{\Gamma}_{4_2} \cdot \mathcal{W}[l_a]^{-1} \cdot \left( R^{\text{Z}}_{4_2} \right)^{-1} \cdot \mathcal{W}[l_a]
    \right) \\
    &= \mathrm{det} \left( 
       R^{\Gamma}_{4_2} \cdot \left( R^{\text{Z}}_{4} \right)^{-1} \right) = \dfrac{\chi_{4_2}(\overline{\Gamma}_5)\chi_{4_2}(\overline{\Gamma}_7)}{\mathrm{i}\chi_{4_2}(\overline{\text{Z}}_5) \mathrm{i} \chi_{4_2}(\overline{\text{Z}}_8)}\\
       &= (-\mathrm{i})^2 = -1,
\end{aligned}
\end{equation} 
where $R^{\Gamma}_{4_2} = \mathrm{e}^{\mathrm{i} C_{4z}\boldsymbol{k}_{\Gamma}\cdot \tau_3} S^{\Gamma}_{4_2} (\overline{\Gamma}_i \overline{\Gamma}_j) = S^{\Gamma}_{4_2} (\overline{\Gamma}_i \overline{\Gamma}_j)$ and $R^{\text{Z}}_{4_2} = \mathrm{e}^{\mathrm{i} C_{4z}\boldsymbol{k}_{\text{Z}}\cdot \tau_3}S^{\text{Z}}_{4_2} (\overline{\text{Z}}_i \overline{\text{Z}}_j) = \mathrm{i} S^{\text{Z}}_{4_2}$ ($C_{4z}\boldsymbol{k}_{\text{Z}}\cdot \tau_3 = \boldsymbol{b}_3/2 \cdot \boldsymbol{a}_3/2 = \pi/2$) are defined in terms of the representation of symmetry $(g\vert\tau_g)$ in the valence band basis $S^{\boldsymbol{k}}_{(g\vert\tau_g)}(\overline{\boldsymbol{k}}_i\overline{\boldsymbol{k}}_j)$ for a coIRREP $\overline{\boldsymbol{k}}_i\overline{\boldsymbol{k}}_j$ at a momentum $\boldsymbol{k}$, and $\chi_{4}(\overline{\boldsymbol{k}}_i)$ is the character of the IRREP $\overline{\boldsymbol{k}}_i$ given in Table \ref{coIRREPs} (we have $\chi_4 (\overline{\text{Z}}_i) = \chi_4 (\overline{\Gamma}_i)$ \cite{Bilbao}). Therefore, $\gamma_B[l] = \pi \mod 2\pi$ and we conclude that the Chern number over the whole sphere, and thus the chirality of the Weyl point inside, is $\mathcal{C}[\mathbb{S}] = 2\mod 4$. Very interestingly, we find a quadratic dispersion in the $(k_x,k_y)$-plane at a fixed $k_z$ from each Weyl point at half-filling in Fig.~\ref{fig:MSG77.18_13_BS}, see also \cite{Vanderbilt_screw}.} 

\textcolor{black}{While the above derivation based on the symmetry reduction of Wilson loop, first developed in Ref.~\cite{Wi2,BBS_nodal_lines} adapting the Wilson loop techniques developed eariler to assess gapped topological phases in \cite{InvTIBernevig,Chenprb2012,Wi1,hourglass,Alex_BerryPhase}, is completely general and can be readily transferred to any other context (i.e.~any other space group, with or without spin-orbit coupling), we note some later alternative approaches in Ref.~\cite{Vanderbilt_screw,2021chiralities}.}

\subsubsection{AFM topological phases}

\begin{figure}[h]
    \centering
    \includegraphics[width=\linewidth]{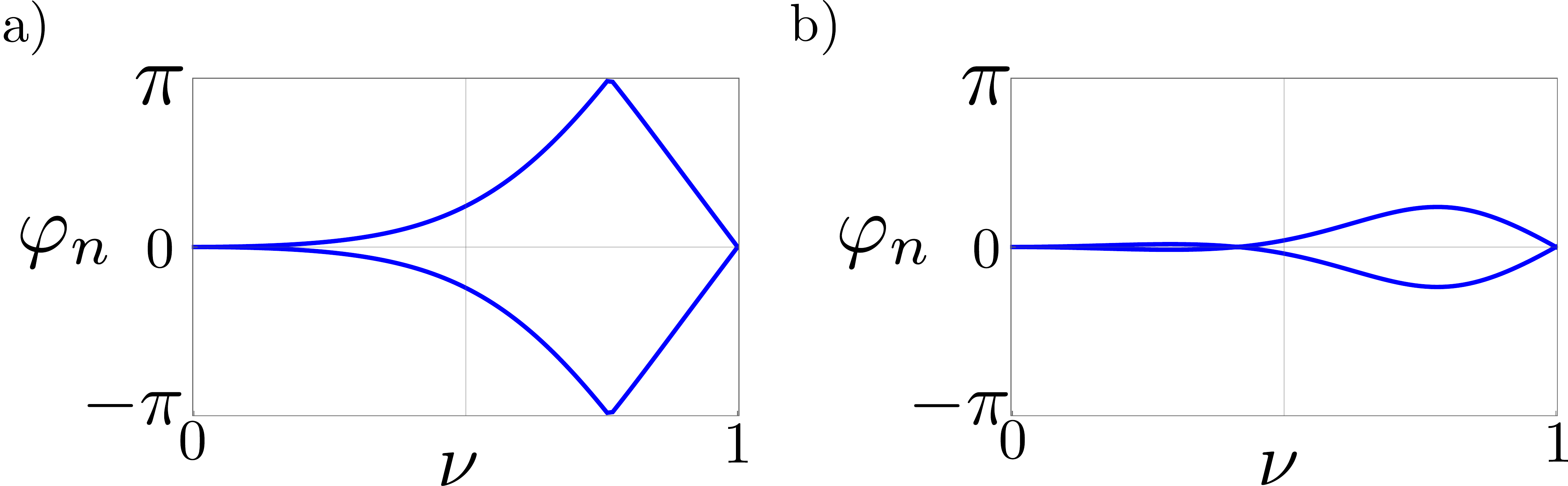}
    \caption{$C_4$-symmetry Wilson loop for the conduction a) and valence b) bands of EBR$^{2a}_{77.18}$ at $k_z=\pi$.}
    \label{fig:MSG77.18_13_C4WL}
\end{figure}

We now discuss the global topology of the AFM topological phases for MSG77.18 ($P_I 4_2$). We first note that EBR$^{2a}_{77.18}$ can be gapped over the planes $k_z=0,\pi$. The 2D topology at $k_z=0$ for EBR$^{2a}_{77.18}$ is the same as the topology discussed for EBR$^{2b}_{75.5}$, that is CEF topology versus stable nodal (CW topology) indicated by $z_2$ in Eq.~(\ref{eq:::indicator}). We therefore can define a sub-dimensional $z_2^0\in \mathbf{Z}_2$ symmetry indicator. We have seen that on the $k_z=\pi$ plane the $C_{2}T$ symmetry squares to $-1$, such that there is no (real) Euler class topology. Nevertheless, we show in Fig.~\ref{fig:MSG77.18_13_C4WL} the $C_4$-symmetric Wilson loop on the plane $k_z=\pi$ for the model for MSG77.18 \textcolor{black}{($P_I4_2$)} Eq.~(\ref{eq:model_77}), over a) the conduction and b) valence bands. The winding of Wilson loop for the conduction bands indicates a crystalline (non-Euler) fragile (CF) topology.

\begin{figure}[t!]
    \centering
   \includegraphics[width=\linewidth]{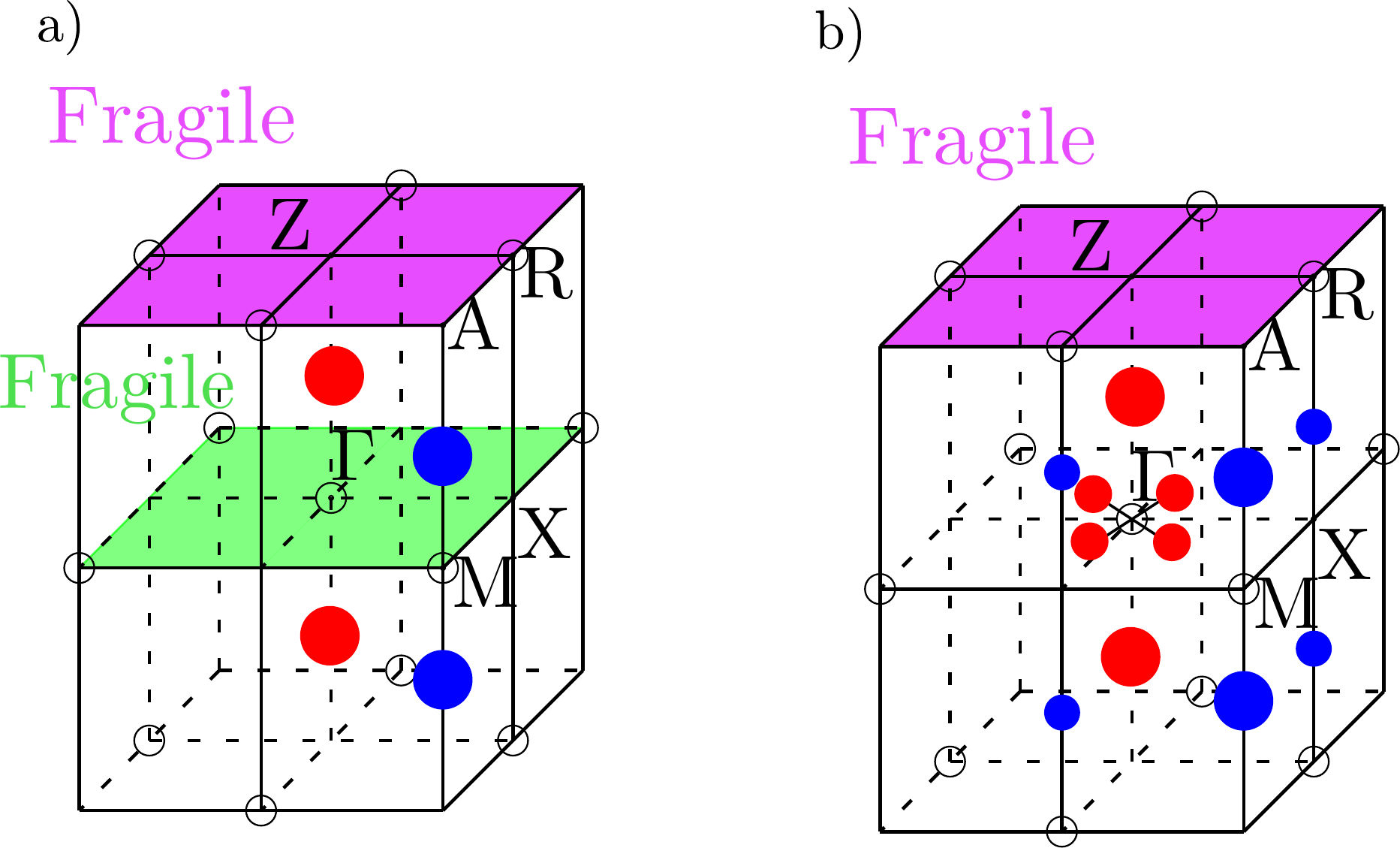}
    \caption{Topology at half-filling of the 3D EBR$^{2a}_{77.18}$ for a) $z_2^0=0$, and b) $z_2^0=1$. The colored dots represent the Weyl points and their chirality, with $\chi=\pm1$ for the small dots, and $\chi=\pm2$ for the big dots, of the crystalline Weyl (CW) phases. The double Weyl points on the vertical axes are protected by the screw $4_2$ symmetry, see Section \ref{weyl_screw}. The plane with crystalline Euler fragile (CEF) topology is colored in green, and the plane with ($C_4$-symmetry protected) crystalline fragile (CF) topology is colored in purple. TRIM (time reversal invariant) momenta are indicated as open circles.}
    \label{fig:weyls_7718}
\end{figure}

\textcolor{black}{We now determine how the subdimensional topologies (at $k_z=0,\pi$) interact with the topology of the $4_2$-symmetry protected Weyl points.} We first note that the Chern number \textcolor{black}{at half-filling} vanishes on the $k_z=0,\pi$-planes as a consequence of $C_2T$ symmetry (see the discussion below Eq.~(\ref{eq:indicated_berry}) for $k_z=0$, and at $k_z=\pi$, we have $\mathcal{F} \equiv 0$ by $[C_2T]^2=-1$). \textcolor{black}{As a consequence, we can deform the sphere $\mathbb{S}$ of Fig.~\ref{fig:weyls_7718}a) into the pyramid $\mathbb{P}$ of Fig.~\ref{fig:weyls_7718}b) while conserving the Chern number, i.e.~$\mathcal{C}[\mathbb{S}] = \mathcal{C}[\mathbb{P}] = 2\mod 4$. This equality can be readily verified through the symmetry reduction of the Wilson loop $\mathcal{W} [l_d \circ l_c \circ l_b \circ l_a] $ similarly to the above derivation for $\mathbb{S}$ but now using both $C_{4z}$ and $C_{2z}$ transformations \cite{Wi2}. We note that even if there are nodal points on the $k_z=0$ plane (for $z_2 = 1$), by $C_4$ symmetry they must contribute to an increase of the Chern number by $\pm4$, which leaves the quantity $\mod4$ unchanged. An other consequence of $C_2T$ symmetry is that any Weyl point above the $k_z=0$ plane must have its mirror symmetric image underneath ($k_z\rightarrow IC_{2z} k_z = -k_z$) with the same chiral charge. Combining the top of the pyramid in Fig.~\ref{fig:Wilson_box}b) with its mirror image in the $k_z$-direction, we obtain an octahedron $\mathbb{O} = \mathbb{P}\cup m_z \mathbb{P}$ that wraps the pair of Weyl points on the $\overline{\Gamma\text{Z}}$ line, and over which there is a total Chern number (chirality) of $\mathcal{C}[\mathbb{O}] = (2+2) \mod 8 = 4\mod 8$. }

\textcolor{black}{We note that $\mathbb{O}$ divides the 3D Brillouin zone in two symmetric halves.} Invoking the Nielsen-Ninomiya cancellation theorem \cite{NIELSEN1981219}, it follows that the total chirality inside $\mathbb{O}$ (say $+4$), must be compensated by the total chirality inside the complement of $\mathbb{O}$ ($-4$), see Fig.~\ref{fig:Wilson_box}. We then arrive for $z_2=0$ to the configuration of Fig.~\ref{fig:weyls_7718}a), where the green (purple) plane hosts a CEF (CF) topology, and where each red (blue) point mark a Weyl point with $\chi=+2$ (rep. $\chi=-2$). We finally conclude that this phase must exhibit large \textcolor{black}{double} Fermi arcs in the surface spectra connecting pairs of projected Weyl nodes of opposite chirality across the surface Brillouin zone. Also, by inverting IRREPs at X, i.e.~setting $z_2^0=1$, we get the octuplet of Weyl points discussed above in the 3D phase of MSG75.5 \textcolor{black}{($P_C4$)}, leading to the configuration of Fig.~\ref{fig:weyls_7718}b). \textcolor{black}{This last case must exhibit an exotic coexistence of Fermi arcs generated by different sets of Weyl points with qualitatively distinct topological origins.} 

\textcolor{black}{Our sub-dimensional analysis has thus allowed us to identify new phases with the coexistence of 2D and 3D nodal topological features.}

\subsubsection{AFM-FM correspondence}

We now consider the effect of breaking the non-symmorphic TRS, i.e.~inducing a transition from the AFM phase of MSG77.18 \textcolor{black}{($P_I4_2$)} to the FM phase of MSG77.13 \textcolor{black}{($P4_2$)}. This can be done by including a Zeeman splitting as we did for MSG75. We show the band structure in Fig.~\ref{fig:MSG77.18_13_BS}b), and along the high-symmetry lines in Fig.~\ref{fig:MSG77.18_13}b), and d) after a band inversion at X. We find that all the Kramers degeneracies are split leaving gapped bands on the $k_z=0$ and $\pi$ planes. Similarly to the case of MSG75.1 \textcolor{black}{($P4$)}, the topology of the gapped bands at fixed $k_z$ are characterized through symmetry indicated Chern numbers (i.e.~with CC topology). Interestingly, for moderate Zeeman coupling, the four bands remain fully connected along the $C_4$-symmetric axes through the persistence of the $C_4$-symmetry protected Weyl nodes, as indicated by the IRREP order at $\Gamma$, Z, M, and A, as it is clearly shown in Fig.~\ref{fig:MSG77.18_13}b) and d).

\textcolor{black}{We note however that, by relaxing the pairing conditions (i.e.~from 2D coIRREPs to 1D IRREPs), there are more combinatorial ways of connecting the bands allowed by the compatibility relations for MSG77.13 $(P4_2)$. In particular, the bands can be ordered at $\{\Gamma,\text{Z},\text{M},\text{A}\}$ as to avoid Weyl points at half-filling (more generally at a filling $\nu \in 4\mathbb{Z}+2$).}

\subsubsection{Stable 3D signatures}
We emphasize that the rational of sub-dimensional topologies thus works in two manners. Firstly, because the total 3D EBR is connected, and thus trivial, these phases are missed by previous schemes. Reciprocally, \textcolor{black}{the presence of in plane topology together} with the nodes to make the EBR globally connected \textcolor{black}{implies the coexistence of topological signatures of qualitatively distinct origins.} \textcolor{black}{On one hand, the connectivity condition} directly induces symmetry indicated Weyl points, and in turn, Fermi arcs in the surface spectra. \textcolor{black}{On the other hand, the subdimensional topology is either gapless, in which case it induces additional Weyl points and thus additional Fermi arcs, or it is gapped fragile topological, in which case it induces corner modes \cite{lange2021subdimensional}}. In this sense this mechanism can thus be used to find new topological signatures that are directly detectable via the usual routes of e.g.~angle resolved photoemission spectroscopy, quantum oscillation techniques or scanning tunneling microscopy.

\subsection{Generalization}

{\def\arraystretch{1.}  
 \begin{table*}[t!]
 \caption{\label{list} Candidate MSGs for the (Euler) fragile/stable-nodal AFM to Chern FM mechanism, including those profiting from the subdimensional topological analysis.
 The table lists the AFM and corresponding FM counterparts as well as their time reversal invariant momenta (TRIM) which host Kramers doublets (i.e.~Weyl nodes). Moreover, it details the topology by enumerating the value of $C_2T=\pm1$ and the two-band subspace (2-BS) characterization on the $k_z=0,\pi$ momentum planes. $[C_2T]^2=+1$ indicates Euler class (real) topology, while $[C_2T]^2=-1$ implies the twofold Kramers degeneracy of the bands. The labels CEF, CF, CW, and CC indicate crystalline Euler fragile (with symmetry-indicated Wilson loop quantization), crystalline fragile (with the winding of $C_4$-symmetric Wilson loop), and crystalline Weyl semimetallic (with a symmetry-indicated $\pi$-Berry phase), respectively.  When the 2-BS Topology is CW, we mean that there must be Weyl nodes connecting adjacent two-band subspaces. When we write CEF/CW, we mean that either of the topologies is realized depending on the ordering of IRREPs. Finally, the EBR column specifies the elementary band representations$^{a}$\footnote[0]{a. The EBRs are defined for a given Wyckoff position and a fixed spin basis. We either take the vertical $\hat{z}$-axis ($C_4$-axis) as the quantization axis for the spin-1/2 (3/2) with the spin basis $(\uparrow_z,\downarrow_z)$, \textcolor{black}{or we take a quantization axis that is perpendicular to $\hat{z}$, e.g. $\hat{y}$ for which the spin basis is $(\rightarrow_y,\leftarrow_y) = (\uparrow_z+\mathrm{i}\downarrow_z,\mathrm{i}\uparrow_z+\downarrow_z)/\sqrt{2} $}.} hosting the topology$^{b}$\footnote[0]{b. In the case of a single EBR, splitable at $k_z=0$ and $k_z=\pi$, we mean that it \textit{must} host one of the listed nontrivial topologies. In the case of a (direct) sum of EBRs, we mean that the topology \textit{can} be achieved through the permutation of IRREPs between the EBRs.}. All FM candidates acquire crystalline Chern (CC) topology when obtained from their AFM parents through Zeeman splitting. The EBR data were retrieved from the Bilbao Crystallographic Server \cite{Bilbao}. } 
\begin{tabular}{|l|l|ccc|c|l|}
 \hline 
 \hline 
 	  ~~AFMSG & ~~TRIM & $~k_z~$ & $[C_2T]^2$  & 2-BS Topology & EBRs  & ~~FMSG\\
 	  \hline
 	  75.4 ($P_c4$) & $\Gamma$,M,X,X' &
 	  $\begin{array}{l}
 	    0 \\
 	    \pi
 	  \end{array}$ & 
 	  $\begin{array}{c}
 	    +1 \\ -1
 	  \end{array}$ &
 	  \begin{tabular}{c}
 	    CEF\\ 
 	    CF
 	  \end{tabular} & 
 	   $\begin{array}{c}
 	   (2a,\uparrow_z)\oplus(2a,\downarrow_z),(2b,\uparrow_z)\oplus(2b,\downarrow_z),\\
 	   (4c,\uparrow_z),(4c,\downarrow_z)
 	   \end{array}$  & 75.1 ($P4$) \\
 	  \hline
 	  75.5 ($P_C4$) & $\Gamma$,M,Z,A &
 	  $\begin{array}{l}
 	    0 \\
 	    \pi
 	  \end{array}$ & 
 	  $\begin{array}{c}
 	    +1 \\ +1
 	  \end{array}$ &
 	  \begin{tabular}{c}
 	    CEF/CW\\ 
 	    CEF/CW
 	  \end{tabular} &
 	  $\begin{array}{c}
 	  (2a,\uparrow_z)\oplus(2a,\downarrow_z),
 	  (2b,\uparrow_z\oplus\downarrow_z),\\
 	  (4c,\rightarrow_y)
 	  \end{array}$  & 75.1 ($P4$) \\
 	  \hline
 	  75.6 ($P_I4$) & $\Gamma$,M,R,R' &
 	  $\begin{array}{l}
 	    0 \\
 	    \pi
 	  \end{array}$ & 
 	  $\begin{array}{c}
 	    +1 \\ -1
 	  \end{array}$ &
 	  \begin{tabular}{c}
 	    CEF/CW\\ 
 	    CF
 	  \end{tabular} &
 	   $\begin{array}{c}
 	   (2a,\uparrow^{1/2}_z)\oplus(2a,\downarrow^{1/2}_z)\oplus (2a,\uparrow^{3/2}_z) \oplus (2a,\uparrow^{3/2}_z),\\
 	   (4b,\uparrow_z),(4b,\downarrow_z)\end{array}$ &  75.1 ($P4$) \\
 	   \hline
 	  76.10 ($P_c4_1$) & $\Gamma$,M,X,X' &
 	  $\begin{array}{l}
 	    0 \\
 	    \pi
 	  \end{array}$ & 
 	  $\begin{array}{c}
 	    +1 \\ +1
 	  \end{array}$ &
 	  \begin{tabular}{c}
 	    CEF\\ 
 	    CEF/CW
 	  \end{tabular} &
 	   $(4a,\rightarrow_y),(4b,\rightarrow_y),
 	  (4c,\rightarrow_y)$ &  76.7 ($P4_1$) \\
 	   \hline
 	  76.11 ($P_C4_1$) & $\Gamma$,M,Z,A &
 	  $\begin{array}{l}
 	    0 \\
 	    \pi
 	  \end{array}$ & 
 	  $\begin{array}{c}
 	    +1 \\ -1
 	  \end{array}$ &
 	  \begin{tabular}{c}
 	    CEF/CW\\ 
 	    CF
 	  \end{tabular} &
 	   $(8a,\rightarrow_y)$ &  76.7 ($P4_1$) \\
 	   \hline
 	  76.12 ($P_I4_1$) & $\Gamma$,M,R,R' &
 	  $\begin{array}{l}
 	    0 \\
 	    \pi
 	  \end{array}$ & 
 	  $\begin{array}{c}
 	    +1 \\ +1
 	  \end{array}$ &
 	  \begin{tabular}{c}
 	    CEF/CW\\ 
 	    CEF/CW
 	  \end{tabular} &
 	   $(4a,\rightarrow_y)$ &  76.7 ($P4_1$) \\
 	   \hline
 	  77.16 ($P_c4_2$) & $\Gamma$,M,X,X' &
 	  $\begin{array}{l}
 	    0 \\
 	    \pi
 	  \end{array}$ & 
 	  $\begin{array}{c}
 	    +1 \\ -1
 	  \end{array}$ &
 	  \begin{tabular}{c}
 	    CEF\\ 
 	    CF
 	  \end{tabular} &
 	   $\begin{array}{c}
 	   (2a,\uparrow_z\oplus\downarrow_z),
 	   (2b,\uparrow_z\oplus\downarrow_z),\\
 	   (4c,\uparrow_z),(4c,\downarrow_z)
 	   \end{array}$ &  77.13 ($P4_2$) \\
 	   \hline
 	  77.17 ($P_C4_2$) & $\Gamma$,M,Z,A &
 	  $\begin{array}{l}
 	    0 \\
 	    \pi
 	  \end{array}$ & 
 	  $\begin{array}{c}
 	    +1 \\ +1
 	  \end{array}$ &
 	  \begin{tabular}{c}
 	    CEF/CW \\ 
 	    CEF/CW
 	  \end{tabular} &
 	   $\begin{array}{c}
 	   (4a,\uparrow_z)\oplus
 	   (4a,\downarrow_z), (4b,\uparrow_z)\oplus
 	   (4b,\downarrow_z), \\
 	  (4c,\rightarrow_y)     
 	   \end{array}$ &  77.13 ($P4_2$) \\
 	   \hline
 	  77.18 ($P_I4_2$) & $\Gamma$,M,R,R' &
 	  $\begin{array}{l}
 	    0 \\
 	    \pi
 	  \end{array}$ & 
 	  $\begin{array}{c}
 	    +1 \\ -1
 	  \end{array}$ &
 	  \begin{tabular}{c}
 	    CEF/CW \\ 
 	    CF
 	  \end{tabular} &
 	   $\begin{array}{c}
 	   (2a,\uparrow_z\oplus\downarrow_z),
 	   (4b,\uparrow_z)\oplus(4b,\downarrow_z)
 	   \end{array}$ &  77.13 ($P4_2$) \\
 	   \hline
 	  78.22 ($P_c4_3$) & $\Gamma$,M,X,X' &
 	  $\begin{array}{l}
 	    0 \\
 	    \pi
 	  \end{array}$ & 
 	  $\begin{array}{c}
 	    +1 \\ +1
 	  \end{array}$ &
 	  \begin{tabular}{c}
 	    CEF \\ 
 	    CEF/CW
 	  \end{tabular} &
 	   $(4a,\rightarrow_y),(4b,\rightarrow_y),
 	   (4c,\rightarrow_y)$ &  78.19 ($P4_3$) \\
 	   \hline
 	  78.23 ($P_C4_3$) & $\Gamma$,M,Z,A &
 	  $\begin{array}{l}
 	    0 \\
 	    \pi
 	  \end{array}$ & 
 	  $\begin{array}{c}
 	    +1 \\ -1
 	  \end{array}$ &
 	  \begin{tabular}{c}
 	    CEF/CW \\ 
 	    CF
 	  \end{tabular} &
 	   $(8a,\rightarrow_y)$ &  78.19 ($P4_3$) \\
 	   \hline
 	  78.24 ($P_I4_3$) & $\Gamma$,M,R,R' &
 	  $\begin{array}{l}
 	    0 \\
 	    \pi
 	  \end{array}$ & 
 	  $\begin{array}{c}
 	    +1 \\ +1
 	  \end{array}$ &
 	  \begin{tabular}{c}
 	    CEF/CW \\ 
 	    CEF/CW
 	  \end{tabular} &
 	   $(4a,\rightarrow_y)$ &  78.19 ($P4_3$) \\
 	   \hline
 	  79.28 ($I_c4$) & $\Gamma$,M,X,X' &
 	  $0$ & 
 	  +1 &
 	  CEF &
 	   $\begin{array}{c}
 	   (4a,\uparrow_z)\oplus(4a,\downarrow_z),(4b,\uparrow_z\oplus\downarrow_z),  \\
 	   (8c,\rightarrow_y)
 	   \end{array}$ &  79.25 ($I4$) \\
 	   \hline
 	  80.32 ($I_c4_1$) & $\Gamma$,M,X,X' &
 	  0 & 
 	  +1 &
 	  CEF &
 	   $\begin{array}{c}
 	   (8a,\rightarrow_y),(8b,\rightarrow_y),  \\
 	   (8c,\uparrow_z),(8c,\downarrow_z)
 	   \end{array}$ &  80.29 ($I4_1$) \\
 	   \hline
 	  81.36 ($P_c\bar{4}$) & $\Gamma$,M,X,X' &
 	  $\begin{array}{l}
 	    0 \\
 	    \pi
 	  \end{array}$ & 
 	  $\begin{array}{c}
 	    +1 \\ -1
 	  \end{array}$ &
 	  \begin{tabular}{c}
 	    CEF \\ 
 	    CF
 	  \end{tabular} &
 	   $\begin{array}{c}
 	   (2a,\uparrow_z)\oplus(2a,\downarrow_z),(2c,\uparrow_z)\oplus(2c,\downarrow_z),\\
 	   (4g,\uparrow_z),(4g,\downarrow_z)
 	   \end{array}$ &  81.33 ($P\bar{4}$) \\
 	   \hline
 	  81.37 ($P_C\bar{4}$) & $\Gamma$,M,Z,A &
 	  $\begin{array}{l}
 	    0 \\
 	    \pi
 	  \end{array}$ & 
 	  $\begin{array}{c}
 	    +1 \\ +1
 	  \end{array}$ &
 	  \begin{tabular}{c}
 	    CEF/CW \\ 
 	    CEF/CW
 	  \end{tabular} &
 	   $\begin{array}{c}
 	   (2a,\uparrow^{1/2}_z)\oplus(2a,\downarrow^{1/2}_z)\oplus (2a,\uparrow^{3/2}_z) \oplus (2a,\uparrow^{3/2}_z),\\
 	   (2b,\uparrow^{1/2}_z)\oplus(2b,\downarrow^{1/2}_z)\oplus (2b,\uparrow^{3/2}_z) \oplus (2b,\uparrow^{3/2}_z),\\
 	   (2c,\uparrow_z\oplus\downarrow_z),(2d,\uparrow_z\oplus\downarrow_z),(4g,\rightarrow_y)
 	   \end{array}$ &  81.33 ($P\bar{4}$) \\
 	   \hline
 	  81.38 ($P_I\bar{4}$) & $\Gamma$,M,R,R' &
 	  $\begin{array}{l}
 	    0 \\
 	    \pi
 	  \end{array}$ & 
 	  $\begin{array}{c}
 	    +1 \\ -1
 	  \end{array}$ &
 	  \begin{tabular}{c}
 	    CEF/CW \\ 
 	    CF
 	  \end{tabular} &
 	   $\begin{array}{c}
 	   (2a,\uparrow^{1/2}_z)\oplus(2a,\downarrow^{1/2}_z)\oplus(2a,\uparrow^{3/2}_z)\oplus (2a,\downarrow^{3/2}_z),\\
 	   (2b,\uparrow^{1/2}_z)\oplus(2b,\downarrow^{1/2}_z)\oplus(2b,\uparrow^{3/2}_z)\oplus (2b,\downarrow^{3/2}_z),\\
 	   (2c,\uparrow_z\oplus\downarrow_z),(2d,\uparrow_z\oplus\downarrow_z)
 	   \end{array}$ &  81.33 ($P\bar{4}$) \\
 	   \hline
 	  82.42 ($I_c\bar{4}$) & $\Gamma$,M,X,X' &
 	  0 & 
 	  +1 &
 	  CEF &
 	   $\begin{array}{c}
 	   (4a,\uparrow_z)\oplus(4a,\downarrow_z),(4d,\uparrow_z)\oplus(4d,\downarrow_z),\\
 	   (8g,\rightarrow_y)
 	   \end{array}$ &  82.39 ($I\bar{4}$) \\
	\hline
	\hline
 \end{tabular}
 \label{table::complete}
\end{table*} 
}

We end by generalizing our sub-dimensional topology scheme and search systematically for candidate MSGs hosting the planar topologies discussed so far$-$ that is, CEF, CF, and CW phases for the AFM cases and CC phases for their FM counterpart$-$and the outlined mechanisms relating them. Remarkably, our results are directly transferable to all tetragonal MSGs with the point groups $C_4$ and $S_4$, i.e.~comprising space group families SG75-SG82. For each family we then consider all the Shubnikov type IV AFM MSGs and the one type I FM MSG. This amounts to a total of 26 MSGs which we list in Table \ref{list}, where we give the type of planar topologies for $k_z=0$ and $k_z=\pi$, and the list of EBRs hosting these. On top of the single EBRs that split on both planes, $k_z=0$ and $k_z=\pi$, and which \textit{must} necessarily host a nontrivial planar topology of type indicated, we have also listed the sums of EBRs that \textit{can} lead to a listed topology upon the permutation of their IRREPs. Whenever there is the choice CEF/CW, the topology is determined by the ordering of IRREPs. We note that $[C_{2}T]^2=\pm1$ indicates CEF or CEF/CW ($+1$) versus CF ($-1$) topology \footnote{In case of CEF, the quantization of the Wilson loop over the base path $\Gamma\text{X}\Gamma'$ follows either from the condition that $ \xi^{\Gamma}_2(1) = \xi^{\Gamma}_2(2)$ (equivalently, $ \xi^{\text{M}}_2(1) = \xi^{\text{M}}_2(2)$), or if both $\Gamma$ and X are TRIMs in which case the Wilson loop phases are also Kramers degenerate at $\{0,0\}$ or $\{\pi,\pi\}$.}\cite{Wi3} as the only alternative since the Chern number must vanish. Finally, all FM candidates acquire CC topology when obtained from their AFM parents through Zeeman splitting. \textcolor{black}{In particular, every nodal AFM phase at half-filling must give rise to a nontrivial Chern FM phase at half-filling}, thus constituting a systematic correspondence between necessarily nontrivial topological phases associated to MSG representations.

\section{Chiral Fermions at high-symmetry momenta and large Fermi arcs}\label{chiral_structure}

\textcolor{black}{All the MSGs of the table with point group $C_4$ (i.e.~the type IV AFM MSG75.4-MSG81.36, and the type I FM MSG75.1-MSG80.29) have only proper symmetries, a consequence of which is that any crystal structure that explicitly breaks all symmetries not included in their MSG must be chiral, i.e.~enantiomorphic (the SGs 75-76-77-78-79-80 are all among the 65 Sohncke space groups with no inversion, no mirror, nor roto-inversion symmetries \cite{Abundance_chiral}). Then, the absence of improper symmetry allows the existence of Weyl nodes at high-symmetry momenta, i.e.~with non-vanishing Chern numbers. This is the rationale for the chiral Fermions found in many (non-magnetic) material candidates with a Sohncke space group \cite{Wi2,Bradly_beyond_DW,Wieder_chiral_RhSi,SCZ_chiral_fermions,Wieder_chiral_natmat,Murakami_chiral_2020}.}

\textcolor{black}{We have noted that MSG75.5 ($P_C4$) and MSG77.18 ($P_I 4_2$) exhibit Weyl nodes at every TRIMP at a filling $2\mathbb{Z}+1$. MSG77.18 must also exhibit Weyl nodes on the $\overline{\Gamma Z}$- and $\overline{M A}$-lines at a filling $4\mathbb{Z}+2$ due to the monodromy of the irreducible representation of the screw axis $4_2$. The same results apply to the FM parents MSG75.1 ($P4$) and MSG77.13 $(P4_2)$. The Weyl points at high-symmetry momenta for all the other MSGs of the table can be found similarly.} 

\textcolor{black}{In Ref.~\cite{Wi2} we have given a detailed analysis of the symmetry indicated higher Chern number generated by the Weyl points locked on a screw axis at half-filling, and derived the necessary existence of large Fermi arcs due to the compensation of chirality across the Brillouin zone. This analysis can be readily transferred to the present situation (e.g.~here $\mathcal{C}=\pm2$ on the $4_2$-axes). Large Fermi arcs has also been reported in other non-magnetic chiral materials \cite{Wieder_chiral_RhSi,SCZ_chiral_fermions}.}

\textcolor{black}{In the next section, we discuss the fate of the Weyl points when extra improper point symmetries are included.}

\textcolor{black}{We now turn to the MSGs with roto-inversion symmetry $IC_{4z}$. Since the chirality of Weyl points is reversed under $IC_{4z}$, the Kramer's degeneracies at the TRIMPs (which are also $IC_{4z}$ invariant momenta) cannot form Weyl nodes at the filling $2\mathbb{Z}+1$. Instead, each double degeneracy at a TRIMP on the $k_z=0$ plane is continued as a nodal line for all values of $k_z$ \cite{Bilbao}.}

\section{Raising and lowering of symmetries}\label{x_sym}

\textcolor{black}{So far, we have focused on the topological correspondence between representations of magnetic space groups with the tetragonal point groups $C_4$ and $S_4$. We now briefly address the effect of adding and removing unitary symmetries.}

\subsection{Magnetic Dirac Fermions}

\textcolor{black}{Let us start by including extra unitary symmetries to MSSG75.5 ($P_C4$). If we include one extra vertical mirror symmetry, say $m_y$, the MSG is promoted to MSG99.169 ($P_C 4mm$) with the (unitary) point group $C_{4v}$ ($4mm$). Remarkably the four bands become all connected through a fourfold magnetic Dirac node at M, similarly to the examples discussed in Ref.~\cite{Wieder_magnetic_Dirac}.}

\textcolor{black}{The same happens for MSG89.93 ($P_C 422$), obtained by including one horizontal $\pi$-rotation symmetry (say $C_{2y}$) leading to the point group $D_4$ ($422$), as well as for MSG83.49 ($P_C 4/m$), obtained by including the basal mirror symmetry $m_z$ leading to the point group $C_{4h}$ ($4/m$).}

\textcolor{black}{The structural chirality is lost for MSG99.169 ($P_C 4mm$) and MSG83.49 ($P_C 4/m$), thus preventing the existence of Weyl nodes at the TRIMPs, i.e.~the Weyl points are absorbed within vertical nodal lines on the $C_{4z}$-axes, and, respectively, within a global twofold Kramer's degeneracy. For MSG89.93 ($P_C 422$) instead, the structural chirality, and thus the Weyl nodes, are preserved.} 

\textcolor{black}{We note that the introduction $m_z$ allows the definition of $C_4$-symmetry indicated mirror Chern numbers \cite{Khalaf_sum_indicators,Chenprb2012,peng2021topological}. The systematic study of the magnetic topological phases for the next magnetic super-space groups, however, lies beyond the scope of the present work.}

%We have discussed above the stability of the $C_4$-symmetric Weyl point quadruplets at half-filling located at generic momenta of the . 

\subsection{\textcolor{black}{Weyl phases protected by $[C_2T]^2=+1$}} 

\textcolor{black}{The symmetry indicator Eq.~(\ref{eq:::indicator}) and its interpretation in terms of a $\mathbb{Z}_2$ quantized Berry phase Eq.~(\ref{eq:indicated_berry}), derived here for MSG75.5 ($P_C4$), can be readily applied to many other MSGs with gapped 2D planes in the Brillouin zone where $[C_2T]^2=+1$. The simplest example MSG3.4 ($P_a2$), obtained by forgetting the $C_{4}$ symmetry, has a $\mathbb{Z}_2$ symmetry indicator that readily corresponds to $z_2$ Eq.~(\ref{eq:::indicator}) \cite{mSI}. For many MSGs though, there are symmetry indicated nodal points between the gapped planes, i.e.~similarly to MSG77.18 ($P_I4_2$), such that they cannot be identified by a (3D) symmetry indicator and they have been listed as trivial \cite{mSI}.}

\subsection{\textcolor{black}{Chern and Weyl phases of type I MSGs}}

\textcolor{black}{We have discussed in detail the transition from the AFM Weyl phase of MSG75.5 ($P_C4$), and MSG77.18 ($P_I4_2$), to the sub-dimensional Chern insulating FM phases of MSG75.1 ($P4$), and MSG77.13 ($P4_2$), respectively, obtained upon the breaking of the non-symmorphic TRS. The effect of breaking TRS is to unlock the Weyl nodes that were pinned on the $C_2T$ planes, so that they move within the Brillouin zone. We conclude that the symmetry indicator $\mathbb{Z}_4$ of the type I MSG75.1 ($P4$) \cite{mSI} indicates a We yl semi-metallic phase, while there is no symmetry indicator for MSG77.13 ($P4_2$) \cite{mSI} with a Weyl semi-metallic phase that must be assessed in terms of sub-dimensional topology.} 
                            
\textcolor{black}{In the same way as we predict many AFM phases with $C_2T$ protected Weyl nodal phases, we predict many FM Weyl nodal phases indicated by sub-dimensional Chern phases upon the breaking of the non-symmorphic TRS.} 

\section{Conclusions}
In conclusion, starting from the specific case study, MSG 75.5 \textcolor{black}{($P_C4$)}, we find that specific Wyckoff positions ($2b$) in this magnetic ground group necessarily results in (fragile) topological bands. 
In  this regard we formulated a first generic model exhibiting fragile topology in the context of magnetic space group symmetries. Breaking the essential symmetry by a Zeeman term then relates the underlying AFM-compatible MSG with a FM counterpart in the same space group family and ensures that the fragile topology gaps into bands with finite Chern number. After translating the $\mathbf{Z}_2$ symmetry indicator of the AFM MSG into a quantized Berry phase of a stable topological semimetallic phase, which originates from the combination of $C_4$ symmetry and $C_2T$-protected Euler class topology, we also discuss a similar correspondence to FM Chern phases. We thus unveil a systematic correspondence between necessarily nontrivial topological phases associated with MSG representations. Moreover, we then promote this mechanism to three spatial dimensions, where we also find novel phases characterized by the concept of {\it subdimensional topologies}. The latter feature the same in-plane mechanism but have 3D elementary band representations that are fully connected. As a result, the non-trivial 2D topology \textcolor{black}{must coexists with nodes} away from the high symmetry planes, e.g.~Weyl points, giving rise to \textcolor{black}{additional topological nodal features, such as} Fermi arcs, that can be diagnosed with established experimental methods. As a result, our work culminates in an exhaustive list of \textcolor{black}{tetragonal} MSGs \textcolor{black}{(with the point groups $C_4$ and $S_4$)} and their EBR content hosting the above correspondence. \textcolor{black}{We have then addressed the effect of adding and removing unitary symmetries that lead to the identification of magnetic Dirac (fourfold) points, and have outlined how the symmetry indicated Weyl semi-metallic phases protected by $C_2T$ can be found in numerous MSGs as a result of our refined subdimensional topological analysis}. Given the generality of these insights and relevance of parameters to access this physics, we hope our results pave the way for new pursuits in topological band structures. In fact, we anticipate that this \textcolor{black}{coexistence} effect, \textcolor{black}{i.e.~of gapped subdimensional topology together with independent topological nodes,} can also occur in the non-magnetic context culminating in novel \textcolor{black}{gapped-nodal topological} phases.

%%TC:ignore
%\newpage
%\clearpage
%\newpage
%\section*{Methods}

\begin{acknowledgments}
	%{\it Acknowledgments.}--
 R.-J.~S.~acknowledges funding from the Marie Sk{\l}odowska-Curie programme under EC Grant No. 842901 and the Winton programme as well as Trinity College at the University of Cambridge. G.F.L acknowledges funding from the Aker Scholarship.
 
 {\it Note added.--} We finally note that our results agree with expressions for magnetic EBRs, compatibility relations and symmetry indicators tabulated in \cite{mtqc}, which was posted during the finalizing stages of this manuscript.
 
\textcolor{black}{ {\it Note added.--} We comment here on Ref.~\cite{peng2021topological} that appeared several months after the first version of this paper and has some overlap with our work. Among many results not covered by our work, they give a complete list of MSGs that have $z_2$ (called $z_4$ in their work) as one of the 3D symmetry indicators, including some of the MSGs discussed here. While the scope of our work was more restricted, our approach based on sub-dimensional topologies, i.e.~allowing symmetry indicated nodes between gapped planes, leads us to predict many more MSGs with $C_2T$ protected Weyl semi-metallic phases. Ref.~\cite{peng2021topological} also considers other Weyl semi-metallic phases in type I MSGs protected by $IC_{4z}$ symmetry, among which are MSG81.33 ($P\bar{4}$) and MSG82.39 ($P\bar{4}$). Alternatively, these FM phases can be readily obtained, upon the breaking of the non-symmorphic TRS, from the (sub-dimensional) AFM Weyl phases of the following type IV MSG81.36 ($P_c\bar{4}$), MSG81.37 ($P_C\bar{4}$), MSG81.38 ($P_I\bar{4}$), and MSG82.42 ($I_c\bar{4}$), all listed in our table.}
 
\end{acknowledgments}

\newpage
\clearpage
\newpage
\appendix

%\section{Analysis of magnetic space group 75.5}

\section{Wilson loop winding for the fragile topological phase}\label{m:patch_WL}

We  algebraically derive the symmetry obstruction on the winding of Wilson loop over one quarter of the Brillouin zone, and following over the whole Brillouin zone, for the valence (conduction) subspace of the split EBR$^{2b}_{75.5}$. For this we choose the patch of the Brillouin zone bounded by $l_{\Gamma\text{X}\Gamma'}$ and $l_{\Gamma\text{M}\Gamma'}$, see Fig.~\ref{fig:wilson75.5}c) (blue dashed). We design a flow starting with the base path $l_{\Gamma\text{X}\Gamma'}$ and ending with $l_{\Gamma\text{M}\Gamma'}$. Defining the Wilson loop phases
\begin{equation}
    \begin{aligned}
\{\varphi_1,\varphi_2\} &=\mathrm{Arg}\left[ \mathrm{eig}\{\mathcal{W}[l_{\Gamma\text{X}\Gamma'}]\}\right],\\ \{\varphi_1',\varphi_2'\}&=\mathrm{Arg}\left[ \mathrm{eig}\{\mathcal{W}[l_{\Gamma\text{M}\Gamma'}]\}\right],
\end{aligned}
\end{equation}
they must extrapolate smoothly between $\{\varphi_1,\varphi_2\}$ and $\{\varphi_1',\varphi_2'\}$ as we smoothly deform the base path from $l_{\Gamma\text{X}\Gamma'}$ to $l_{\Gamma\text{M}\Gamma'}$. The Wilson loop over an oriented base path $l:k_1\rightarrow k_2$ is defined through $\mathcal{W}_{k_2\leftarrow k_1}=\langle \boldsymbol{u} ,k_2\vert \hat{W} \vert \boldsymbol{u}, k_1 \rangle$ where $ \vert \boldsymbol{u}, k_1 \rangle$ is the matrix of Bloch eigenvectors of the band subspace under consideration, and $\hat{W} = \prod_{k}^{k_2\leftarrow k_1} \mathcal{P}(k)$ with the projector $\mathcal{P}(k) =  \vert \boldsymbol{u}, k\rangle \langle \boldsymbol{u} ,k\vert$.

We now show that the crystal symmetries act as an obstruction imposing the quantization of Wilson loop phases. First, we find
\begin{equation}
\begin{aligned}
    \mathcal{W}[l_{\Gamma\text{X}\Gamma'}] &= \mathcal{W}_{\Gamma'\leftarrow \text{X}} \cdot\mathcal{W}_{ \text{X}\leftarrow\Gamma}\\
    &= R^{\Gamma}_2\cdot \mathcal{W}^{-1}_{\text{X}\leftarrow \Gamma} \cdot (R^{\text{X}}_2)^{-1} \cdot \mathcal{W}_{\text{X}\leftarrow \Gamma}  , 
\end{aligned}
\end{equation}
where $R^{\bar{\boldsymbol{k}}}_2 = \langle \boldsymbol{u}, D_{\pi}\bar{\boldsymbol{k}} \vert \hat{U}(C_{2z}) \vert \boldsymbol{u}, \bar{\boldsymbol{k}} \rangle$ is the sewing matrix of symmetry $C_{2z}$ at the high symmetry point $\bar{\boldsymbol{k}}$ in the valence Bloch eigenvectors basis (with $\hat{U}(C_{2z})=\mathbb{1}\otimes -\mathrm{i}\sigma_z$ the representation of $C_{2z}$ in the orbital basis $\vert \boldsymbol{\varphi},\boldsymbol{k} \rangle$, and $D_{\pi}$ is the rotation matrix by $\pi$ around the $z$-axis). Writing $\vert \boldsymbol{u}, \bar{\boldsymbol{k}}+\boldsymbol{K} \rangle = \hat{T}^{\dagger}(\boldsymbol{K}) \vert \boldsymbol{u}, \bar{\boldsymbol{k}} \rangle $ where $\hat{T}(\boldsymbol{K}) = \mathrm{diag}(\mathrm{e}^{\mathrm{i} \boldsymbol{r}_A \boldsymbol{K}},\mathrm{e}^{\mathrm{i} \boldsymbol{r}_A \boldsymbol{K}},\mathrm{e}^{\mathrm{i} \boldsymbol{r}_B \boldsymbol{K}},\mathrm{e}^{\mathrm{i} \boldsymbol{r}_B \boldsymbol{K}})$ accounts for the phase factors due to the displacements of the sub-lattice sites with respect to the origin of the unit cell, we can rewrite $R^{\bar{\boldsymbol{k}}}_2 = \langle \boldsymbol{u}, \bar{\boldsymbol{k}} \vert \hat{T}(D_{\pi} \bar{\boldsymbol{k}} - \bar{\boldsymbol{k}}) \hat{U}(C_{2z}) \vert \boldsymbol{u}, \bar{\boldsymbol{k}} \rangle$ which guarantees that the arbitrary gauge phase factors are removed, and use $\vert u,D_{\pi} \bar{\boldsymbol{k}} \rangle = \hat{U}(C_{2z})\vert u, \bar{\boldsymbol{k}} \rangle \left.R^{\bar{\boldsymbol{k}}}_2\right.^{\dagger} $ in the following. We note that at $C_{2z}$-invariant momenta, i.e.~$D_{\pi} \bar{\boldsymbol{k}} = \bar{\boldsymbol{k}} + \boldsymbol{K}$ with $\boldsymbol{K}$ a reciprocal lattice vector, we have $[\hat{T}(\boldsymbol{K}) \hat{U}(C_{2z}),H(\bar{\boldsymbol{k}})] = 0$ and $R^{\bar{\boldsymbol{k}}}_2$ is diagonal. Since for the fragile phase ($z_2=0\mod 2$, see Eq.~\ref{eq:::indicator}) we have $R^{\text{X}}_2 = (\pm) \mathrm{diag}(\mathrm{i},\mathrm{i})$ and $R^{\Gamma}_2 = \mathrm{diag}(\mathrm{i},-\mathrm{i})$ (see Table \ref{coIRREPs}), we readily find
\begin{equation}
    \{\varphi_1,\varphi_2\} = \{0,\pi\}\mod 2\pi.
\end{equation}

We emphasize that the quantization of the Wilson loop over $\Gamma\text{X}\Gamma'$ comes from the repetition of the IRREPs at X (equivalently at $\Gamma$). An alternative source of Wilson loop quantization is when both $\Gamma$ and X are TRIMs in which case the Wilson loop phases are Kramers degenerated at $\{0,0\}$ or $\{\pi,\pi\}$.

Considering now the base path $l_{\Gamma\text{M}\Gamma'}$, we have 
\begin{equation}
\begin{aligned}
    \mathcal{W}[l_{\Gamma\text{M}\Gamma'}] &= \mathcal{W}_{\Gamma'\leftarrow \text{M}} \cdot\mathcal{W}_{ \text{M}\leftarrow\Gamma}\\
    &= R^{\Gamma}_4\cdot \mathcal{W}^{-1}_{ \text{M}\leftarrow\Gamma} \cdot (R^{\text{M}}_4)^{-1} \cdot \mathcal{W}_{\text{M}\leftarrow\Gamma}  , 
\end{aligned}
\end{equation}
where $R^{\bar{\boldsymbol{k}}}_4= \langle \boldsymbol{u}, D_{\pi/2}\bar{\boldsymbol{k}} \vert \hat{U}(C_{4z}) \vert \boldsymbol{u}, \bar{\boldsymbol{k}} \rangle = \langle \boldsymbol{u}, \bar{\boldsymbol{k}} \vert \hat{T}(C_{4z}\bar{\boldsymbol{k}} -\bar{\boldsymbol{k}} ) \hat{U}(C_{4z}) \vert \boldsymbol{u}, \bar{\boldsymbol{k}} \rangle $ is the representation in the basis of Bloch eigenvectors of $C_{4z}$ at $\bar{\boldsymbol{k}}$, with, in the orbital basis $\vert \boldsymbol{\varphi},\boldsymbol{k}\rangle$, $\hat{U}(C_{4z}) = \sigma_x\otimes M_4$ where $M_4 = \mathrm{diag}(\mathrm{e}^{-\mathrm{i} \pi/4},\mathrm{e}^{\mathrm{i} \pi/4})$, and we have used $\vert u,D_{\pi/2} \bar{\boldsymbol{k}} \rangle = \hat{U}(C_{4z})\vert u, \bar{\boldsymbol{k}} \rangle \left.R^{\bar{\boldsymbol{k}}}_4\right.^{\dagger} $. 

Using parallel transported Bloch eigenvectors the Wilson loop becomes diagonal, and we write $\widetilde{\mathcal{W}}_{\mathrm{M} \leftarrow \Gamma} = \mathrm{diag}(\mathrm{e}^{\text{i} \varphi_a},\mathrm{e}^{\text{i} \varphi_b})$. The above expression thus reduces to
\begin{equation}
\begin{aligned}
    \widetilde{\mathcal{W}}[l_{\Gamma\text{M}\Gamma'}] 
    &= \widetilde{R}^{\Gamma}_4\cdot \left(\subalign{\mathrm{e}^{-\text{i} \varphi_a} &~0\\
    0 \quad &\mathrm{e}^{-\text{i} \varphi_b}}\right) \cdot (\widetilde{R}^{\text{M}}_4)^{-1} \cdot \left(\subalign{\mathrm{e}^{\text{i} \varphi_a} &~0\\
    0 \quad &\mathrm{e}^{\text{i} \varphi_b}}\right)  .
\end{aligned}
\end{equation}
Since at $C_{4z}$-symmetric momenta the parallel transported Bloch eigenvectors are also eigenvectors of the $C_{4z}$ operator, we retrieve irreducible representations $\widetilde{R}^{\Gamma}_4 = (\pm) \mathrm{diag}(\omega,\omega^*)$ and $\widetilde{R}^{\text{M}}_4 = (\pm) \mathrm{diag}(\omega,-\omega^*)$, where the signs depend on the coIRREPs realized at $\Gamma$ and M (see Table \ref{coIRREPs}). 

The quantization of the Wilson loop $\widetilde{W}[l_{\Gamma\text{M}\Gamma'}]$ depends on the relative spin-$z$ components of the parallel transported Bloch eigenstates at $\Gamma$ and M, see the discussion in Ref.~\cite{bouhon2019wilson}. Writing $\widetilde{R}^{\bar{\boldsymbol{k}}}_4 = \mathrm{diag}(\xi^{\bar{\boldsymbol{k}}}_4(\Gamma_a) , \xi^{\bar{\boldsymbol{k}}}_4(\Gamma_b))$, we find  
\begin{equation}
\begin{aligned}
    \widetilde{\mathcal{W}}[l_{\Gamma\text{M}\Gamma'}] 
    &= \left(\begin{array}{cc}
         \xi^{\Gamma}_4(\overline{\Gamma}_a)/\xi^{\text{M}}_4(\overline{\text{M}}_a) &  0\\
        0 & \xi^{\Gamma}_4(\overline{\Gamma}_b)/\xi^{\text{M}}_4(\overline{\text{M}}_b)
    \end{array} \right)  .
\end{aligned}
\end{equation}
Assuming the same spin-$z$ component at $\Gamma$ and M, e.g.~with $(\overline{\Gamma}_a,\overline{\text{M}}_a,\overline{\Gamma}_b,\overline{\text{M}}_b)=(\overline{\Gamma}_5,\overline{\text{M}}_8,\overline{\Gamma}_7,\overline{\text{M}}_5)$ (see Table \ref{coIRREPs}), we find
\begin{equation}
    \{\varphi_1',\varphi_2'\} = \{\pi/4,\pi/4\}\mod 2\pi.
\end{equation}
This matches exactly with the direct numerical evaluation of the Wilson loop shown in Fig.~\ref{fig:wilson75.5}e) in the main text. If instead we assume opposite spin-$z$ components at $\Gamma$ and M, e.g.~with $(\overline{\Gamma}_a,\overline{\text{M}}_a,\overline{\Gamma}_b,\overline{\text{M}}_b)=(\overline{\Gamma}_5,\overline{\text{M}}_5,\overline{\Gamma}_7,\overline{\text{M}}_8)$ (see Table \ref{coIRREPs}), we find
\begin{equation}
    \{\varphi_1',\varphi_2'\} = \{0,\pi\}\mod 2\pi.
\end{equation} 
This later quantization thus corresponds to a system where there is a twisting spin texture from $\Gamma$ to M, which would require strong Rashba-type spin-orbit coupling.

It thus follows that, in the absence of a twisted spin texture, the Wilson loop phases must wind from $\{\varphi_1,\varphi_2\} = \{0,
\pi\}$ to $\{\varphi_1',\varphi_2'\} = \{\pi/4,\pi/4\}\mod 2\pi$, as we scan over one quarter of the Brillouin zone through the deformation of the base path from $l_{\Gamma\text{X}\Gamma'}$ to $l_{\Gamma\text{M}\Gamma'}$. There is thus a minimal winding of $(\Delta\varphi_1,\Delta\varphi_2) = (\varphi_1',\varphi_2')-(\varphi_1,\varphi_2)= (+\pi/2,-\pi/2)$. By the action of $C_4$ symmetry we can recover the whole Brillouin zone for which we predict a minimal winding of the Wilson loop phases of $4(\Delta\varphi_1,\Delta\varphi_2) = (+2\pi,-2\pi)$. This precisely predicts algebraically the numerical evaluation of the Wilson loop over the whole Brillouin zone shown in Fig.~\ref{fig:wilson75.5}d).

We finally conclude that the valence (conduction) subspace of the split EBR$^{2b}_{75.5}$ is topologically non-trivial as indicated by the finite winding of Wilson loop phases.  

\section{Derivation of formula Eq.~(\ref{eq:sym_chern}) and Eq.~(\ref{eq:indicated_berry})}\label{m:derivation_eq4}

We give here the algebraic derivation of the Wilson loop over $l_q = \Gamma \text{M'}\text{X}\text{M}\Gamma$ (red loop in Fig.~\ref{fig:wilson75.5}c)) as given in Eq.~(\ref{eq:indicated_berry}), from which Eq.~(\ref{eq:sym_chern}) readily follows by reducing to a single band-subspace. We use the algebraic Wilson loop techniques developed in \cite{Chenprb2012,hourglass} and \cite{Wi2,BBS_nodal_lines,bouhon2019wilson}. 

It is convenient to decompose the Wilson loop into the contributions of each segment that connects two successive high-symmetry points, i.e.~$\mathcal{W}[l_q] = \mathcal{W}_d\mathcal{W}_c\mathcal{W}_b\mathcal{W}_a$, with
\begin{equation}
    \begin{array}{rclrcl}
        \mathcal{W}_a & =& \langle \boldsymbol{u},\Gamma\vert \hat{W} \vert \boldsymbol{u}, \text{M} \rangle, & \mathcal{W}_c & =& \langle \boldsymbol{u},\text{X}\vert \hat{W} \vert \boldsymbol{u}, \text{M}' \rangle,\\
        \mathcal{W}_b & =& \langle \boldsymbol{u},\text{M}'\vert \hat{W} \vert \boldsymbol{u}, \Gamma \rangle, &
        \mathcal{W}_d & =& \langle \boldsymbol{u},\text{M}\vert \hat{W} \vert \boldsymbol{u}, \text{X} \rangle ,
    \end{array}
\end{equation}
where $\text{M'}=\text{M}-\boldsymbol{b}_2$. 

We now use symmetries to rewrite $\mathcal{W}_a$ and $\mathcal{W}_d$, as 
\begin{equation}
    \begin{aligned}
        \mathcal{W}_a & = \langle \boldsymbol{u},\Gamma\vert \hat{W} \vert \boldsymbol{u}, \text{M} \rangle 
        = \langle \boldsymbol{u},C_{4z}\Gamma\vert \hat{W} \vert \boldsymbol{u}, C_{4z}\text{M'} \rangle \\
        & = R^{\Gamma}_4 \cdot \langle \boldsymbol{u},\Gamma\vert \hat{U}^{\dagger}(C_{4z})  \hat{W} \hat{U}(C_{4z})  \vert \boldsymbol{u}, \text{M'} \rangle \cdot (R^{\text{M'}}_4)^{\dagger} \\
        & = R^{\Gamma}_4 \cdot \mathcal{W}^{-1}_b \cdot  (R^{\text{M'}}_4)^{\dagger},\\
    \mathcal{W}_d &= \langle \boldsymbol{u},\text{M}\vert \hat{W} \vert \boldsymbol{u}, \text{X} \rangle = 
        \langle \boldsymbol{u},C_{2z}\text{M''}\vert \hat{W} \vert \boldsymbol{u}, C_{2z}\text{X'} \rangle \\
        &= R^{\text{M''}}_2 \cdot\langle \boldsymbol{u},\text{M''}\vert \hat{U}^{\dagger}(C_{2z})\hat{W} \hat{U}(C_{2z})\vert \boldsymbol{u},\text{X'} \rangle \cdot (R^{\text{X'}}_2)^{\dagger}\\
        &= R^{\text{M''}}_2 \cdot \langle \boldsymbol{u},\text{M'}\vert\hat{T}(-\boldsymbol{b}_1) \hat{U}^{\dagger}(C_{2z})\hat{W} \\
        & \quad\quad\quad\quad\quad\quad\hat{U}(C_{2z}) \vert\hat{T}^{\dagger}(-\boldsymbol{b}_1)  \boldsymbol{u},\text{X'} \rangle \cdot (R^{\text{X'}}_2)^{\dagger}\\
        &= R^{\text{M''}}_2 \cdot \mathcal{W}^{-1}_c\cdot (R^{\text{X'}}_2)^{\dagger},
    \end{aligned}
\end{equation}
where $\text{M''} = \text{M'}-\boldsymbol{b}_1$ and $\text{X'} = \text{X}-\boldsymbol{b}_1$. We thus have, 
\begin{equation}
    \mathrm{Det}\,\mathcal{W}[l_q] = \mathrm{Det}\left[ 
    R^{\text{M''}}_2 \cdot (R^{\text{X'}}_2)^{\dagger} \cdot R^{\Gamma}_4 \cdot (R^{\text{M'}}_4)^{\dagger} \right].
\end{equation}

Defining the irreducible representation of the symmetry $(g\vert\tau_g)$ in the basis of Bloch eigenstates as
\begin{equation}
    S^{\bar{\boldsymbol{k}}}_{g} = 
    \mathrm{e}^{-\mathrm{i}\, g\bar{\boldsymbol{k}}\cdot \tau_g } R^{\bar{\boldsymbol{k}}}_{g},
\end{equation}
and substituting in the above expression, we get
\begin{equation}
\begin{aligned}
    \mathrm{Det}\,\mathcal{W}[l_q] &= \mathrm{e}^{\mathrm{i}\, N_{\mathrm{occ}} \;[ (\text{M}-\text{X})\cdot \tau_{C_{2z}} +
    (\Gamma-\text{M})\cdot \tau_{C_{4z}}] } \\
    &\quad \quad\quad\mathrm{Det}\left[ 
    S^{\text{M}}_2 \cdot (S^{\text{X}}_2)^{\dagger} \cdot S^{\Gamma}_4 \cdot (S^{\text{M}}_4)^{\dagger} \right]\\
    &= \mathrm{e}^{\mathrm{i}\, N_{\mathrm{occ}} \;[ (\text{M}-\text{X})\cdot \tau_{C_{2z}} +
    (\Gamma-\text{M})\cdot \tau_{C_{4z}}] } \prod\limits_{i=1}^{N_{\mathrm{occ}}} 
    \dfrac{\xi^{\Gamma}_4(i) \xi^{\text{M}}_2(i)  }{\xi^{\text{M}}_4(i) \xi^{\text{X}}_2(i)},
\end{aligned}
\end{equation}
where $N_{\mathrm{occ}}$ is the number of occupied bands, and $\xi^{\bar{\boldsymbol{k}}}_g(i)$ is the eigenvalue of the symmetry $(g\vert\tau_g)$ of the $i$-th band at the high-symmetry momentum $\bar{\boldsymbol{k}}$. For magnetic space groups with symmorphic $C_{4z}$ and $C_{2z}$ symmetries (i.e.~$\tau_{C_{4z}} = \tau_{C_{2z}} = 0$) as MSG75.5 (and MSG75.1), it simplifies to  
\begin{equation}
    \mathrm{Det}\,\mathcal{W}[l_q] = \prod\limits_{i=1}^{N_{\mathrm{occ}}} 
    \dfrac{\xi^{\Gamma}_4(i) \xi^{\text{M}}_2(i)  }{\xi^{\text{M}}_4(i) \xi^{\text{X}}_2(i)}.
\end{equation}

\section{Numerical computation of Berry phase flows for MSG75.1 \textcolor{black}{($P4$)}}\label{ap:numerical_appendix_75.1}
%Finally, we also present extra details on the band structure of the MSG75.1 cases. First, in Fig.~\ref{fig:msg75.1bs} we detail the gapped band structure as defined by gapping the fragile topological MSG75.5 model, see main text.
%\begin{figure}[h!]
%    \centering
%    \includegraphics[width=\linewidth]{fig6_combined_V2.pdf}
%    \caption{Numerical evaluation for MSG75.1 model a) Band structure of the phase with broken nonsymmorphic time reversal, i.e.~MSG75.1. The Zeeman coupling gaps all the Kramers degeneracies. b) Band structure for MSG75.1 along the high-symmetry lines.}
%    \label{fig:msg75.1bs}
%\end{figure}

\begin{figure}%[th!]
    \centering
   \includegraphics[width=0.49\textwidth]{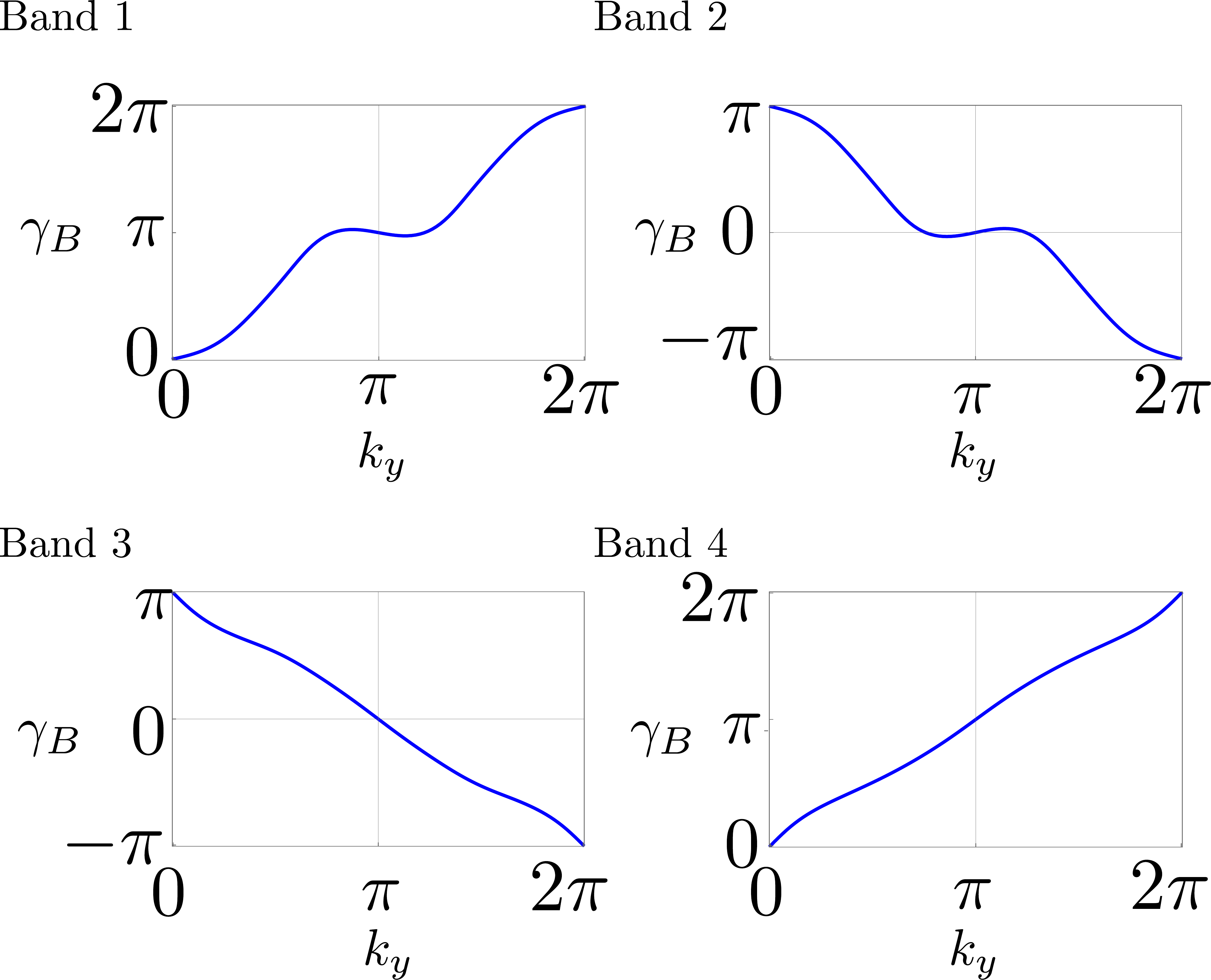}
    \caption{Berry phase flows for the four split bands of MSG75.1 \textcolor{black}{($P4$)} as obtained  from gapping the fragile topological MSG75.5 \textcolor{black}{($P_C4$)} phase. Band 1 and 4 have Chern number ${\cal C}=+1$. Band 2 and 3 exhibit Chern number ${\cal C}=-1$.}
    \label{fig:berry}
\end{figure}

We present here the numerical evaluation of the nontriviality of the case MSG75.1 \textcolor{black}{($P4$)}, i.e.~for the model Eq.~(\ref{model_75_5}) with an additional Zeeman coupling.
In Fig.~\ref{fig:berry}. we show the numerically obtained Berry phase for the individual bands of the band structure in Fig.~\ref{fig:MSG75.1} obtained for the model of MSG75.1 \textcolor{black}{($P4$)}. These evaluations corroborate the analytical results. That is, each band shows a finite Chern number $\cal{C}$. While bands $2$ and $3$ exhibit a value of ${\cal C}=1$, the other two bands have opposite Chern number. 
\begin{figure}%[th!]
    \centering
        \includegraphics[width=\linewidth]{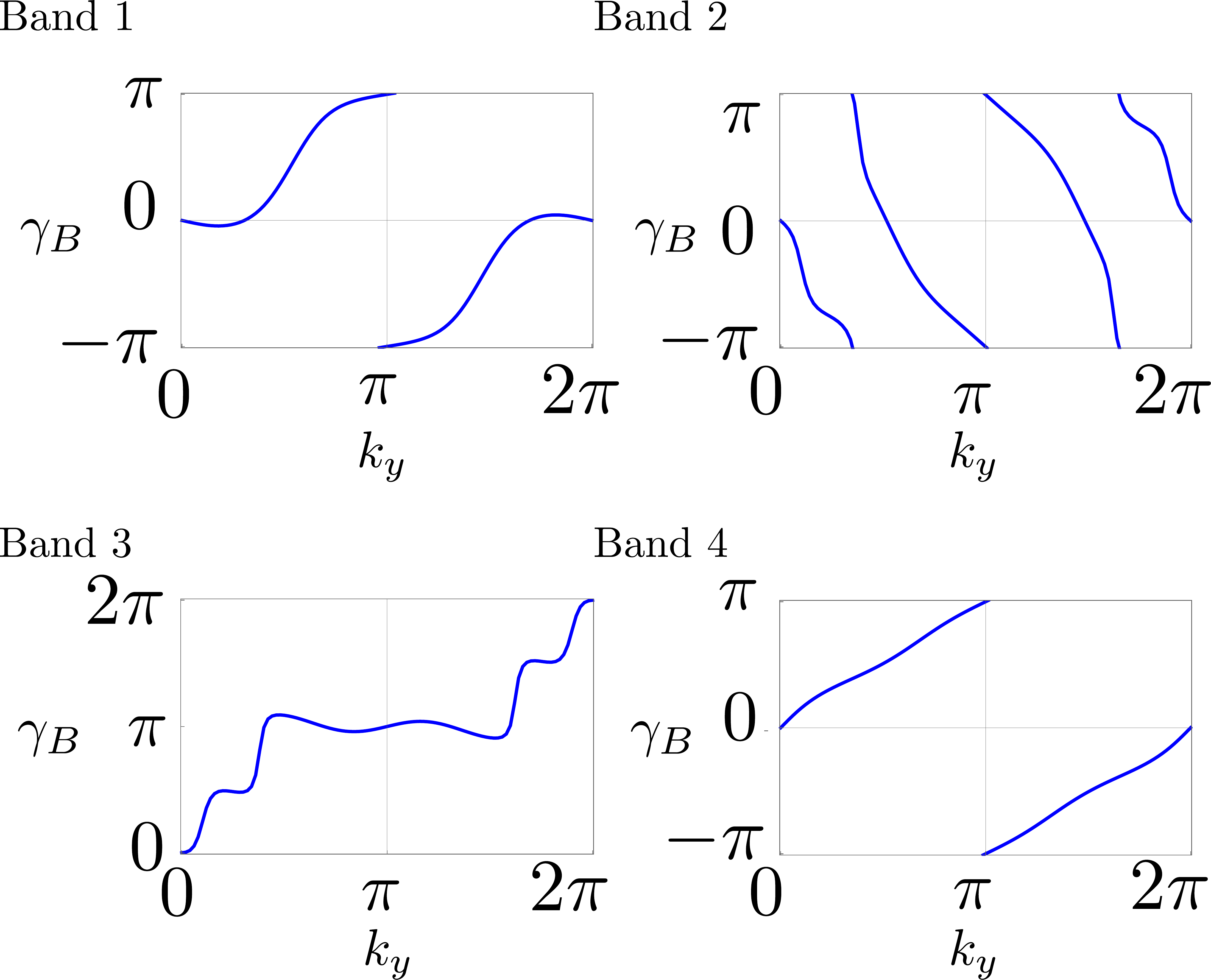}
    \caption{Berry phase flows for the four split bands of MSG75.1 \textcolor{black}{($P4$)} obtained from the gapping of the stable nodal phase of MSG75.5 \textcolor{black}{($P_C4$)}. Band 1, 3 and 4 have Chern number ${\cal C}=+1$. Band 2 has has higher Chern number ${\cal C}=-3$.}
    \label{fig:berry_stable}
\end{figure}
Finally, we also show in Fig.~\ref{fig:berry_stable} the Berry phase of the bands as obtained from gapping the stable nodal topological MSG75.5 \textcolor{black}{($P_C4$)} band structure of Fig.~\ref{fig:stable_nodal}. As described in the main text, the resulting spectrum features a single band (band number two in this case)
with Chern number ${\cal C}=-3$, whereas the others exhibit a Chern number ${\cal C}=1$. 

%\subsection{Further 3D embedding of MSG75.5 model}\label{ap:numerical_appendix_75.5_embedding}
%As stated in the main text, we here discuss alternative Weyl nodal phases for the 3D embedding of the MSG75.5. In Fig.~\ref{fig:additional75.5}a) we show the configuration of Weyl points for $(z_2^0,z_2^{\pi})=(1,1)$ when the Weyl points of the planes $k_z=0,\pi$ have opposite chiralities and such that the vertical Weyl nodes are not annihilated (this can be compared with Fig.~\ref{fig:weyls_755}a)). In Fig.~\ref{fig:additional75.5}b) we show the configuration of Weyl points, still for $(z_2^0,z_2^{\pi})=(1,1)$, when the Weyl points of the planes $k_z=0,\pi$ have the same chirality.

%\begin{figure}
%    \centering
%    \includegraphics[width=\linewidth]{Figure5_appendix.pdf}
%    \caption{Additional scenarios pertaining to  Fig.~\ref{fig:weyls_755} for MSG75.5 with $(z_2^0,z_2^{\pi}) = (1,1)$. a) Same as Fig.~\ref{fig:weyls_755}b) before the annihilation of the Weyl points on the vertical axes. b) Alternative to Fig.~\ref{fig:weyls_755}b) when the horizontal (vertical) Weyl points all have equal chirality. }
%    \label{fig:additional75.5}
%\end{figure}

%\section{Further details on MSG77.18 \textcolor{black}{($P_I4_2$)} and MSG77.13 \textcolor{black}{($P4_2$)}}\label{ap:numerical_appendix_77.18_77.13}

\section{Details on symmetry indicator analysis}\label{ap:SI_analysis}
We here give further detail on the symmetry indicator analysis for some of the MSGs considered.

\subsection{Symmetry indicators for MSG75.5 \textcolor{black}{($P_C4$)}}\label{ap:SI_analysis_75.5}
MSG75.5 \textcolor{black}{($P_C4$)} is derived from SG75 by including the operator $(E|\tau)'$ with $\tau = \frac{1}{2}(\mathbf{a}_1+\mathbf{a}_2)$. We can thus classify the possible band structures for our model by considering the IRREPs of SG75, and pair them appropriately to form coIRREPs of MSG75.5 \textcolor{black}{($P_C4$)}, as described in \cite{mSI}.  
The full compatibility relations\cite{Clas3} and EBRs for SG75 can be found on the \texttt{Bilbao Crystallographic Server}\cite{Bilbao}. Using their notation (see also Table \ref{coIRREPs}), the compatibility relations for the $\Gamma$ point are given by:
\begin{equation}
    \begin{split}
        \overline{\Gamma}_5 \rightarrow \overline{Z}_5\\
        \overline{\Gamma}_6 \rightarrow \overline{Z}_6\\
        \overline{\Gamma}_7 \rightarrow \overline{Z}_7\\
        \overline{\Gamma}_7 \rightarrow \overline{Z}_8\\
    \end{split}
\end{equation}
The compatibility relations between M and A display a similar structure. Finally, the compatibility relations between X and R are:
\begin{equation}
    \begin{split}
        \overline{X}_3 \rightarrow \overline{R}_3\\
        \overline{X}_4 \rightarrow \overline{R}_4\\
    \end{split}
\end{equation}
From the above relations it is evident that it suffices to consider the points $\Gamma$, X and M as the compatibility relations uniquely link IRREPs at these points to all IRREPs at all other high-symmetry points in the 3D BZ \cite{Clas3}. To form the coIRREPs, we consider the additional constraints imposed by the antiunitary symmetries. These can be determined from the Herring rule \cite{BradCrack} applied at each high-symmetry point. We only consider spinful IRREPs. At $\Gamma$ and M (and A and Z), this gives pairing of inequivalent IRREPs.  At X (and R), no additional pairing is required. To determine which IRREPs are paired to form the coIRREPs at $\Gamma$ and M, we pair representations of $g$ with representations of $AgA^{-1}$, where $g$ is an element of the unitary little group at $\Gamma$ or M and  $A = (E|\tau)'$. The allowed pairings at $\Gamma$ are $\overline{\Gamma}_5\overline{\Gamma}_7,\overline{\Gamma}_6\overline{\Gamma}_8$ and the pairings at M are $\overline{M}_6\overline{M}_7$, $\overline{M}_5\overline{M}_8$.

The magnetic EBRs for the MSG75.5 \textcolor{black}{($P_C4$)} can be found from the EBRs for SG75 \cite{Bilbao}. We note that the vector $\tau$ relates WP $1a$ $(0,0)$ and $1b$ $(1/2,1/2)$ in SG75, so the EBRs from these WPs are paired. Similarly, $\tau$ maps WP $2c$ $(0,1/2),(1/2,0)$ in SG75 to itself, so we pair EBRs at  WP $2c$ directly.  We can uniquely determine which EBRs are paired, by realizing that the magnetic EBRs must satisfy the magnetic compatibility relations detailed above. This gives the following magnetic EBRs for MSG75.5 \textcolor{black}{($P_C4$)} (using the WP labels from SG75):
\begin{widetext}
\begin{align*}
       (1a:\ ^1\overline{E}_1+1b:\ ^2\overline{E}_2) \uparrow G & =  (\overline{\Gamma}_6\overline{\Gamma}_8,\overline{M}_6\overline{M}_7,2\overline{X}_3)\\
      (1a:\ ^1\overline{E}_2+1b:\ ^2\overline{E}_1)\uparrow G &=  (\overline{\Gamma}_5\overline{\Gamma}_7,\overline{M}_5\overline{M}_8,2\overline{X}_3)\\
     (1a:\ ^2\overline{E}_1+1b:\ ^1\overline{E}_2)\uparrow G  &=  (\overline{\Gamma}_6\overline{\Gamma}_8,\overline{M}_5\overline{M}_8,2\overline{X}_4)\\
    (1a:\ ^2\overline{E}_2+1b:\ ^1\overline{E}_1)\uparrow G &=  (\overline{\Gamma}_5\overline{\Gamma}_6,\overline{M}_6\overline{M}_7,2\overline{X}_4)\\
     (2c:\ ^1\overline{E}+2c:\ ^2\overline{E})\uparrow G &=  (\overline{\Gamma}_5\overline{\Gamma}_6\overline{\Gamma}_7\overline{\Gamma}_8,\overline{M}_5\overline{M}_6\overline{M}_7\overline{M}_8,2\overline{X}_42\overline{X}_3)\\
\end{align*}
\end{widetext}
To compute the symmetry indicators, as described in \cite{mSI}, we compute the Smith normal form of the matrix of EBRs. This gives the indicator group for MSG75.5 \textcolor{black}{($P_C4$)} as $\mathbb{Z}_2$, in agreement with \cite{mSI}.

From our magnetic EBRs, we see that it is not possible to construct a band structure with an odd number of bands in the $\overline{X}_3$ (or equivalently the $\overline{X}_4$) IRREP from an integer combination of magnetic EBRs. Every other combination consistent with the compatibility relations can be constructed from the magnetic EBRs. Thus, the $\mathbb{Z}_2$ indicator can be conveniently computed as:
\begin{equation}
n_{\overline{X}_3}\ \mathrm{mod}\ 2,
\end{equation}
in agreement with the expression in the main text.

\subsection{Symmetry indicators for MSG77.18 \textcolor{black}{($P_I4_2$)}}\label{ap:SI_analysis_77.18}
MSG77.18 \textcolor{black}{($P_I4_2$)} is formed from SG77, by including the operator $(E|\tau)'$ with $\tau = \frac{1}{2}(\mathbf{a}_1+\mathbf{a}_2+\mathbf{a}_3)$. For the $C_4$ symmetric points ($\Gamma$,Z,M,A), the Herring test gives the same result as for 75.5. At X, the Herring test gives that the antiunitary symmetries impose no further degeneracies, whereas at R, the Herring test gives that each IRREP must be doubly degenerate. Additional degeneracies are imposed by the non-symmorphic symmetry elements of SG77. For the $C_4$ symmetric points, the non-symmorphic symmetries constrain that the IRREPs must switch partners when moving through the BZ. This enforces gap closings along the $k_z$ direction, which acts as an obstruction to defining two-band subspaces across the entire BZ, see figure \ref{fig:MSG77.18_13_BS}a) and c). Thus the minimal connectivity of bands in the BZ is 4.\\
\linebreak
The symmetry indicator group can be computed as before. WP $2c$ $(0,1/2,z),(0,1/2,z+1/2)$ of SG77 goes into WP $2a$ of MSG77.18 \textcolor{black}{($P_I4_2$)}, and the two spinful site-symmetry IRREPs glue together. Similarly, WP $2a$ $(0,0,z)$,$(0,0,z+1/2)$ and $2b$ $(1/2,1/2,z)$,$(1/2,1/2,z+1/2)$ of SG77 go into WP $4b$ of MSG77.18 \textcolor{black}{($P_I4_2$)}, and by checking the compatibility relations, we realize that IRREPs $^1\overline{E}$ pairs with $^2\overline{E}$ at the different sites. Thus, we get three spinful magnetic EBRs, which are given by (using the WP labels from SG77):
\begin{widetext}
\begin{align*}
       (2c:\ \overline{E}_1+2c:\ \overline{E}_2) \uparrow G &=  (\overline{\Gamma}_5\overline{\Gamma}_6 \overline{\Gamma}_7\overline{\Gamma}_8,\overline{Z}_5\overline{Z}_6 \overline{Z}_7\overline{Z}_8,\overline{M}_5\overline{M}_6 \overline{M}_7\overline{M}_8,\overline{A}_5\overline{A}_6 \overline{A}_7\overline{A}_8,2\overline{X}_32\overline{X}_4, 2\overline{R}_32\overline{R}_4)\\
      (2a:\ \overline{E}_1+2b:\ \overline{E}_2)\uparrow G &=  (\overline{\Gamma}_5\overline{\Gamma}_6 \overline{\Gamma}_7\overline{\Gamma}_8,\overline{Z}_5\overline{Z}_6 \overline{Z}_7\overline{Z}_8,\overline{M}_5\overline{M}_6 \overline{M}_7\overline{M}_8,\overline{A}_5\overline{A}_6 \overline{A}_7\overline{A}_8,4\overline{X}_3, 4\overline{R}_3)\\
            (2a:\ \overline{E}_2+2b:\ \overline{E}_1)\uparrow G &=  (\overline{\Gamma}_5\overline{\Gamma}_6 \overline{\Gamma}_7\overline{\Gamma}_8,\overline{Z}_5\overline{Z}_6 \overline{Z}_7\overline{Z}_8,\overline{M}_5\overline{M}_6 \overline{M}_7\overline{M}_8,\overline{A}_5\overline{A}_6 \overline{A}_7\overline{A}_8,4\overline{X}_4, 4\overline{R}_4)\\
\end{align*}
\end{widetext}
Using the Smith normal form decomposition as before, we see that there's no nontrivial indicator in  this MSG, in agreement with \cite{mSI}.

\subsection{Symmetry indicator for MSG81.36 \textcolor{black}{($P_c\bar{4}$)}}\label{ap:MSG81.36_sym_ind}
As a final example, we here compute the symmetry indicators for MSG81.36 \textcolor{black}{($P_c\bar{4}$)}, as an example of an MSG with $S_4$ rotoinversion symmetry. This MSG is generated from SG81 by including the operator $(E|\tau)'$ with $\tau = \frac{1}{2}\mathbf{a}_3$. SG81 is similar to SG75, but the $C_4$ rotations are combined with inversion. The Herring test gives that the $C_2$ IRREPs must glue together at the $C_2$ symmetric points X and R. At the $C_4$ symmetric points, the Herring test glues together $\overline{\Gamma}_5\overline{\Gamma}_7$ and $\overline{\Gamma}_6\overline{\Gamma}_8$ respectively, and similarly at M. At Z, $\overline{Z}_5\overline{Z}_8$ and $\overline{Z}_6\overline{Z}_7$ glue together respectively, and similarly at A. This assignment satisfies the compatibility relations, so the minimal connectivity of bands in the BZ is 2, and two-band subspaces can be defined throughout the BZ.\\
To compute the symmetry indicator group, we note that $\tau$ connects WP $1a$ $(0,0,0)$ and $1b$ $(0,0,1/2)$ of SG81 to form magnetic WP $2a$. Similarly, WP $1c$ $(1/2,1/2,0)$ and $1d$ $(1/2,1/2,1/2)$ of SG81 connect to form magnetic WP $2c$. Some of the non-maximal WPs of SG81 become maximal WPs for MSG81.36 \textcolor{black}{($P_c\bar{4}$)}, e.g. WP $2e$ $(0,0,z),(0,0,-z)$ of SG81 goes into maximal magnetic WP $2b$ $(0,0,1/4),(0,0,3/4)$, and similarly for non-magnetic WP $2f$ going into magnetic WP $2d$. As the EBRs at these WPs have to satisfy compatibility relations, however, it is straightforward to investigate which site-symmetry IRREPs to pair. This gives the following magnetic EBRs (labelled using the WPs of SG81):
\begin{widetext}
\begin{align*}
       (1a:\ ^1\overline{E}_1+1b:\ ^2\overline{E}_1) \uparrow G &= (\overline{\Gamma}_6\overline{\Gamma}_8,\overline{Z}_6\overline{Z}_7,\overline{M}_6\overline{M}_8,\overline{A}_6\overline{A}_7,\overline{X}_3\overline{X}_4, \overline{R}_3\overline{R}_4)\\
       (1a:\ ^1\overline{E}_2+1b:\ ^2\overline{E}_2) \uparrow G &= (\overline{\Gamma}_5\overline{\Gamma}_7,\overline{Z}_5\overline{Z}_8,\overline{M}_5\overline{M}_7,\overline{A}_5\overline{A}_8,\overline{X}_3\overline{X}_4, \overline{R}_3\overline{R}_4)\\
       (1a:\ ^2\overline{E}_1+1b:\ ^1\overline{E}_1) \uparrow G &= (\overline{\Gamma}_6\overline{\Gamma}_8,\overline{Z}_5\overline{Z}_8,\overline{M}_6\overline{M}_8,\overline{A}_5\overline{A}_8,\overline{X}_3\overline{X}_4, \overline{R}_3\overline{R}_4)\\
      (1a:\ ^2\overline{E}_2+1b:\ ^1\overline{E}_2) \uparrow G &= (\overline{\Gamma}_5\overline{\Gamma}_7,\overline{Z}_6\overline{Z}_7,\overline{M}_5\overline{M}_7,\overline{A}_6\overline{A}_7,\overline{X}_3\overline{X}_4, \overline{R}_3\overline{R}_4)\\
     (1c:\ ^1\overline{E}_1+1d:\ ^2\overline{E}_1) \uparrow G &= (\overline{\Gamma}_6\overline{\Gamma}_8,\overline{Z}_6\overline{Z}_7,\overline{M}_5\overline{M}_7,\overline{A}_5\overline{A}_8,\overline{X}_3\overline{X}_4, \overline{R}_3\overline{R}_4)\\
    (1c:\ ^1\overline{E}_2+1d:\ ^2\overline{E}_2) \uparrow G &= (\overline{\Gamma}_5\overline{\Gamma}_7,\overline{Z}_5\overline{Z}_8,\overline{M}_6\overline{M}_8,\overline{A}_6\overline{A}_7,\overline{X}_3\overline{X}_4, \overline{R}_3\overline{R}_4)\\
    (1c:\ ^2\overline{E}_1+1b:\ ^1\overline{E}_1) \uparrow G &= (\overline{\Gamma}_6\overline{\Gamma}_8,\overline{Z}_5\overline{Z}_8,\overline{M}_5\overline{M}_7,\overline{A}_6\overline{A}_7,\overline{X}_3\overline{X}_4, \overline{R}_3\overline{R}_4)\\
    (1c:\ ^2\overline{E}_2+1d:\ ^1\overline{E}_2) \uparrow G &= (\overline{\Gamma}_5\overline{\Gamma}_7,\overline{Z}_6\overline{Z}_7,\overline{M}_6\overline{M}_8,\overline{A}_5\overline{A}_8,\overline{X}_3\overline{X}_4, \overline{R}_3\overline{R}_4)\\
     (2e:\ ^1\overline{E}+2e:\ ^2\overline{E}) \uparrow G &= (\overline{\Gamma}_5\overline{\Gamma}_6\overline{\Gamma}_7\overline{\Gamma}_8,\overline{Z}_5\overline{Z}_6\overline{Z}_7\overline{Z}_8,\overline{M}_5\overline{M}_6\overline{M}_7\overline{M}_8,\overline{A}_5\overline{A}_6\overline{A}_7\overline{A}_8,2\overline{X}_32\overline{X}_4, 2\overline{R}_32\overline{R}_4)\\
    (2f:\ ^1\overline{E}+2f:\ ^2\overline{E}) \uparrow G &= (\overline{\Gamma}_5\overline{\Gamma}_6\overline{\Gamma}_7\overline{\Gamma}_8,\overline{Z}_5\overline{Z}_6\overline{Z}_7\overline{Z}_8,\overline{M}_5\overline{M}_6\overline{M}_7\overline{M}_8,\overline{A}_5\overline{A}_6\overline{A}_7\overline{A}_8,2\overline{X}_32\overline{X}_4, 2\overline{R}_32\overline{R}_4)\\
\end{align*}
\end{widetext}
Computing the Smith normal form gives a single $\mathbb{Z}_2$ factor. Inspecting the solution space, we see that the corresponding indicator is given by whether the IRREPs at the $C_4$ invariant points contain an odd or an even number of representations with subscript 5 (or any other subscript). Thus, an explicit expression for the symmetry indicator is:
\begin{equation}
\label{eq:SI_axion}
z_2' = n_{\overline{\Gamma}_5}+n_{\overline{Z}_5}+n_{\overline{M}_5}+n_{\overline{A}_5}\ \mathrm{mod}\ 2
\end{equation}
\textcolor{black}{It has been shown in \cite{mtqc} for MSG81.33 ($P\bar{4}$) that this symmetry indicator relates to a 3D axion topological insulating phase.}

%\linebreak
We finally note that our results in this Appendix agree with general expressions for magnetic EBRs, compatibility relations and symmetry indicators in \cite{mtqc}, which we became aware of during the finalizing stages of this manuscript.\\

\bibliography{references}

\end{document}